\documentstyle[12pt]{report}
\newtheorem{theorem}{Theorem}
\newtheorem{proposition}{Proposition}
\newtheorem{definition}{Definition}

\begin{document}

\date{October 27, 1998}

%{-} \vskip180pt
\centerline{\Large \bf GAUGE GRAVITY  }
\vskip4pt \centerline{\Large \bf AND CONSERVATION LAWS }
\vskip4pt \centerline{\Large \bf IN HIGHER ORDER  }
\vskip4pt \centerline{\Large \bf ANISOTROPIC SPACES} \vskip%
40pt \centerline{\large \sf Sergiu I. Vacaru} \vskip18pt
\centerline{\noindent{\em Institute of Applied Physics, Academy of Sciences,}}
\centerline{\noindent{\em 5 Academy str., Chi\c sin\v au 2028,
Republic of Moldova}} \vskip8pt
\centerline{\noindent{ Fax: 011-3732-738149, E-mail: vacaru@lises.as.md}}
\vskip15pt
%\newpage
%{-}
%\vskip140pt
 { {\bf Abstract.}
 We propose an approach to the theory of higher order anisotropic
field interactions and  curved spaces (in brief, ha--field,
 ha--in\-te\-rac\-ti\-ons and
ha--spaces). The concept of ha--space generaliSes various types of Lagrange
and Finsler spaces and higher dimension (Kaluza--Klein) spaces. This work
consists from two parts. In the first we outline the theory of Yang--Mills
ha--fields and two gauge models of higher order anisotropic gravity are
analyzed. The second is devoted to the theory of nearly autoparallel maps
(na--maps) of locally anisotropic spaces (la--spaces) and to the problem of
 formulation of conservation laws for la--field interactions.
 By defining invariants of na--map
transforms we present a systematic classification of la--spaces.}

\vskip15pt
%\vskip30pt {\sl Short title:}
%{\sf Anisotropic Gauge Fields and Conservation Laws}
\newpage
\tableofcontents
%\newpage
\section{Introduction}

\subsubsection*{Introductory remarks}

This paper presents some applications of the higher order vector bundle
formalism in physics: the theory of higher order anisotropic interactions of
Yang Mills fields is formulated, there are developed two models of higher
order anisotropic gauge gravity and found solutions of field equations
describing anisotropic gravitational instantons and there are proposed two
variants of definition the conservation laws for field interactions on
 locally anisotropic spaces.

This work as well fuses several areas of modern differential geometry, not
all of which are familiar to theoretical and mathematical physicists. Here
we cite our previous papers on nonlinear connections in vector
(super)bundles \cite{VacaruSupergravity}, the Sinyukov's theory of nearly
geodesic maps \cite{sin} and the theory of nearly autoparallel maps
(generalizing the class of conformal transforms) of Einstein--Cartan--Weyl
spaces \cite{vk,vb12,vrjp} and of generalized Lagrange and Finsler spaces
\cite{voarm,voa,vcl96}. We emphasize that we are inspired by the
Yano--Ishihara \cite{YanoIshihara} and Miron--Anastasiei--Atanasiu \cite
{MironAnastasiei1987,MironAnastasiei1994,MironAtanasiu} geometric ideas and
we consider that their constructions on modeling of locally anisotropic
spaces on (in general higher order) tangent and vector bundles appears to be
particularly promising in elaboration of some new divisions of quantum field
theory and gravity.

We follow our program \cite{VacaruJMP,VacaruGoncharenko,VacaruAP,VacaruNP}
of formulation of self--consistent field theories incorporating various
possible anisotropic, inhomogeneous and stochastic manifestations of
classical and quantum interactions on locally anisotropic and higher order
anisotropic spaces (respectively, in brief, la- and ha--spaces). To evolve
this new type of physical theories we shall also use the geometric
background and methods developed in monographs \cite
{YanoIshihara,Matsumoto,MironAnastasiei1987,MironAnastasiei1994,
AntonelliMiron,Bejancu,Asanov,Asanov1988} and papers \cite{MironAtanasiu,
Ikeda1994}.

\subsubsection*{Synopsis}

An approach to the theory of higher order anisotropic superspaces and
superstrings and an analysis of low energy dynamics of supersymmetric
nonlinear and anisotropic sigma models have been recently elaborated in our
works \cite{VacaruNP,VacaruSupergravity,VacaruAP}.
 This contribution is in a series of papers in which we more
 specifically focus on the (non
supersymmetric) field theory of higher order anisotropic interactions. In
our works \cite{Vacaru1998,VacaruM} we defined higher order anisotropic
spinors on ha--spaces (in brief, ha--spinors) and presented a detailed study
of the relationship between Clifford, spinor and nonlinear and distinguished
connections structures on higher order extensions of vector bundles and
tangent bundles and on prolongations of generalized Lagrange and Finsler
spaces \cite{MironAtanasiu}. As a next step (in this paper) we consider the
topics of higher order anisotropic gauge field interactions and of
definition of conservation laws on locally anisotropic spaces.

By convention the contents of this paper can be devided into two parts:
gauge ha--models and nearly autoparallel maps.

The aim of the first part is twofold. The first objective is to extend our
results on locally anisotropic gauge theories \cite
{VacaruGoncharenko,v295a,v295b} in order to consider Yang-Mills fields (with
semisimple structural groups) on spaces with higher order anisotropy. The
second objective is to propose a geometric formalism for gauge theories with
nonsemisimple structural groups which permit a unique fiber bundle treatment
for both higher order anisotropic Yang--Mills field and gravitational
interactions. In general lines, we shall follow the ideas and methods
proposed in Refs. \cite{ts,p,pd,pon,bis}, but we shall apply them in a form
convenient for introducing into consideration higher order anisotropic
physical theories.

There is a number of works on gauge models of interactions on Finsler spaces
\cite{Finsler,car35,run} and theirs extensions (see, for instance, \cite
{Asanov,Asanov1988,Asanov1989,bei,mirbal}). One has introduced different
variants of generalized gauge transforms, postulated corresponding
Lagrangians for gravitational, gauge and matter field interactions and
formulated variational calculus (here we note the approaches developed by A.
Bejancu \cite{bej89h,bej91a,Bejancu} and Gh. Munteanu and S. Ikeda \cite
{MunteanuIkeda}). The main problem of such models is the dependence of the
basic equations on chosen definition of gauge "compensation" symmetries and
on type of space and field interactions anisotropy. In order to avoid the
ambiguities connected with particular characteristics of possible ha--gauge
theories we consider a "pure" geometric approach to gauge theories (on both
locally isotropic and anisotropic spaces) in the framework of the theory of
fiber bundles provided in general with different types of nonlinear and
linear multiconnection and metric structures. This way developed by using
global geometric methods holds also good for nonvariational, in the total
spaces of bundles, gauge theories (in the case of gauge gravity one
considers the Poincare or affine gauge groups); physical values and motion
(field) equations have adequate geometric interpretation and do not depend
on the type of local anisotropy of space--time background.

The elaboration of models with higher order anisotropic field interactions
entails great difficulties because of problematical character of the
possibility and manner of definition of conservation laws on ha--spaces. It
will be recalled that, for instance, in special relativity the conservation
laws of energy--momentum type are defined by the global group (the Poincare
group) of automorphisms of the fundamental Mikowski spaces. For
(pseudo)Riemannian spaces one has tangent space's automorphisms and for
particular cases there are symmetries generated by Killing vectors. No global
or local automorphisms exist on generic ha--spaces and in result of this
fact the formulation of ha--conservation laws is sophisticate and full of
ambiguities. R. Miron and M. Anastasiei firstly pointed out the nonzero
divergence of the matter energy--momentum d--tensor, the source in Einstein
equations on la--spaces, and considered an original approach to the geometry
of time--dependent Lagrangians \cite
{anask,MironAnastasiei1987,MironAnastasiei1994}. Nevertheless, the rigorous
definition of energy--momentum values for locally anisotropic gravitational
and matter fields and the form of conservation laws for such values have not
been considered in present--day studies of the mentioned problem.

The aim of the second part of this paper is to develop a necessary geometric
background (the theory of nearly autoparallel maps,in brief na--maps, and
tensor integral formalism on  multispaces) for
formulation and a detailed investigation of conservation laws on locally
isotropic and anisotropic curved spaces.For simplicity,the explicit
 constructions will be presented only for local anisotropies
 the transition to higher order anisotropies being considered
 as straightforward extensions.We shall develop for generic
 locally anisotropic spaces our previous results on the theory of na--maps for
generalized affine spaces \cite{vk,vth,vob12},Einstein-Cartan and Einstein
spaces \cite{vrjp,vod,voarm},fibre bundles \cite{vb12,vog} and different
subclasses of la--spaces \cite{vodg,voa,gv,vcl96}.

The question of definition of tensor integration as the inverse operation of
covariant derivation was posed and studied by A.Mo\'or \cite{moo}.
Tensor--integral and bitensor formalisms turned out to be very useful in
solving certain problems connected with conservation laws in general
relativity \cite{goz,vk}. In order to extend tensor--integral constructions
we have proposed \cite{vob12,vog} to take into consideration nearly
autoparallel \cite{voarm,vk,vb12} and nearly geodesic \cite{sin} maps (in
brief, we shall write ng--maps, ng--theory) which forms a subclass of local
1--1 maps of curved spaces with deformation of the connection and metric
structures. A generalization of the Sinyukov's ng--theory for spaces with
local anisotropy was proposed by considering maps with deformation of
connection for Lagrange spaces (on Lagrange spaces see \cite
{ker,MironAnastasiei1987,MironAnastasiei1994}) and generalized Lagrange
spaces \cite{vcl96,vodg,voa,vod,gvd}. Tensor integration formalism for
generalized Lagrange spaces was developed in \cite{v295a,gv,vcl96}. One of
the main purposes of this work is to synthesize the results on na--maps and
multispace tensor integrals and to reformulate them for a very general class
of locally anisotropic spaces. As the final step the problem of formulation
of conservation laws on spaces with local anisotropy and definition
of enegry--momentum type value for la--gravity is considered.

\subsubsection*{Outline of the article}

The paper is organized as follows:\ Section 2 contains a brief summary of
the geometry of higher order anisotropic spaces. In Section 3 we formulate
the theory of gauge (Yang-Mills) fields on generic ha--spaces and define a
higher order anisotropic variant of Yang-Mills equations; the
variational proof of gauge field equations is considered in connection with
the "pure" geometrical method of definition of field equations. In Section 4
the ha--gravity is reformulated as a gauge theory for nonsemisimple groups.
A model of nonlinear de Sitter gauge gravity with higher order anisotropy is
formulated in Section 5. We study the gravitational gauge instantons with
trivial local anisotropy in Section 6. Section 7 is devoted to the
  theory of nearly autoparallel maps of la--spaces. The
classification of na--maps and corresponding invariant conditions are given
in Section 8. In Section 9 we define the nearly autoparallel
tensor--integral on locally anisotropic multispaces. The problem of
formulation of conservation laws on spaces with local anisotropy is studied
in Section 10. We present a definition of conservation laws for
la--gravitational fields on na--images of la--spaces in Section 11.
 Concluding remarks are given in Section 12.

\section{Higher Order Anisotropic Spaces}

In this section we present the necessary results on higher order vector
bundles provided with nonlinear and distinguished connections and metric
structures which are used for modeling of spaces with higher order
anisotropy. The  denotations we are following are that from
 \cite{Vacaru1998,VacaruM}.

As a geometric background for our further constructions we use a locally
trivial distinguished vector bundle, dv--bundle, ${\cal E}^{<z>} =$ $%
( E^{<z>},p,M,Gr,$ $F^{<z>}) $ where $F^{<z>}={\cal R}^{m_1}\oplus
...\oplus {\cal R}^{m_z}$ (a real vector space of dimension $%
m=m_1+...+m_z,\dim F=m,$ ${\cal R\ }$ denotes the real number field) is the
typical fibre, the structural group is chosen to be the group of
automorphisms of ${\cal R}^m$ , i.e. $Gr=GL\left( m,{\cal R}\right) ,$ and
$p:E^{<z>}\rightarrow M$ (defined by intermediary projections
 $$
p_{<z,z-1>}:E^{<z>}\rightarrow E^{<z-1>},p_{<z-1,z-2>}:E^{<z-1>}\rightarrow
E^{<z-2>},...$$
  $$p:E^{<1>}\rightarrow M)$$
 is a differentiable surjection of a
differentiable manifold $E$ (total space, $\dim E=n+m)$ to a differentiable
manifold $M$ (base space, $\dim M=n).$ Local coordinates on ${\cal E}^{<z>}$
are denoted as
$$
u^{<\alpha {\bf >}}=\left( x^i,y^{<a{\bf >\ }}\right) =\left( x^i\doteq
y^{a_0},y^{a_1},....,y^{a_z}\right)$$ $$
=(...,y^{a_{(p)}},...)=\{y_{(p)}^{a_{(p)}}\}=\{y^{a_{(p)}}\},
$$

A local coordinate parametrization of ${\cal E}^{<z>}$ naturally defines a
coordinate basis of the module of d--vector fi\-elds $\Xi \left( {\cal E}%
^{<z>}\right) ,$%
$$
\partial _{<\alpha >}=(\partial _i,\partial _{<a>})=(\partial _i,\partial
_{a_1},...,\partial _{a_p},...,\partial _{a_z})=\eqno(1)
$$
$$
\frac \partial {\partial u^{<\alpha >}}=\left( \frac \partial {\partial
x^i},\frac \partial {\partial y^{<a>}}\right) =\left( \frac \partial
{\partial x^i},\frac \partial {\partial y^{a_1}},...\frac \partial {\partial
y^{a_p}},...,\frac \partial {\partial y^{a_z}}\right) ,
$$
and the reciprocal to (1) coordinate basis
$$
d^{<\alpha >}=(d^i,d^{<a>})=(d^i,d^{a_1},...,d^{a_p},...,d^{a_z})=\eqno(2)
$$
$$
du^{<\alpha >}=(dx^i,dy^{<a>})=(dx^i,dy^{a_1},...,dy^{a_p},...,dy^{a_z}),
$$
which is uniquely defined from the equations%
$$
d^{<\alpha >}\circ \partial _{<\beta >}=\delta _{<\beta >}^{<\alpha >},
$$
where $\delta _{<\beta >}^{<\alpha >}$ is the Kronecher symbol and by ''$%
\circ $$"$ we denote the inner (scalar) product in the tangent bundle ${\cal %
TE}^{<z>}.$

A {\bf nonlinear connection,} in brief, N--connection, (see a detailed study
and basic references in
\cite{MironAnastasiei1987,MironAnastasiei1994,MironAtanasiu})
 in a dv--bundle ${\cal E}%
^{<z>}$ can be defined as a distribution $\{N:E_u\rightarrow
H_uE,T_uE=H_uE\oplus V_u^{(1)}E\oplus ...\oplus V_u^{(p)}E...\oplus
V_u^{(z)}E\}$ on $E^{<z>}$ being a global decomposition, as a Whitney sum,
into horizontal,${\cal HE,\ }$ and vertical, ${\cal VE}^{<p>}{\cal ,}%
p=1,2,...,z$ subbundles of the tangent bundle ${\cal TE:}$
$$
{\cal TE}=H{\cal E}\oplus V{\cal E}^{<1>}\oplus ...\oplus V{\cal E}%
^{<p>}\oplus ...\oplus V{\cal E}^{<z>}.
$$

Locally a N-connection in ${\cal E}^{<z>}$ is given by it components $N_{<%
{\bf a}_f>}^{<{\bf a}_p>}({\bf u}),z\geq p>f\geq 0$ (in brief we shall write
$N_{<a_f>}^{<a_p>}(u)$ ) with respect to bases (1) and (2)):

$$
{\bf N}=N_{<a_f>}^{<a_p>}(u)\delta ^{<a_f>}\otimes \delta _{<a_p>},(z\geq
p>f\geq 0),
$$

To coordinate locally geometric constructions with the global splitting of $%
{\cal E}^{<z>}$ defined by a N-connection structure, we have to introduce a
lo\-cal\-ly adap\-ted bas\-is ( la---basis, la---frame ):%
$$
\delta _{<\alpha >}=(\delta _i,\delta _{<a>})=(\delta _i,\delta
_{a_1},...,\delta _{a_p},...,\delta _{a_z}),\eqno(3)
$$
with components parametrized as

$$
\delta _i=\partial _i-N_i^{a_1}\partial _{a_1}-...-N_i^{a_z}\partial _{a_z},
$$
$$
\delta _{a_1}=\partial _{a_1}-N_{a_1}^{a_2}\partial
_{a_2}-...-N_{a_1}^{a_z}\partial _{a_z},
$$
$$
................
$$
$$
\delta _{a_p}=\partial _{a_p}-N_{a_p}^{a_{p+1}}\partial
_{a_{p+1}}-...-N_{a_p}^{a_z}\partial _{a_z},
$$
$$
...............
$$
$$
\delta _{a_z}=\partial _{a_z}
$$
and it dual la--basis%
$$
\delta ^{<\alpha >}=(\delta ^i,\delta ^{<a>})=\left( \delta ^i,\delta
^{a_1},...,\delta ^{a_p},...,\delta ^{a_z}\right) ,\eqno(4)
$$
$$
\delta x^i=dx^i,
$$
$$
\delta y^{a_1}=dy^{a_1}+M_i^{a_1}dx^i,
$$
$$
\delta y^{a_2}=dy^{a_2}+M_{a_1}^{a_2}dy^{a_1}+M_i^{a_2}dx^i,
$$
$$
.................
$$
$$
\delta
y^{a_p}=dy^{a_p}+M_{a_{p-1}}^{a_p}dy^{p-1}+M_{a_{p-2}}^{a_p}dy^{a_{p-2}}+...+M_i^{a_p}dx^i,
$$
$$
...................
$$
$$
\delta
y^{a_z}=dy^{a_z}+M_{a_{z-1}}^{a_z}dy^{z-1}+M_{a_{z-2}}^{a_z}dy^{a_{z-2}}+...+M_i^{a_z}dx^i.
$$

The interrelation between $N-$ and $M-$ coefficients from (3) and (4) is
considered in \cite{Vacaru1998,VacaruM}.

We emphasize that on a dv--bundle ${\cal E}^{<z>}\,$the higher order
anisotropic operators (3) and (4) substitutes respectively the local
operators of partial derivation (1) and of differentials (2).

The {\bf algebra of tensorial distinguished fields} $DT\left( {\cal E}%
^{<z>}\right) $ (d--fields, d--tensors, d--objects) on ${\cal E}^{<z>}$ is
introduced as the tensor algebra\\ ${\cal T}=\{{\cal T}%
_{qs_1...s_p...s_z}^{pr_1...r_p...r_z}\}$ of the dv-bundle ${\cal E}_{\left(
d\right) }^{<z>},$
$$
p_d:{\cal HE}^{<z>}{\cal \oplus V}^1{\cal E}^{<z>}{\oplus }...{\oplus V}^p%
{\cal E}^{<z>}{\oplus }...{\oplus V}^z{\cal E}^{<z>}{\cal \rightarrow E}%
^{<z>},
$$
${\ }$ An element ${\bf t}\in {\cal T}_{qs_1...s_z}^{pr_1...r_z},$ d-tensor
field of type $\left(
\begin{array}{cccccc}
p & r_1 & ... & r_p & ... & r_z \\
q & s_1 & ... & s_p & ... & s_z
\end{array}
\right) ,$ is written in local form as%
$$
{\bf t}%
=t_{j_1...j_qb_1^{(1)}...b_{r_1}^{(1)}...b_1^{(p)}...b_{r_p}^{(p)}...b_1^{(z)}...b_{r_z}^{(z)}}^{i_1...i_pa_1^{(1)}...a_{r_1}^{(1)}...a_1^{(p)}...a_{r_p}^{(p)}...a_1^{(z)}...a_{r_z}^{(z)}}\left( u\right) \delta _{i_1}\otimes ...\otimes \delta _{i_p}\otimes d^{j_1}\otimes ...\otimes d^{j_q}\otimes
$$
$$
\delta _{a_1^{(1)}}\otimes ...\otimes \delta _{a_{r_1}^{(1)}}\otimes \delta
^{b_1^{(1)}}...\otimes \delta ^{b_{s_1}^{(1)}}\otimes ...\otimes \delta
_{a_1^{(p)}}\otimes ...\otimes \delta _{a_{r_p}^{(p)}}\otimes ...\otimes
$$
$$
\delta ^{b_1^{(p)}}...\otimes \delta ^{b_{s_p}^{(p)}}\otimes \delta
_{a_1^{(z)}}\otimes ...\otimes \delta _{a_{rz}^{(z)}}\otimes \delta
^{b_1^{(z)}}...\otimes \delta ^{b_{s_z}^{(z)}}.
$$

One use respectively denotations $X{\cal (E}^{<z>})$ (or $X{\left( M\right)
),\ }\Lambda ^p\left( {\cal E}^{<z>}\right) $ (or $\Lambda ^p\left( M\right)
)$ and ${\cal F(E}^{<z>})$ (or ${\cal F}$ $\left( M\right) $) for the module
of d-vector fields on ${\cal E}^{<z>}$ (or $M$ ), the exterior algebra of
p-forms on ${\cal E}^{<z>}$ (or $M)$ and the set of real functions on ${\cal %
E}^{<z>}$(or $M).$

The geometric objects with various group and coordinate transforms
coordinated with the N--connection structure on ${\cal E}^{<z>}$ are called
in brief d--objects on ${\cal E}^{<z>}.$

For a N-connection structure in ${\cal E}^{<z>}$ it is defined a
corresponding decomposition of d-tensors into sums of horizontal and
vertical parts, for example, for every d-vector $X\in {\cal X(E}^{<z>})$ and
1-form $\widetilde{X}\in \Lambda ^1\left( {\cal E}^{<z>}\right) $ we have
respectively
$$
X=hX+v_1X+...+v_zX{\bf \ \quad }\mbox{and \quad }\widetilde{X}=h\widetilde{X}%
+v_1\widetilde{X}+...~v_z\widetilde{X}.
$$
In consequence, we can associate to every d-covariant derivation along a
d-vector $X,$\ $D_X=X\circ D,$ two new operators of h- and v-covariant
derivations defined respectively as
$$
D_X^{(h)}Y=D_{hX}Y
$$
and
$$
D_X^{\left( v_1\right) }Y=D_{v_1X}Y{\bf ,...,D_X^{\left( v_z\right)
}Y=D_{v_zX}Y\quad }\forall Y{\bf \in }{\cal X(E}^{<z>}),
$$
for which the following conditions hold:%
$$
D_XY{\bf =}D_X^{(h)}Y+D_X^{(v_1)}Y+...+D_X^{(v_z)}Y,
$$
$$
D_X^{(h)}f=(hX{\bf )}f
$$
and
$$
D_X^{(v_p)}f=(v_pX{\bf )}f,\quad X,Y{\bf \in }{\cal X\left( E\right) ,}f\in
{\cal F}\left( M\right) ,p=1,2,...z.
$$

A {\bf metric structure }${\bf G\ }$ in the total space $E^{<z>}$ of a
dv--bundle ${\cal E}^{<z>} = $ $\left( E^{<z>},p,M\right) $ over a
connected and paracompact base $M$ is introduced as a symmetrical covariant
tensor field of type $\left( 0,2\right) $, $G_{<\alpha ><\beta >,}$ being
nondegenerate and of constant signature on $E^{<z>}.$

Nonlinear connection ${\bf N}$ and metric ${\bf G}$ structures on ${\cal E}%
^{<z>}$ are mutually compatible it there are satisfied the conditions:
$$
{\bf G}\left( \delta _{a_f},\delta _{a_p}\right) =0,\mbox{or equivalently, }%
G_{a_fa_p}\left( u\right) -N_{a_f}^{<b>}\left( u\right) h_{a_f<b>}\left(
u\right) =0,
$$
where $h_{a_pb_p}={\bf G}\left( \partial _{a_p},\partial _{b_p}\right) $ and
$G_{b_fa_p}={\bf G}\left( \partial _{b_f},\partial _{a_p}\right) ,0\leq
f<p\leq z,\,$ which gives%
$$
N_{c_f}^{b_p}\left( u\right) =h^{<a>b_p}\left( u\right) G_{c_f<a>}\left(
u\right)
$$
(the matrix $h^{a_pb_p}$ is inverse to $h_{a_pb_p}).$ With respect to
la---basis (4) a d--metric is written as%
$$
{\bf G}=g_{<\alpha ><\beta >}\left( u\right) \delta ^{<\alpha >}\otimes
\delta ^{<\beta >}=g_{ij}\left( u\right) d^i\otimes d^j+h_{<a><b>}\left(
u\right) \delta ^{<a>}\otimes \delta ^{<b>},\eqno(5)
$$
where $g_{ij}={\bf G}\left( \delta _i,\delta _j\right) .$

The torsion ${\bf T}$ of a d--connection ${\bf D\ }$ in ${\cal E}^{<z>}$ is
defined by the equation%
$$
{\bf T\left( X,Y\right) =XY_{\circ }^{\circ }T\doteq }D_X{\bf Y-}D_Y{\bf X\
-\left[ X,Y\right] .}
$$
One holds the following h- and v$_{(p)}$--decompositions%
$$
{\bf T\left( X,Y\right) =T\left( hX,hY\right) +T\left( hX,vY\right) +T\left(
vX,hY\right) +T\left( vX,vY\right) .}
$$
We consider the projections: ${\bf hT\left( X,Y\right) ,v}_{(p)}{\bf T\left(
hX,hY\right) ,hT\left( hX,hY\right) ,...}$ and say that, for instance, ${\bf %
hT\left( hX,hY\right) }$ is the h(hh)-torsion of ${\bf D},$ \\ ${\bf v}_{(p)}%
{\bf T\left( hX,hY\right) \ }$ is the v$_p$(hh)-torsion of ${\bf D}$ and so
on.

A torsion ${\bf T\left( X,Y\right) }$ is determined by local d--tensor
fields, torsions, defined as
$$
T_{jk}^i={\bf hT}\left( \delta _k,\delta _j\right) \cdot d^i,\quad
T_{jk}^{a_p}={\bf v}_{(p)}{\bf T}\left( \delta _k,\delta _j\right) \cdot
\delta ^{a_p},
$$
$$
P_{jb_p}^i={\bf hT}\left( \delta _{b_p},\delta _j\right) \cdot d^i,\quad
P_{jb_f}^{a_p}={\bf v}_{(p)}{\bf T}\left( \delta _{b_f},\delta _j\right)
\cdot \delta ^{a_p},
$$
$$
S_{b_fc_f}^{a_p}={\bf v}_{(p)}{\bf T}\left( \delta _{c_f},\delta
_{b_f}\right) \cdot \delta ^{a_p}.
$$
and with components computed in the form (see \cite{MironAtanasiu} and Paper
I):
$$
T_{.jk}^i=T_{jk}^i=L_{jk}^i-L_{kj}^i,\quad
T_{j<a>}^i=C_{.j<a>}^i,T_{<a>j}^i=-C_{j<a>}^i,\eqno(6)
$$
$$
T_{.j<a>}^i=0,T_{.<b><c>}^{<a>}=S_{.<b><c>}^{<a>}=C_{<b><c>}^{<a>}-C_{<c><b>}^{<a>},
$$
$$
T_{.b_fc_f}^{a_p}=\frac{\delta N_{c_f}^{a_p}}{\partial y^{b_f}}-\frac{\delta
N_{b_f}^{a_p}}{\partial y^{c_f}},T_{.<b>i}^{<a>}=P_{.<b>i}^{<a>}=\frac{%
\delta N_i^{<a>}}{\partial y^{<b>}}%
-L_{.<b>j}^{<a>},T_{.i<b>}^{<a>}=-P_{.<b>i}^{<a>}.
$$

The curvature ${\bf R}$ of d--connection in ${\cal E}^{<z>}$ is introduced
as an operator
$$
{\bf R\left( X,Y\right) Z=XY_{\bullet }^{\bullet }R\bullet Z}=D_XD_Y{\bf Z}%
-D_YD_X{\bf Z-}D_{[X,Y]}{\bf Z.}
$$
One holds the next properties for the h- and v-decompositions of curvature:%
$$
{\bf v}_{(p)}{\bf R\left( X,Y\right) hZ=0,\ hR\left( X,Y\right) v}_{(p)}{\bf %
Z=0,~v}_{(f)}{\bf R\left( X,Y\right) v}_{(p)}{\bf Z=0,}
$$
$$
{\bf R\left( X,Y\right) Z=hR\left( X,Y\right) hZ+vR\left( X,Y\right) vZ,}
$$
where ${\bf v=v}_1+...+{\bf v}_z.$ The curvature of a d-con\-nec\-ti\-on $%
{\bf D}$ in ${\cal E}^{<z>}$ is completely determined by the following
d-tensor fields (see \cite{MironAtanasiu} and \cite{Vacaru1998,VacaruM}):%
$$
R_{h.jk}^{.i}=\delta ^i\cdot {\bf R}\left( \delta _k,\delta _j\right) \delta
_h,~R_{<b>.jk}^{.<a>}=\delta ^{<a>}\cdot {\bf R}\left( \delta _k,\delta
_j\right) \delta _{<b>},
$$
$$
P_{j.k<c>}^{.i}=d^i\cdot {\bf R}\left( \delta _{<c>},\delta _{<k>}\right)
\delta _j,~P_{<b>.<k><c>}^{.<a>}=\delta ^{<a>}\cdot {\bf R}\left( \delta
_{<c>},\delta _{<k>}\right) \delta _{<b>},
$$
$$
S_{j.<b><c>}^{.i}=d^i\cdot {\bf R}\left( \delta _{<c>},\delta _{<b>}\right)
\delta _j,~S_{<b>.<c><d>}^{.<a>}=\delta ^{<a>}\cdot {\bf R}\left( \delta
_{<d>},\delta _{<c>}\right) \delta _{<b>}.
$$
with components
$$
R_{h.jk}^{.i}=\frac{\delta L_{.hj}^i}{\delta x^h}-\frac{\delta L_{.hk}^i}{%
\delta x^j}+L_{.hj}^mL_{mk}^i-L_{.hk}^mL_{mj}^i+C_{.h<a>}^iR_{.jk}^{<a>},%
\eqno(7)
$$
$$
R_{<b>.jk}^{.<a>}=\frac{\delta L_{.<b>j}^{<a>}}{\delta x^k}-\frac{\delta
L_{.<b>k}^{<a>}}{\delta x^j}%
+L_{.<b>j}^{<c>}L_{.<c>k}^{<a>}-L_{.<b>k}^{<c>}L_{.<c>j}^{<a>}+C_{.<b><c>}^{<a>}R_{.jk}^{<c>},
$$
$$
P_{j.k<a>}^{.i}=\frac{\delta L_{.jk}^i}{\partial y^{<a>}}%
+C_{.j<b>}^iP_{.k<a>}^{<b>}-
$$
$$
\left( \frac{\partial C_{.j<a>}^i}{\partial x^k}%
+L_{.lk}^iC_{.j<a>}^l-L_{.jk}^lC_{.l<a>}^i-L_{.<a>k}^{<c>}C_{.j<c>}^i\right)
,
$$
$$
P_{<b>.k<a>}^{.<c>}=\frac{\delta L_{.<b>k}^{<c>}}{\partial y^{<a>}}%
+C_{.<b><d>}^{<c>}P_{.k<a>}^{<d>}-
$$
$$
\left( \frac{\partial C_{.<b><a>}^{<c>}}{\partial x^k}+L_{.<d>k}^{<c>%
\,}C_{.<b><a>}^{<d>}-L_{.<b>k}^{<d>}C_{.<d><a>}^{<c>}-L_{.<a>k}^{<d>}C_{.<b><d>}^{<c>}\right) ,
$$
$$
S_{j.<b><c>}^{.i}=\frac{\delta C_{.j<b>}^i}{\partial y^{<c>}}-\frac{\delta
C_{.j<c>}^i}{\partial y^{<b>}}+C_{.j<b>}^hC_{.h<c>}^i-C_{.j<c>}^hC_{h<b>}^i,
$$
$$
S_{<b>.<c><d>}^{.<a>}=\frac{\delta C_{.<b><c>}^{<a>}}{\partial y^{<d>}}-%
\frac{\delta C_{.<b><d>}^{<a>}}{\partial y^{<c>}}+
$$
$$
C_{.<b><c>}^{<e>}C_{.<e><d>}^{<a>}-C_{.<b><d>}^{<e>}C_{.<e><c>}^{<a>}.
$$

The components of the Ricci d--tensor
$$
R_{<\alpha ><\beta >}=R_{<\alpha >.<\beta ><\tau >}^{.<\tau >}
$$
with respect to a locally adapted frame (4) are as follows:%
$$
R_{ij}=R_{i.jk}^{.k},\quad R_{i<a>}=-^2P_{i<a>}=-P_{i.k<a>}^{.k},\eqno(8)
$$
$$
R_{<a>i}=^1P_{<a>i}=P_{<a>.i<b>}^{.<b>},\quad
R_{<a><b>}=S_{<a>.<b><c>}^{.<c>}.
$$
We note that on ha--spaces the Ricci d--tensor is non symmetrical because, in
general, $^1P_{<a>i}\neq ~^2P_{i<a>}$.

The scalar curvature of d--connection ${\bf D}$ in ${\cal E}^{<z>}$ is
introduced by using a d--metric of type (5) :
$$
{\overleftarrow{R}}=G^{<\alpha ><\beta >}R_{<\alpha ><\beta >}=R+S,\eqno(9)
$$
where $R=g^{ij}R_{ij}$ and $S=h^{<a><b>}S_{<a><b>}.$

The Einstein equations on ha--spaces are postulated in a standard manner
$$
\overleftarrow{G}_{<\alpha ><\beta >}+\lambda g_{<\alpha ><\beta >}=\kappa
E_{<\alpha ><\beta >},\eqno(10)
$$
where%
$$
\overleftarrow{G}_{<\alpha ><\beta >}=R_{<\alpha ><\beta >}-\frac 12%
\overleftarrow{R}g_{<\alpha ><\beta >}
$$
is the Einstein d--tensor, $\lambda $ and $\kappa $ are correspondingly the
cosmological and gravitational constants and by $E_{<\alpha ><\beta >}$ is
denoted the locally anisotropic energy--momentum d--tensor. By using
 the formulas (6)--(9) we can write down in explicit form
 the h-  and $v_{(p)}$--components of the Einstein equations (10) (we shall
 not consider such formulas in this paper).

Because ha--spaces generally have nonzero torsions we shall add to (10) a
system of algebraic d--field equations with the source $S_{~<\beta ><\gamma
>}^{<\alpha >}$ being the locally anisotropic spin density of matter (if we
consider a variant of higher order anisotropic Einstein--Cartan theory ):
$$
T_{~<\alpha ><\beta >}^{<\gamma >}+2\delta _{~[<\alpha >}^{<\gamma
>}T_{~<\beta >]<\delta >}^{<\delta >}=\kappa S_{~<\alpha ><\beta
>.}^{<\gamma >}
$$
 In a more general case we must introduced some constraints on torsions
 and nonlinear connections induced from string theory \cite{VacaruAP},
 locally (and/ or higher order) anisotropic supergravity
 \cite{VacaruNP,VacaruSupergravity}. A form of gauge dynamical field equations
 for torsions  will be considered in the Section 5 of this paper.

\section{Gauge Fields on Ha--Spaces}

This section presents a geometrical background for
gauge field theories on spaces with higher order anisotropy.

\subsection{Bundles on ha--spaces}

Let $\left( P,\pi ,Gr,{\cal E}^{<z>}\right) $ be a principal bundle ${\cal E}%
^{<z>}$ (being a ha-space) with structural group $Gr$ and surjective map $%
\pi :P\rightarrow {\cal E}^{<z>}. $At every point $u=\left(
x,y_{(1)},...,y_{(z)}\right) \in {\cal E}^{<z>}$ there is a vicinity ${\cal %
U\subset E}^{<z>}, u\in {\cal U,}$ with trivializing $P$
diffeomorphisms $f$ and $\varphi :$%
$$
f_{{\cal U}}:\pi ^{-1}\left( {\cal U}\right) \rightarrow {\cal U\times }%
Gr,\qquad f\left( p\right) =\left( \pi \left( p\right) ,\varphi \left(
p\right) \right) ,
$$
$$
\varphi _{{\cal U}}:\pi ^{-1}\left( {\cal U}\right) \rightarrow Gr,\varphi
(pq)=\varphi \left( p\right) q,\quad \forall q\in Gr,~p\in P.
$$
We remark that in the general case for two open regions%
$$
{\cal U,V}\subset {\cal E}^{<z>}{\cal ,U\cap V}\neq \emptyset ,f_{{\cal U|}%
_p}\neq f_{{\cal V|}_p},\mbox{ even }p\in {\cal U\cap V.}
$$

Transition functions $g_{{\cal UV}}$ are defined as
$$
g_{{\cal UV}}:{\cal U\cap V\rightarrow }Gr,g_{{\cal UV}}\left( u\right)
=\varphi _{{\cal U}}\left( p\right) \left( \varphi _{{\cal V}}\left(
p\right) ^{-1}\right) ,\pi \left( p\right) =u.
$$

Hereafter we shall omit, for simplicity, the specification of trivializing
regions of maps and denote, for example, $f\equiv f_{{\cal U}},\varphi
\equiv \varphi _{{\cal U}},$ $s\equiv s_{{\cal U}},$ if this will not give
rise to ambiguities.

Let $\theta \,$ be the canonical left invariant 1-form on $Gr$ with values
in algebra Lie ${\cal G}$ of group $Gr$ uniquely defined from the relation $%
\theta \left( q\right) =q,\forall q\in {\cal G,}$ and consider a 1-form $%
\omega $ on ${\cal U\subset E}^{<z>}$ with values in ${\cal G.}$ Using $%
\theta $ and $\omega ,$ we can locally define the connection form $\Omega $
in $P$ as a 1-form:%
$$
\Omega =\varphi ^{*}\theta +Ad~\varphi ^{-1}\left( \pi ^{*}\omega \right)
\eqno(11)
$$
where $\varphi ^{*}\theta $ and $\pi ^{*}\omega $ are, respectively, forms
induced on $\pi ^{-1}\left( {\cal U}\right) $ and $P$ by maps $\varphi $ and
$\pi $ and $\omega =s^{*}\Omega .$ The adjoint action on a form $\lambda $
with values in ${\cal G}$ is defined as
$$
\left( Ad~\varphi ^{-1}\lambda \right) _p=\left( Ad~\varphi ^{-1}\left(
p\right) \right) \lambda _p
$$
where $\lambda _p$ is the value of form $\lambda $ at point $p\in P.$

Introducing a basis $\{\Delta _{\widehat{a}}\}$ in ${\cal G}$ (index $%
\widehat{a}$ enumerates the generators making up this basis), we write the
1-form $\omega $ on ${\cal E}^{<z>}$ as
$$
\omega =\Delta _{\widehat{a}}\omega ^{\widehat{a}}\left( u\right) ,~\omega ^{%
\widehat{a}}\left( u\right) =\omega _{<\mu >}^{\widehat{a}}\left( u\right)
\delta u^{<\mu >}\eqno(12)
$$
where $\delta u^{<\mu >}=\left( dx^i,\delta y^{<a>}\right) $ and the
Einstein summation rule on indices $\widehat{a}$ and $<\mu >$ is used.
Functions $\omega _{<\mu >}^{\widehat{a}}\left( u\right) $ from (12) will
be called the components of Yang-Mills fields on ha-space ${\cal E}^{<z>}.$
 Gauge transforms of $\omega $ can be geometrically interpreted as
transition relations for $\omega _{{\cal U}}$ and $\omega _{{\cal V}},$ when
$u\in {\cal U\cap V,}$%
$$
\left( \omega _{{\cal U}}\right) _u=\left( g_{{\cal UV}}^{*}\theta \right)
_u+Ad~g_{{\cal UV}}\left( u\right) ^{-1}\left( \omega _{{\cal V}}\right) _u.%
\eqno(13)
$$

To relate $\omega _{<\mu >}^{\widehat{a}}$ with a covariant derivation we
shall consider a vector bundle $\Upsilon $ associated to $P.$ Let $\rho
:Gr\rightarrow GL\left( {\cal R}^s\right) $ and $\rho ^{\prime }:{\cal G}%
\rightarrow End\left( E^s\right) $ be, respectively, linear representations
of group $Gr$ and Lie algebra ${\cal G}$ (in a more general case we can
consider ${\cal C}^s$ instead of ${\cal R}^s).$ Map $\rho $ defines a left
action on $Gr$ and associated vector bundle $\Upsilon =P\times {\cal R}%
^s/Gr,~\pi _E:E\rightarrow {\cal E}^{<z>}{\cal .}$ Introducing the standard
basis $\xi _{\underline{i}}=\{\xi _{\underline{1}},\xi _{\underline{2}%
},...,\xi _{\underline{s}}\}$ in ${\cal R}^s,$ we can define the right
action on $P\times $ ${\cal R}^s,\left( \left( p,\xi \right) q=\left(
pq,\rho \left( q^{-1}\right) \xi \right) ,q\in Gr\right) ,$ the map induced
from $P$%
$$
p:{\cal R}^s\rightarrow \pi _E^{-1}\left( u\right) ,\quad \left( p\left( \xi
\right) =\left( p\xi \right) Gr,\xi \in {\cal R}^s,\pi \left( p\right)
=u\right)
$$
and a basis of local sections $e_{\underline{i}}:U\rightarrow \pi
_E^{-1}\left( U\right) ,~e_{\underline{i}}\left( u\right) =s\left( u\right)
\xi _{\underline{i}}.$ Every section $\varsigma :{\cal E}^{<z>}{\cal %
\rightarrow }\Upsilon $ can be written locally as $\varsigma =\varsigma
^ie_i,\varsigma ^i\in C^\infty \left( {\cal U}\right) .$ To every vector
field $X$ on ${\cal E}^{<z>}$ and Yang-Mills field $\omega ^{\widehat{a}}$
on $P$ we associate operators of covariant derivations:%
$$
\nabla _X\zeta =e_{\underline{i}}\left[ X\zeta ^{\underline{i}}+B\left(
X\right) _{\underline{j}}^{\underline{i}}\zeta ^{\underline{j}}\right]
$$
$$
B\left( X\right) =\left( \rho ^{\prime }X\right) _{\widehat{a}}\omega ^{%
\widehat{a}}\left( X\right) .\eqno(14)
$$
Transformation laws (13) and operators (14) are interrelated by these
transition transforms for values $e_{\underline{i}},\zeta ^{\underline{i}},$
and $B_{<\mu >}:$%
$$
e_{\underline{i}}^{{\cal V}}\left( u\right) =\left[ \rho g_{{\cal UV}}\left(
u\right) \right] _{\underline{i}}^{\underline{j}}e_{\underline{i}}^{{\cal U}%
},~\zeta _{{\cal U}}^{\underline{i}}\left( u\right) =\left[ \rho g_{{\cal UV}%
}\left( u\right) \right] _{\underline{i}}^{\underline{j}}\zeta _{{\cal V}}^{%
\underline{i}},
$$
$$
B_{<\mu >}^{{\cal V}}\left( u\right) =\left[ \rho g_{{\cal UV}}\left(
u\right) \right] ^{-1}\delta _{<\mu >}\left[ \rho g_{{\cal UV}}\left(
u\right) \right] +\left[ \rho g_{{\cal UV}}\left( u\right) \right]
^{-1}B_{<\mu >}^{{\cal U}}\left( u\right) \left[ \rho g_{{\cal UV}}\left(
u\right) \right] ,\eqno(15)
$$
where $B_{<\mu >}^{{\cal U}}\left( u\right) =B^{<\mu >}\left( \delta
/du^{<\mu >}\right) \left( u\right) .$

Using (15), we can verify that the operator $\nabla _X^{{\cal U}},$ acting
on sections of $\pi _\Upsilon :\Upsilon \rightarrow {\cal E}^{<z>}$
according to definition (14), satisfies the properties%
$$
\begin{array}{c}
\nabla _{f_1X+f_2Y}^{
{\cal U}}=f_1\nabla _X^{{\cal U}}+f_2\nabla _X^{{\cal U}},~\nabla _X^{{\cal U%
}}\left( f\zeta \right) =f\nabla _X^{{\cal U}}\zeta +\left( Xf\right) \zeta
, \\ \nabla _X^{{\cal U}}\zeta =\nabla _X^{{\cal V}}\zeta ,\quad u\in {\cal %
U\cap V,}f_1,f_2\in C^\infty \left( {\cal U}\right) .
\end{array}
$$

So, we can conclude that the Yang--Mills connection in the vector bundle $%
\pi _\Upsilon :\Upsilon \rightarrow {\cal E}^{<z>}$ is not a general one,
but is induced from the principal bundle $\pi :P\rightarrow {\cal E}^{<z>}$
with structural group $Gr.$

The curvature ${\cal K}$ of connection $\Omega $ from (11) is defined as%
$$
{\cal K}=D\Omega ,~D=\widehat{H}\circ d\eqno(16)
$$
where $d$ is the operator of exterior derivation acting on ${\cal G}$-valued
forms as\\ $d\left( \Delta _{\widehat{a}}\otimes \chi ^{\widehat{a}}\right)
=\Delta _{\widehat{a}}\otimes d\chi ^{\widehat{a}}$ and $\widehat{H}\,$ is
the horizontal projecting operator actin, for example, on the 1-form $%
\lambda $ as $\left( \widehat{H}\lambda \right) _P\left( X_p\right) =\lambda
_p\left( H_pX_p\right) ,$ where $H_p$ projects on the horizontal subspace
$$
{\cal H}_p\in P_p\left[ X_p\in {\cal H}_p\mbox{ is equivalent to }\Omega
_p\left( X_p\right) =0\right] .%
$$
We can express (16) locally as
$$
{\cal K}=Ad~\varphi _{{\cal U}}^{-1}\left( \pi ^{*}{\cal K}_{{\cal U}%
}\right) \eqno(17)
$$
where
$$
{\cal K}_{{\cal U}}=d\omega _{{\cal U}}+\frac 12\left[ \omega _{{\cal U}%
},\omega _{{\cal U}}\right] .\eqno(18)
$$
The exterior product of ${\cal G}$-valued form (18) is defined as
$$
\left[ \Delta _{\widehat{a}}\otimes \lambda ^{\widehat{a}},\Delta _{\widehat{%
b}}\otimes \xi ^{\widehat{b}}\right] =\left[ \Delta _{\widehat{a}},\Delta _{%
\widehat{b}}\right] \otimes \lambda ^{\widehat{a}}\bigwedge \xi ^{\widehat{b}%
},
$$
where the antisymmetric tensorial product is%
$$
\lambda ^{\widehat{a}}\bigwedge \xi ^{\widehat{b}}=\lambda ^{\widehat{a}}\xi
^{\widehat{b}}-\xi ^{\widehat{b}}\lambda ^{\widehat{a}}.
$$

Introducing structural coefficients $f_{\widehat{b}\widehat{c}}^{\quad
\widehat{a}}$ of ${\cal G}$ satisfying
$$
\left[ \Delta _{\widehat{b}},\Delta _{\widehat{c}}\right] =f_{\widehat{b}%
\widehat{c}}^{\quad \widehat{a}}\Delta _{\widehat{a}}
$$
we can rewrite (18) in a form more convenient for local considerations:%
$$
{\cal K}_{{\cal U}}=\Delta _{\widehat{a}}\otimes {\cal K}_{<\mu ><\nu >}^{%
\widehat{a}}\delta u^{<\mu >}\bigwedge \delta u^{<\nu >}\eqno(19)
$$
where%
$$
{\cal K}_{<\mu ><\nu >}^{\widehat{a}}=\frac{\delta \omega _{<\nu >}^{%
\widehat{a}}}{\partial u^{<\mu >}}-\frac{\delta \omega _{<\mu >}^{\widehat{a}%
}}{\partial u^{<\nu >}}+\frac 12f_{\widehat{b}\widehat{c}}^{\quad \widehat{a}%
}\left( \omega _{<\mu >}^{\widehat{b}}\omega _{<\nu >}^{\widehat{c}}-\omega
_{<\nu >}^{\widehat{b}}\omega _{<\mu >}^{\widehat{c}}\right) .
$$

This subsection ends by considering the problem of reduction of the local
an\-i\-sot\-rop\-ic gauge symmetries and gauge fields to isotropic ones. For
local trivial considerations we can consider that the vanishing of
dependencies on $y$ variables leads to isotropic Yang-Mills fields with the
same gauge group as in the anisotropic case, Global geometric constructions
require a more rigorous topological study of possible obstacles for
reduction of total spaces and structural groups on anisotropic bases to
their analogous on isotropic (for example, pseudo-Riemannian) base spaces.

\subsection{Yang-Mills equations on ha-spaces}

Interior gauge (nongravitational) symmetries are associated to semisimple
structural groups. On the principal bundle $\left( P,\pi ,Gr,{\cal E}%
^{<z>}\right) $ with nondegenerate Killing form for semisimple group $Gr$ we
can define the generalized Lagrange metric%
$$
h_p\left( X_p,Y_p\right) =G_{\pi \left( p\right) }\left( d\pi _PX_P,d\pi
_PY_P\right) +K\left( \Omega _P\left( X_P\right) ,\Omega _P\left( X_P\right)
\right) ,\eqno(20)
$$
where $d\pi _P$ is the differential of map $\pi :P\rightarrow {\cal E}^{<z>}%
{\cal ,}$ $G_{\pi \left( p\right) }$ is locally generated as the ha-metric
(5), and $K$ is the Killing form on ${\cal G:}$%
$$
K\left( \Delta _{\widehat{a}},\Delta _{\widehat{b}}\right) =f_{\widehat{b}%
\widehat{d}}^{\quad \widehat{c}}f_{\widehat{a}\widehat{c}}^{\quad \widehat{d}%
}=K_{\widehat{a}\widehat{b}}.
$$

Using the metric $G_{<\alpha ><\beta >}$ on ${\cal E}^{<z>}$ $\left[
h_P\left( X_P,Y_P\right) \mbox{ on }P\right] ,$ we can introduce operators $%
*_G$ and $\widehat{\delta }_G$ acting in the space of forms on ${\cal E}%
^{<z>}$ ($*_H$ and $\widehat{\delta }_H$ acting on forms on ${\cal E}^{<z>}%
{\cal ).}$ Let $e_{<\mu >}$ be orthonormalized frames on ${\cal U\subset E}%
^{<z>}$ and $e^{<\mu >}$ the adjoint coframes. Locally%
$$
G=\sum\limits_{<\mu >}\eta \left( <\mu >\right) e^{<\mu >}\otimes e^{<\mu
>},
$$
where $\eta _{<\mu ><\mu >}=\eta \left( <\mu >\right) =\pm 1,$ $<\mu
>=1,2,...,n_E,n_E=1,...,n+m_1+...+m_z,$ and the Hodge operator $*_G$ can be
defined as $*_G:\Lambda ^{\prime }\left( {\cal E}^{<z>}\right) \rightarrow
\Lambda ^{n+m_1+...+m_z}\left( {\cal E}^{<z>}\right) ,$ or, in explicit
form, as%
$$
*_G\left( e^{<\mu _1>}\bigwedge ...\bigwedge e^{<\mu _r>}\right) =\eta
\left( \nu _1\right) ...\eta \left( \nu _{n_E-r}\right) \times \eqno(21)
$$
$$
sign\left(
\begin{array}{ccccc}
1 & 2 & ...r & r+1 & ...n_E \\
<\mu _{1>} & <\mu _2> & ...<\mu _r> & <\nu _1> & ...\nu _{n_E-r}
\end{array}
\right) \times
$$
$$
e^{<\nu _1>}\bigwedge ...\bigwedge e^{<\nu _{n_E-r}>}.
$$
Next, define the operator%
$$
*_{G^{}}^{-1}=\eta \left( 1\right) ...\eta \left( n_E\right) \left(
-1\right) ^{r\left( n_E-r\right) }*_G
$$
and introduce the scalar product on forms $\beta _1,\beta _2,...\subset
\Lambda ^r\left( {\cal E}^{<z>}\right) $ with compact carrier:%
$$
\left( \beta _1,\beta _2\right) =\eta \left( 1\right) ...\eta \left(
n_E\right) \int \beta _1\bigwedge *_G\beta _2.
$$
The operator $\widehat{\delta }_G$ is defined as the adjoint to $d$
associated to the scalar product for forms, specified for $r$-forms as
$$
\widehat{\delta }_G=\left( -1\right) ^r*_{G^{}}^{-1}\circ d\circ *_G.%
\eqno(22)
$$

We remark that operators $*_H$ and $\delta _H$ acting in the total space of $%
P$ can be defined similarly to (21) and (22), but by using metric
(20). Both these operators also act in the space of ${\cal G}$-valued
forms:%
$$
*\left( \Delta _{\widehat{a}}\otimes \varphi ^{\widehat{a}}\right) =\Delta _{%
\widehat{a}}\otimes (*\varphi ^{\widehat{a}}),
$$
$$
\widehat{\delta }\left( \Delta _{\widehat{a}}\otimes \varphi ^{\widehat{a}%
}\right) =\Delta _{\widehat{a}}\otimes (\widehat{\delta }\varphi ^{\widehat{a%
}}).
$$

The form $\lambda $ on $P$ with values in ${\cal G}$ is called horizontal if
$\widehat{H}\lambda =\lambda $ and equivariant if $R^{*}\left( q\right)
\lambda =Ad~q^{-1}\varphi ,~\forall g\in Gr,R\left( q\right) $ being the
right shift on $P.$ We can verify that equivariant and horizontal forms also
satisfy the conditions%
$$
\lambda =Ad~\varphi _{{\cal U}}^{-1}\left( \pi ^{*}\lambda \right) ,\qquad
\lambda _{{\cal U}}=S_{{\cal U}}^{*}\lambda ,
$$
$$
\left( \lambda _{{\cal V}}\right) _{{\cal U}}=Ad~\left( g_{{\cal UV}}\left(
u\right) \right) ^{-1}\left( \lambda _{{\cal U}}\right) _u.
$$

Now, we can define the field equations for curvature (17) and connection
(11) :%
$$
\Delta {\cal K}=0,\eqno(23)
$$
$$
\nabla {\cal K}=0,\eqno(24)
$$
where $\Delta =\widehat{H}\circ \widehat{\delta }_H.$ Equations (23) are
similar to the well-known Maxwell equations and for non-Abelian gauge fields
are called Yang-Mills equations. The structural equations (24) are called
Bianchi identities.

The field equations (23) do not have a physical meaning because they are
written in the total space of bundle $\Upsilon $ and not on the base
anisotropic space-time ${\cal E}^{<z>}.$ But this difficulty may be obviated
by projecting the mentioned equations on the base. The 1-form $\Delta {\cal K%
}$ is horizontal by definition and its equivariance follows from the right
invariance of metric (20). So, there is a unique form $(\Delta {\cal K})_{%
{\cal U}}$ satisfying
$$
\Delta {\cal K=}Ad~\varphi _{{\cal U}}^{-1}\pi ^{*}(\Delta {\cal K})_{{\cal U%
}}.
$$
Projection of (23) on the base can be written as $(\Delta {\cal K})_{{\cal %
U}}=0.$ To calculate $(\Delta {\cal K})_{{\cal U}},$ we use the equality
\cite{bis,pd}%
$$
d\left( Ad~\varphi _{{\cal U}}^{-1}\lambda \right) =Ad~~\varphi _{{\cal U}%
}^{-1}~d\lambda -\left[ \varphi _{{\cal U}}^{*}\theta ,Ad~\varphi _{{\cal U}%
}^{-1}\lambda \right]
$$
where $\lambda $ is a form on $P$ with values in ${\cal G.}$ For r-forms we
have
$$
\widehat{\delta }\left( Ad~\varphi _{{\cal U}}^{-1}\lambda \right)
=Ad~\varphi _{{\cal U}}^{-1}\widehat{\delta }\lambda -\left( -1\right)
^r*_H\{\left[ \varphi _{{\cal U}}^{*}\theta ,*_HAd~\varphi _{{\cal U}%
}^{-1}\lambda \right]
$$
and, as a consequence,%
$$
\widehat{\delta }{\cal K}=Ad~\varphi _{{\cal U}}^{-1}\{\widehat{\delta }%
_H\pi ^{*}{\cal K}_{{\cal U}}+*_H^{-1}[\pi ^{*}\omega _{{\cal U}},*_H\pi ^{*}%
{\cal K}_{{\cal U}}]\}-
$$
$$
-*_H^{-1}\left[ \Omega ,Ad~\varphi _{{\cal U}}^{-1}*_H\left( \pi ^{*}{\cal K}%
\right) \right] .\eqno(25)
$$
By using straightforward calculations in the adapted dual basis on $\pi
^{-1}\left( {\cal U}\right) $ we can verify the equalities%
$$
\left[ \Omega ,Ad~\varphi _{{\cal U}}^{-1}~*_H\left( \pi ^{*}{\cal K}_{{\cal %
U}}\right) \right] =0,\widehat{H}\delta _H\left( \pi ^{*}{\cal K}_{{\cal U}%
}\right) =\pi ^{*}\left( \widehat{\delta }_G{\cal K}\right) ,
$$
$$
*_H^{-1}\left[ \pi ^{*}\omega _{{\cal U}},*_H\left( \pi ^{*}{\cal K}_{{\cal U%
}}\right) \right] =\pi ^{*}\{*_G^{-1}\left[ \omega _{{\cal U}},*_G{\cal K}_{%
{\cal U}}\right] \}.\eqno(26)
$$
From (25) and (26) it follows that
$$
\left( \Delta {\cal K}\right) _{{\cal U}}=\widehat{\delta }_G{\cal K}_{{\cal %
U}}+*_G^{-1}\left[ \omega _{{\cal U}},*_G{\cal K}_{{\cal U}}\right] .%
\eqno(27)
$$

Taking into account (27) and (22), we prove that projection on ${\cal E}$
of equations (23) and (24) can be expressed respectively as
$$
*_G^{-1}\circ d\circ *_G{\cal K}_{{\cal U}}+*_G^{-1}\left[ \omega _{{\cal U}%
},*_G{\cal K}_{{\cal U}}\right] =0.\eqno(28)
$$
$$
d{\cal K}_{{\cal U}}+\left[ \omega _{{\cal U}},{\cal K}_{{\cal U}}\right] =0.%
$$

Equations (28) (see (27)) are gauge--invariant because%
$$
\left( \Delta {\cal K}\right) _{{\cal U}}=Ad~g_{{\cal UV}}^{-1}\left( \Delta
{\cal K}\right) _{{\cal V}}.
$$

By using formulas (19)-(22) we can rewrite (28) in coordinate form%
$$
D_{<\nu >}\left( G^{<\nu ><\lambda >}{\cal K}_{~<\lambda ><\mu >}^{\widehat{a%
}}\right) +f_{\widehat{b}\widehat{c}}^{\quad \widehat{a}}G^{<v><\lambda
>}\omega _{<\lambda >}^{~\widehat{b}}{\cal K}_{~<\nu ><\mu >}^{\widehat{c}%
}=0,\eqno(29)
$$
where $D_{<\nu >}$ is, for simplicity, a compatible with metric covariant
derivation on ha-space ${\cal E}^{<z>}.$

We point out that for our bundles with semisimple structural groups the
Yang-Mills equations (23) (and, as a consequence, their horizontal
projections (28) or (29)) can be obtained by variation of the action%
$$
I=\int {\cal K}_{~<\mu ><\nu >}^{\widehat{a}}{\cal K}_{~<\alpha ><\beta >}^{%
\widehat{b}}G^{<\mu ><\alpha >}G^{<\nu ><\beta >}K_{\widehat{a}\widehat{b}%
}\times \eqno(30)
$$
$$
\left| G_{<\alpha ><\beta >}\right| ^{1/2}dx^1...dx^n\delta
y_{(1)}^1...\delta y_{(1)}^{m_1}...\delta y_{(z)}^1...\delta y_{(z)}^{m_z}.
$$
Equations for extremals of (30) have the form
$$
K_{\widehat{r}\widehat{b}}G^{<\lambda ><\alpha >}G^{<\kappa ><\beta
>}D_{<\alpha >}{\cal K}_{~<\lambda ><\beta >}^{\widehat{b}}-
$$
$$
K_{\widehat{a}\widehat{b}}G^{<\kappa ><\alpha >}G^{<\nu ><\beta >}f_{%
\widehat{r}\widehat{l}}^{\quad \widehat{a}}\omega _{<\nu >}^{\widehat{l}}%
{\cal K}_{~<\alpha ><\beta >}^{\widehat{b}}=0,
$$
which are equivalent to ''pure'' geometric equations (29) (or (28)) due
to nondegeneration of the Killing form $K_{\widehat{r}\widehat{b}}$ for
semisimple groups.

To take into account gauge interactions with matter fields (section of
vector bundle $\Upsilon $ on ${\cal E}$ ) we have to introduce a source
1--form ${\cal J}$ in equations (23) and to write them as%
$$
\Delta {\cal K}={\cal J}\eqno(31)
$$

Explicit constructions of ${\cal J}$ require concrete definitions of the
bundle $\Upsilon ;$ for example, for spinor fields an invariant formulation
of the Dirac equations on la--spaces is necessary. We omit spinor
considerations in this section (see \cite{Vacaru1998,VacaruM}).

\section{Gauge Higher Order Anisotropic Gravity}

A considerable body of work on the formulation of gauge gravitational models
on isotropic spaces is based on using nonsemisimple groups, for example,
Poincare and affine groups, as structural gauge groups (see critical
analysis and original results in \cite{wal,ts,lue80,pon}). The main
impediment to developing such models is caused by the degeneration of
Killing forms for nonsemisimple groups, which make it impossible to
construct consistent variational gauge field theories (functional (30) and
extremal equations are degenerate in these cases). There are at least two
possibilities to get around the mentioned difficulty.\ The first is to
realize a minimal extension of the nonsemisimple group to a semisimple one,
similar to the extension of the Poincare group to the de Sitter group
considered in \cite{p,pd,ts} (in the next section we shall use this
operation for the definition of locally anisotropic gravitational
instantons). The second possibility is to introduce into consideration the
bundle of adapted affine frames on la-space ${\cal E}^{<z>},$ to use an
auxiliary nondegenerate bilinear form $a_{\widehat{a}\widehat{b}}$ instead
of the degenerate Killing form $K_{\widehat{a}\widehat{b}}$ and to consider
a ''pure'' geometric method, illustrated in the previous section, of
defining gauge field equations. Projecting on the base ${\cal E}^{<z>},$ we
shall obtain gauge gravitational field equations on ha-space having a form
similar to Yang-Mills equations.

The goal of this section is to prove that a specific parametrization of
components of the Cartan connection in the bundle of adapted affine frames
on ${\cal E}^{<z>}$ establishes an equivalence between Yang-Mills equations
(31) and Einstein equations (10) on ha--spaces.

\subsection{Bundles of linear ha--frames}

Let $\left( X_{<\alpha >}\right) _u=\left( X_i,X_{<a>}\right) _u=\left(
X_i,X_{a_1},...,X_{a_z}\right) _u$ be an adapted frame (see (14) at point $%
u\in {\cal E}^{<z>}.$ We consider a local right distinguished action of
matrices%
$$
A_{<\alpha ^{\prime }>}^{\quad <\alpha >}=\left(
\begin{array}{cccc}
A_{i^{\prime }}^{\quad i} & 0 & ... & 0 \\
0 & B_{a_1^{\prime }}^{\quad a_1} & ... & 0 \\
... & ... & ... & ... \\
0 & 0 & ... & B_{a_z^{\prime }}^{\quad a_z}
\end{array}
\right) \subset GL_{n_E}=
$$
$$
GL\left( n,{\cal R}\right) \oplus GL\left( m_1,{\cal R}\right) \oplus
...\oplus GL\left( m_z,{\cal R}\right) .
$$
Nondegenerate matrices $A_{i^{\prime }}^{\quad i}$ and $B_{j^{\prime
}}^{\quad j}$ respectively transforms linearly $X_{i|u}$ into $X_{i^{\prime
}|u}=A_{i^{\prime }}^{\quad i}X_{i|u}$ and $X_{a_p^{\prime }|u}$ into $%
X_{a_p^{\prime }|u}=B_{a_p^{\prime }}^{\quad a_p}X_{a_p|u},$ where $%
X_{<\alpha ^{\prime }>|u}=A_{<\alpha ^{\prime }>}^{\quad <\alpha
>}X_{<\alpha >}$ is also an adapted frame at the same point $u\in {\cal E}%
^{<z>}.$ We denote by $La\left( {\cal E}^{<z>}\right) $ the set of all
adapted frames $X_{<\alpha >}$ at all points of ${\cal E}^{<z>}$ and
consider the surjective map $\pi $ from $La\left( {\cal E}^{<z>}\right) $ to
${\cal E}^{<z>}$ transforming every adapted frame $X_{\alpha |u}$ and point $%
u$ into point $u.$ Every $X_{<\alpha ^{\prime }>|u}$ has a unique
representation as $X_{<\alpha ^{\prime }>}=A_{<\alpha ^{\prime }>}^{\quad
<\alpha >}X_{<\alpha >}^{\left( 0\right) },$ where $X_{<\alpha >}^{\left(
0\right) }$ is a fixed distinguished basis in tangent space $T\left( {\cal E}%
^{<z>}\right) .$ It is obvious that $\pi ^{-1}\left( {\cal U}\right) ,{\cal U%
}\subset {\cal E}^{<z>},$ is bijective to ${\cal U}\times GL_{n_E}\left(
{\cal R}\right) .$ We can transform $La\left( {\cal E}^{<z>}\right) $ in a
differentiable manifold taking $\left( u^{<\beta >},A_{<\alpha ^{\prime
}>}^{\quad <\alpha >}\right) $ as a local coordinate system on $\pi
^{-1}\left( {\cal U}\right) .$ Now, it is easy to verify that
$$
{\cal {L}}a({\cal E}^{<z>})=(La({\cal E}^{<z>},{\cal E}^{<z>},GL_{n_E}({\cal %
R})))%
$$
is a principal bundle. We call ${\cal {L}}a({\cal E}^{<z>})$ the bundle of
linear adapted frames on ${\cal E}^{<z>}.$

The next step is to identify the components of, for simplicity, compatible
d-connection $\Gamma _{<\beta ><\gamma >}^{<\alpha >}$ on ${\cal E}^{<z>}:$%
$$
\Omega _{{\cal U}}^{\widehat{a}}=\omega ^{\widehat{a}}=\{\omega _{\quad
<\lambda >}^{\widehat{\alpha }\widehat{\beta }}\doteq \Gamma _{<\beta
><\gamma >}^{<\alpha >}\}.\eqno(32)
$$
Introducing (32) in (27), we calculate the local 1-form%
$$
\left( \Delta {\cal R}^{(\Gamma )}\right) _{{\cal U}}=\Delta _{\widehat{%
\alpha }\widehat{\alpha }_1}\otimes (G^{<\nu ><\lambda >}D_{<\lambda >}{\cal %
R}_{\qquad <\nu ><\mu >}^{<\widehat{\alpha }><\widehat{\gamma }>}+
$$
$$
f_{\qquad <\widehat{\beta }><\widehat{\delta }><\widehat{\gamma }><\widehat{%
\varepsilon }>}^{<\widehat{\alpha }><\widehat{\gamma }>}G^{<\nu ><\lambda
>}\omega _{\qquad <\lambda >}^{<\widehat{\beta }><\widehat{\delta }>}{\cal R}%
_{\qquad <\nu ><\mu >}^{<\widehat{\gamma }><\widehat{\varepsilon }>})\delta
u^{<\mu >},\eqno(33)
$$
where
$$
\Delta _{\widehat{\alpha }\widehat{\beta }}=\left(
\begin{array}{cccc}
\Delta _{\widehat{i}\widehat{j}} & 0 & ... & 0 \\
0 & \Delta _{\widehat{a}_1\widehat{b}_1} & ... & 0 \\
... & ... & ... & ... \\
0 & 0 & ... & \Delta _{\widehat{a}_z\widehat{b}_z}
\end{array}
\right)
$$
is the standard distinguished basis in Lie algebra of matrices ${{\cal {G}}l}%
_{n_E}\left( {\cal R}\right) $ with $\left( \Delta _{\widehat{i}\widehat{k}%
}\right) _{jl}=\delta _{ij}\delta _{kl}$ and $\left( \Delta _{\widehat{a}_p%
\widehat{c}_p}\right) _{b_pd_p}=\delta _{a_pb_p}\delta _{c_pd_p}$ be\-ing
res\-pec\-ti\-ve\-ly the stand\-ard bas\-es in ${\cal {G}}l\left( {\cal R}%
^{n_E}\right) .$ We have denoted the curvature of connection (32),
considered in (33), as
$$
{\cal R}_{{\cal U}}^{(\Gamma )}=\Delta _{\widehat{\alpha }\widehat{\alpha }%
_1}\otimes {\cal R}_{\qquad <\nu ><\mu >}^{\widehat{\alpha }\widehat{\alpha }%
_1}X^{<\nu >}\bigwedge X^{<\mu >},
$$
where ${\cal R}_{\qquad <\nu ><\mu >}^{\widehat{\alpha }\widehat{\alpha }%
_1}=R_{<\alpha _1>\quad <\nu ><\mu >}^{\quad <\alpha >}$ (see curvatures
(7)).

\subsection{Bundles of affine ha--frames and Einstein equations}

Besides ${\cal {L}}a\left( {\cal E}^{<z>}\right) $ with ha-space ${\cal E}%
^{<z>},$ another bundle is naturally related, the bundle of adapted affine
frames with structural group $Af_{n_E}\left( {\cal R}\right) =GL_{n_E}\left(
{\cal E}^{<z>}\right)$ $\otimes {\cal R}^{n_E}.$ Because as linear space the
Lie Algebra $af_{n_E}\left( {\cal R}\right) $ is a direct sum of ${{\cal {G}}%
l}_{n_E}\left( {\cal R}\right) $ and ${\cal R}^{n_E},$ we can write forms on
${\cal {A}}a\left( {\cal E}^{<z>}\right) $ as $\Theta =\left( \Theta
_1,\Theta _2\right) ,$ where $\Theta _1$ is the ${{\cal {G}}l}_{n_E}\left(
{\cal R}\right) $ component and $\Theta _2$ is the ${\cal R}^{n_E}$
component of the form $\Theta .$ Connection (32), $\Omega $ in ${{\cal {L}}%
a}\left( {\cal E}^{<z>}\right) ,$ induces the Cartan connection $\overline{%
\Omega }$ in ${{\cal {A}}a}\left( {\cal E}^{<z>}\right) ;$ see the isotropic
case in \cite{p,pd,bis}. This is the unique connection on ${{\cal {A}}a}%
\left( {\cal E}^{<z>}\right) $ represented as $i^{*}\overline{\Omega }%
=\left( \Omega ,\chi \right) ,$ where $\chi $ is the shifting form and $i:{%
{\cal {A}}a}\rightarrow {{\cal {L}}a}$ is the trivial reduction of bundles.
If $s_{{\cal U}}^{(a)}$ is a local adapted frame in ${{\cal {L}}a}\left(
{\cal E}^{<z>}\right) ,$ then $\overline{s}_{{\cal U}}^{\left( 0\right)
}=i\circ s_{{\cal U}}$ is a local section in ${{\cal {A}}a}\left( {\cal E}%
^{<z>}\right) $ and
$$
\left( \overline{\Omega }_{{\cal U}}\right) =s_{{\cal U}}\Omega =\left(
\Omega _{{\cal U}},\chi _{{\cal U}}\right) ,\eqno(34)
$$
$$
\left( \overline{{\cal R}}_{{\cal U}}\right) =s_{{\cal U}}\overline{{\cal R}}%
=\left( {\cal R}_{{\cal U}}^{(\Gamma )},T_{{\cal U}}\right) ,
$$
where $\chi =e_{\widehat{\alpha }}\otimes \chi _{\quad <\mu >}^{\widehat{%
\alpha }}X^{<\mu >},G_{<\alpha ><\beta >}=\chi _{\quad <\alpha >}^{\widehat{%
\alpha }}\chi _{\quad <\beta >}^{\widehat{\beta }}\eta _{\widehat{\alpha }%
\widehat{\beta }}\quad (\eta _{\widehat{\alpha }\widehat{\beta }}$ is
diagonal with $\eta _{\widehat{\alpha }\widehat{\alpha }}=\pm 1)$ is a frame
decomposition of metric (5) on ${\cal E}^{<z>},e_{\widehat{\alpha }}$ is
the standard distinguished basis on ${\cal R}^{n_E},$ and the projection of
torsion , $T_{{\cal U}},$ on base ${\cal E}^{<z>}$ is defined as
$$
T_{{\cal U}}=d\chi _{{\cal U}}+\Omega _{{\cal U}}\bigwedge \chi _{{\cal U}%
}+\chi _{{\cal U}}\bigwedge \Omega _{{\cal U}}=\eqno(35)
$$
$$
e_{\widehat{\alpha }}\otimes \sum\limits_{<\mu ><<\nu >}T_{\quad <\mu ><\nu
>}^{\widehat{\alpha }}X^{<\mu >}\bigwedge X^{<\nu >}.
$$
For a fixed local adapted basis on ${\cal U}\subset {\cal E}^{<z>}$ we can
identify components $T_{\quad <\mu ><\nu >}^{\widehat{a}}$ of torsion (35)
with components of torsion (6) on ${\cal E}^{<z>},$ i.e. $T_{\quad <\mu
><\nu >}^{\widehat{\alpha }}=T_{\quad <\mu ><\nu >}^{<\alpha >}.$ By
straightforward calculation we obtain
$$
{(\Delta \overline{{\cal R}})}_{{\cal U}}=[{(\Delta {\cal R}^{(\Gamma )})}_{%
{\cal U}},\ {(R\tau )}_{{\cal U}}+{(Ri)}_{{\cal U}}],\eqno(36)
$$
where%
$$
\left( R\tau \right) _{{\cal U}}=\widehat{\delta }_GT_{{\cal U}%
}+*_G^{-1}\left[ \Omega _{{\cal U}},*_GT_{{\cal U}}\right] ,\quad \left(
Ri\right) _{{\cal U}}=*_G^{-1}\left[ \chi _{{\cal U}},*_G{\cal R}_{{\cal U}%
}^{(\Gamma )}\right] .
$$
Form $\left( Ri\right) _{{\cal U}}$ from (36) is locally constructed by
using components of the Ricci tensor (see (10)) as follows from
decomposition on the local adapted basis $X^{<\mu >}=\delta u^{<\mu >}:$
$$
\left( Ri\right) _{{\cal U}}=e_{\widehat{\alpha }}\otimes \left( -1\right)
^{n_E+1}R_{<\lambda ><\nu >}G^{\widehat{\alpha }<\lambda >}\delta u^{<\mu >}%
$$

We remark that for isotropic torsionless pseudo-Riemannian spaces the
requirement that $\left( \Delta \overline{{\cal R}}\right) _{{\cal U}}=0,$
i.e., imposing the connection (32) to satisfy Yang-Mills equations (23)
(equivalently (28) or (29) we obtain \cite{p,pd,ald} the equivalence of
the mentioned gauge gravitational equations with the vacuum Einstein
equations $R_{ij}=0.\,$ In the case of ha--spaces with arbitrary given
torsion, even considering vacuum gravitational fields, we have to introduce
a source for gauge gravitational equations in order to compensate for the
contribution of torsion and to obtain equivalence with the Einstein
equations.

Considerations presented in this section constitute the proof
of the following result

\begin{theorem}
The Einstein equations (10) for ha--gravity are equivalent to Yang-Mills
equations%
$$
\left( \Delta \overline{{\cal R}}\right) =\overline{{\cal J}}\eqno(37)
$$
for the induced Cartan connection $\overline{\Omega }$ (see (32), (34))
in the bundle of local adapted affine frames ${\cal A}a\left( {\cal E}%
\right) $ with source $\overline{{\cal J}}_{{\cal U}}$ constructed locally
by using the same formulas (36) (for $\left( \Delta \overline{{\cal R}}%
\right) $), where $R_{<\alpha ><\beta >}$ is changed by the matter source ${%
\tilde E}_{<\alpha ><\beta >}-\frac 12G_{<\alpha ><\beta >}{\tilde E},$
where ${\tilde E}_{<\alpha ><\beta >}=kE_{<\alpha ><\beta >}-\lambda
G_{<\alpha ><\beta >}.$
\end{theorem}

We note that this theorem is an extension for higher order anisotropic
spaces of the Popov and Dikhin results \cite{pd} with respect to a possible
gauge like treatment of the Einstein gravity. Similar theorems have been
proved for locally anisotropic gauge gravity
\cite{v295a,v295b,VacaruGoncharenko} and in
the framework of some variants of locally (and higher order) anisotropic
supergravity \cite{VacaruSupergravity}.

\section{Nonlinear De Sitter Gauge Ha--Gravity}

The equivalent reexpression of the Einstein theory as a gauge like theory
implies, for both locally isotropic and anisotropic space--times, the
nonsemisimplicity of the gauge group, which leads to a nonvariational theory
in the total space of the bundle of locally adapted affine frames. A
variational gauge gravitational theory can be formulated by using a minimal
extension of the affine structural group ${{\cal A}f}_{n_E}\left( {\cal R}%
\right) $ to the de Sitter gauge group $S_{n_E}=SO\left( n_E\right) $ acting
on distinguished ${\cal R}^{n_E+1}$ space.

\subsection{Nonlinear gauge theories of de Sitter group}

Let us consider the de Sitter space $\Sigma ^{n_E}$ as a hypersurface given
by the equations $\eta _{AB}u^Au^B=-l^2$ in the (n+m)--dimensional spaces
enabled with diagonal metric $\eta _{AB},\eta _{AA}=\pm 1$ (in this
subsection $A,B,C,...=1,2,...,n_E+1),(n_E=n+m_1+...+m_z),$ where $\{u^A\}$
are global Cartesian coordinates in ${\cal R}^{n_E+1};l>0$ is the curvature
of de Sitter space. The de Sitter group $S_{\left( \eta \right) }=SO_{\left(
\eta \right) }\left( n_E+1\right) $ is defined as the isometry group of $%
\Sigma ^{n_E}$--space with $\frac{n_E}2\left( n_E+1\right) $ generators of
Lie algebra ${{\it s}o}_{\left( \eta \right) }\left( n_E+1\right) $
satisfying the commutation relations%
$$
\left[ M_{<AB},M_{CD}\right] =\eta _{AC}M_{BD}-\eta _{BC}M_{AD}-\eta
_{AD}M_{BC}+\eta _{BD}M_{AC}.\eqno(38)
$$

Decomposing indices $A,B,...$ as $A=\left( \widehat{\alpha },n_E+1\right)
,B=\left( \widehat{\beta },n_E+1\right) ,$ \\ $...,$ the metric $\eta _{AB}$
as $\eta _{AB}=\left( \eta _{\widehat{\alpha }\widehat{\beta }},\eta
_{\left( n_E+1\right) \left( n_E+1\right) }\right) ,$ and operators $M_{AB}$
as $M_{\widehat{\alpha }\widehat{\beta }}={\cal F}_{\widehat{\alpha }%
\widehat{\beta }}$ and $P_{\widehat{\alpha }}=l^{-1}M_{n_E+1,\widehat{\alpha
}},$ we can write (38) as
$$
\left[ {\cal F}_{\widehat{\alpha }\widehat{\beta }},{\cal F}_{\widehat{%
\gamma }\widehat{\delta }}\right] =\eta _{\widehat{\alpha }\widehat{\gamma }}%
{\cal F}_{\widehat{\beta }\widehat{\delta }}-\eta _{\widehat{\beta }\widehat{%
\gamma }}{\cal F}_{\widehat{\alpha }\widehat{\delta }}+\eta _{\widehat{\beta
}\widehat{\delta }}{\cal F}_{\widehat{\alpha }\widehat{\gamma }}-\eta _{%
\widehat{\alpha }\widehat{\delta }}{\cal F}_{\widehat{\beta }\widehat{\gamma
}},
$$
$$
\left[ P_{\widehat{\alpha }},P_{\widehat{\beta }}\right] =-l^{-2}{\cal F}_{%
\widehat{\alpha }\widehat{\beta }},\quad \left[ P_{\widehat{\alpha }},{\cal F%
}_{\widehat{\beta }\widehat{\gamma }}\right] =\eta _{\widehat{\alpha }%
\widehat{\beta }}P_{\widehat{\gamma }}-\eta _{\widehat{\alpha }\widehat{%
\gamma }}P_{\widehat{\beta }},
$$
where we have indicated the possibility to decompose ${{\it s}o}_{\left(
\eta \right) }\left( n_E+1\right) $ into a direct sum, ${{\it s}o}_{\left(
\eta \right) }\left( n_E+1\right) ={{\it s}o}_{\left( \eta \right)
}(n_E)\oplus V_{n_E},$ where $V_{n_E}$ is the vector space stretched on
vectors $P_{\widehat{\alpha }}.$ We remark that $\Sigma ^{n_E}=S_{\left(
\eta \right) }/L_{\left( \eta \right) },$ where $L_{\left( \eta \right)
}=SO_{\left( \eta \right) }\left( n_E\right) .$ For $\eta _{AB}=diag\left(
1,-1,-1,-1\right) $ and $S_{10}=SO\left( 1,4\right) ,L_6=SO\left( 1,3\right)
$ is the group of Lorentz rotations.

Let $W\left( {\cal E},{\cal R}^{n_E+1},S_{\left( \eta \right) },P\right) $
be the vector bundle associated with principal bundle $P\left( S_{\left(
\eta \right) },{\cal E}\right) $ on la-spaces. The action of the structural
group $S_{\left( \eta \right) }$ on $E\,$ can be realized by using $\left(
n_E\right) \times \left( n_E\right) $ matrices with a parametrization
distinguishing subgroup $L_{\left( \eta \right) }:$%
$$
B=bB_L,\eqno(39)
$$
where%
$$
B_L=\left(
\begin{array}{cc}
L & 0 \\
0 & 1
\end{array}
\right) ,
$$
$L\in L_{\left( \eta \right) }$ is the de Sitter bust matrix transforming
the vector $\left( 0,0,...,\rho \right) \in {\cal R}^{n_E+1}$ into the
arbitrary point $\left( V^1,V^2,...,V^{n_E+1}\right) \in \Sigma _\rho
^{n_E}\subset {\cal R}^{n_E+1}$ with curvature $\rho \quad \left(
V_AV^A=-\rho ^2,V^A=t^A\rho \right) .$ Matrix $b$ can be expressed as
$$
b=\left(
\begin{array}{cc}
\delta _{\quad \widehat{\beta }}^{\widehat{\alpha }}+\frac{t^{\widehat{%
\alpha }}t_{\widehat{\beta }}}{\left( 1+t^{n_E+1}\right) } & t^{
\widehat{\alpha }} \\ t_{\widehat{\beta }} & t^{n_E+1}
\end{array}
\right) .
$$

The de Sitter gauge field is associated with a linear connection in $W$,
i.e., with a ${{\it s}o}_{\left( \eta \right) }\left( n_E+1\right) $-valued
connection 1--form on ${\cal E}^{<z>}:$
$$
\widetilde{\Omega }=\left(
\begin{array}{cc}
\omega _{\quad \widehat{\beta }}^{\widehat{\alpha }} & \widetilde{\theta }^{%
\widehat{\alpha }} \\ \widetilde{\theta }_{\widehat{\beta }} & 0
\end{array}
\right) ,\eqno(40)
$$
where $\omega _{\quad \widehat{\beta }}^{\widehat{\alpha }}\in
so(n_E)_{\left( \eta \right) },$ $\widetilde{\theta }^{\widehat{\alpha }}\in
{\cal R}^{n_E},\widetilde{\theta }_{\widehat{\beta }}\in \eta _{\widehat{%
\beta }\widehat{\alpha }}\widetilde{\theta }^{\widehat{\alpha }}.$

Because $S_{\left( \eta \right) }$-transforms mix $\omega _{\quad \widehat{%
\beta }}^{\widehat{\alpha }}$ and $\widetilde{\theta }^{\widehat{\alpha }}$
fields in (40) (the introduced para\-met\-ri\-za\-ti\-on is invariant on
action on $SO_{\left( \eta \right) }\left( n_E\right) $ group we cannot
identify $\omega _{\quad \widehat{\beta }}^{\widehat{\alpha }}$ and $%
\widetilde{\theta }^{\widehat{\alpha }},$ respectively, with the connection $%
\Gamma _{~<\beta ><\gamma >}^{<\alpha >}$ and the fundamental form $\chi
^{<\alpha >}$ in ${\cal E}^{<z>}$ (as we have for (32) and (34)). To
avoid this difficulty we consider \cite{ts,pon} a nonlinear gauge
realization of the de Sitter group $S_{\left( \eta \right) },$ namely, we
introduce into consideration the nonlinear gauge field%
$$
\Omega =b^{-1}\Omega b+b^{-1}db=\left(
\begin{array}{cc}
\Gamma _{~\widehat{\beta }}^{\widehat{\alpha }} & \theta ^{
\widehat{\alpha }} \\ \theta _{\widehat{\beta }} & 0
\end{array}
\right) ,\eqno(41)
$$
where
$$
\Gamma _{\quad \widehat{\beta }}^{\widehat{\alpha }}=\omega _{\quad \widehat{%
\beta }}^{\widehat{\alpha }}-\left( t^{\widehat{\alpha }}Dt_{\widehat{\beta }%
}-t_{\widehat{\beta }}Dt^{\widehat{\alpha }}\right) /\left(
1+t^{n_E+1}\right) ,
$$
$$
\theta ^{\widehat{\alpha }}=t^{n_E+1}\widetilde{\theta }^{\widehat{\alpha }%
}+Dt^{\widehat{\alpha }}-t^{\widehat{\alpha }}\left( dt^{n_e+1}+\widetilde{%
\theta }_{\widehat{\gamma }}t^{\widehat{\gamma }}\right) /\left(
1+t^{n_E+1}\right) ,
$$
$$
Dt^{\widehat{\alpha }}=dt^{\widehat{\alpha }}+\omega _{\quad \widehat{\beta }%
}^{\widehat{\alpha }}t^{\widehat{\beta }}.
$$

The action of the group $S\left( \eta \right) $ is nonlinear, yielding
transforms $\Gamma ^{\prime }=L^{\prime }\Gamma \left( L^{\prime }\right)
^{-1}+L^{\prime }d\left( L^{\prime }\right) ^{-1},\theta ^{\prime }=L\theta
, $ where the nonlinear matrix-valued function\\ $L^{\prime }=L^{\prime
}\left( t^{<\alpha >},b,B_T\right) $ is defined from $B_b=b^{\prime
}B_{L^{\prime }}$ (see parametrization (39)).

Now, we can identify components of (41) with components of $\Gamma
_{~<\beta ><\gamma >}^{<\alpha >}$ and $\chi _{\quad <\alpha >}^{\widehat{%
\alpha }}$ on ${\cal E}^{<z>}$ and induce in a consistent manner on the base
of bundle $W\left( {\cal E},{\cal R}^{n_E+1},S_{\left( \eta \right)
},P\right) $ the ha--geometry.

\subsection{Dynamics of the nonlinear $S\left( \eta \right) $ ha--gravity}

Instead of the gravitational potential (32), we introduce the
gravitational connection (similar to (41))
$$
\Gamma =\left(
\begin{array}{cc}
\Gamma _{\quad \widehat{\beta }}^{\widehat{\alpha }} & l_0^{-1}\chi ^{
\widehat{\alpha }} \\ l_0^{-1}\chi _{\widehat{\beta }} & 0
\end{array}
\right) \eqno(42)
$$
where
$$
\Gamma _{\quad \widehat{\beta }}^{\widehat{\alpha }}=\Gamma _{\quad \widehat{%
\beta }<\mu >}^{\widehat{\alpha }}\delta u^{<\mu >},
$$
$$
\Gamma _{\quad \widehat{\beta }<\mu >}^{\widehat{\alpha }}=\chi _{\quad
<\alpha >}^{\widehat{\alpha }}\chi _{\quad <\beta >}^{\widehat{\beta }%
}\Gamma _{\quad <\beta ><\gamma >}^{<\alpha >}+\chi _{\quad <\alpha >}^{%
\widehat{\alpha }}\delta _{<\mu >}\chi _{\quad \widehat{\beta }}^{<\alpha
>},
$$
$\chi ^{\widehat{\alpha }}=\chi _{\quad \mu }^{\widehat{\alpha }}\delta
u^\mu ,$ and $G_{\alpha \beta }=\chi _{\quad \alpha }^{\widehat{\alpha }%
}\chi _{\quad \beta }^{\widehat{\beta }}\eta _{\widehat{\alpha }\widehat{%
\beta }},$ and $\eta _{\widehat{\alpha }\widehat{\beta }}$ is parametrized
as
$$
\eta _{\widehat{\alpha }\widehat{\beta }}=\left(
\begin{array}{cccc}
\eta _{ij} & 0 & ... & 0 \\
0 & \eta _{a_1b_1} & ... & 0 \\
... & ... & ... & ... \\
0 & 0 & ... & \eta _{a_zb_z}
\end{array}
\right) ,
$$
$\eta _{ij}=\left( 1,-1,...,-1\right) ,...\eta _{ij}=\left( \pm 1,\pm
1,...,\pm 1\right) ,...,l_0$ is a dimensional constant.

The curvature of (42), ${\cal R}^{(\Gamma )}=d\Gamma +\Gamma \bigwedge
\Gamma ,$ can be written as%
$$
{\cal R}^{(\Gamma )}=\left(
\begin{array}{cc}
{\cal R}_{\quad \widehat{\beta }}^{\widehat{\alpha }}+l_0^{-1}\pi _{\widehat{%
\beta }}^{\widehat{\alpha }} & l_0^{-1}T^{
\widehat{\alpha }} \\ l_0^{-1}T^{\widehat{\beta }} & 0
\end{array}
\right) ,\eqno(43)
$$
where
$$
\pi _{\widehat{\beta }}^{\widehat{\alpha }}=\chi ^{\widehat{\alpha }%
}\bigwedge \chi _{\widehat{\beta }},{\cal R}_{\quad \widehat{\beta }}^{%
\widehat{\alpha }}=\frac 12{\cal R}_{\quad \widehat{\beta }<\mu ><\nu >}^{%
\widehat{\alpha }}\delta u^{<\mu >}\bigwedge \delta u^{<\nu >},
$$
and
$$
{\cal R}_{\quad \widehat{\beta }<\mu ><\nu >}^{\widehat{\alpha }}=\chi _{%
\widehat{\beta }}^{\quad <\beta >}\chi _{<\alpha >}^{\quad \widehat{\alpha }%
}R_{\quad <\beta ><\mu ><\nu >}^{<\alpha >}
$$
(see (7), the components of d-curvatures). The de Sitter gauge
group is semisimple and we are able to construct a variational gauge
gravitational locally anisotropic theory (bundle metric (20) is
nondegenerate). The Lagrangian of the theory is postulated as
$$
L=L_{\left( G\right) }+L_{\left( m\right) }
$$
where the gauge gravitational Lagrangian is defined as
$$
L_{\left( G\right) }=\frac 1{4\pi }Tr\left( {\cal R}^{(\Gamma )}\bigwedge *_G%
{\cal R}^{(\Gamma )}\right) ={\cal L}_{\left( G\right) }\left| G\right|
^{1/2}\delta ^{n_E}u,
$$
$$
{\cal L}_{\left( G\right) }=\frac 1{2l^2}T_{\quad <\mu ><\nu >}^{\widehat{%
\alpha }}T_{\widehat{\alpha }}^{\quad <\mu ><\nu >}+\eqno(44)
$$
$$
\frac 1{8\lambda }{\cal R}_{\quad \widehat{\beta }<\mu ><\nu >}^{\widehat{%
\alpha }}{\cal R}_{\quad \widehat{\alpha }}^{\widehat{\beta }\quad <\mu
><\nu >}-\frac 1{l^2}\left( {\overleftarrow{R}}\left( \Gamma \right)
-2\lambda _1\right) ,
$$
$T_{\quad <\mu ><\nu >}^{\widehat{\alpha }}=\chi _{\quad <\alpha >}^{%
\widehat{\alpha }}T_{\quad <\mu ><\nu >}^{<\alpha >}$ (the gravitational
constant $l^2$ in (44) satisfies the relations $l^2=2l_0^2\lambda ,\lambda
_1=-3/l_0],\quad Tr$ denotes the trace on $\widehat{\alpha },\widehat{\beta }
$ indices, and the matter field Lagrangian is defined as%
$$
L_{\left( m\right) }=-1\frac 12Tr\left( \Gamma \bigwedge *_G{\cal I}\right) =%
{\cal L}_{\left( m\right) }\left| G\right| ^{1/2}\delta ^{n_E}u,
$$
$$
{\cal L}_{\left( m\right) }=\frac 12\Gamma _{\quad \widehat{\beta }<\mu >}^{%
\widehat{\alpha }}S_{\quad <\alpha >}^{\widehat{\beta }\quad <\mu
>}-t_{\quad \widehat{\alpha }}^{<\mu >}l_{\quad <\mu >}^{\widehat{\alpha }}.%
\eqno(45)
$$
The matter field source ${\cal I}$ is obtained as a variational derivation
of ${\cal L}_{\left( m\right) }$ on $\Gamma $ and is parametrized as
$$
{\cal I}=\left(
\begin{array}{cc}
S_{\quad \widehat{\beta }}^{\widehat{\alpha }} & -l_0t^{
\widehat{\alpha }} \\ -l_0t_{\widehat{\beta }} & 0
\end{array}
\right) \eqno(46)
$$
with $t^{\widehat{\alpha }}=t_{\quad <\mu >}^{\widehat{\alpha }}\delta
u^{<\mu >}$ and $S_{\quad \widehat{\beta }}^{\widehat{\alpha }}=S_{\quad
\widehat{\beta }<\mu >}^{\widehat{\alpha }}\delta u^{<\mu >}$ being
respectively the canonical tensors of energy-momentum and spin density.
Because of the contraction of the ''interior'' indices $\widehat{\alpha },%
\widehat{\beta }$ in (44) and (45) we used the Hodge operator $*_G$
instead of $*_H$ (hereafter we consider $*_G=*).$

Varying the action
$$
S=\int \left| G\right| ^{1/2}\delta ^{n_E}u\left( {\cal L}_{\left( G\right)
}+{\cal L}_{\left( m\right) }\right)
$$
on the $\Gamma $-variables (36), we obtain the gauge--gravitational field
equations:%
$$
d\left( *{\cal R}^{(\Gamma )}\right) +\Gamma \bigwedge \left( *{\cal R}%
^{(\Gamma )}\right) -\left( *{\cal R}^{(\Gamma )}\right) \bigwedge \Gamma
=-\lambda \left( *{\cal I}\right) .\eqno(47)
$$

Specifying the variations on $\Gamma _{\quad \widehat{\beta }}^{\widehat{%
\alpha }}$ and $l^{\widehat{\alpha }}$-variables, we rewrite (47) as
$$
\widehat{{\cal D}}\left( *{\cal R}^{(\Gamma )}\right) +\frac{2\lambda }{l^2}%
\left( \widehat{{\cal D}}\left( *\pi \right) +\chi \bigwedge \left(
*T^T\right) -\left( *T\right) \bigwedge \chi ^T\right) =-\lambda \left(
*S\right) ,\eqno(48)
$$
$$
\widehat{{\cal D}}\left( *T\right) -\left( *{\cal R}^{(\Gamma )}\right)
\bigwedge \chi -\frac{2\lambda }{l^2}\left( *\pi \right) \bigwedge \chi =%
\frac{l^2}2\left( *t+\frac 1\lambda *\tau \right) ,\eqno(49)
$$
where
$$
T^t=\{T_{\widehat{\alpha }}=\eta _{\widehat{\alpha }\widehat{\beta }}T^{%
\widehat{\beta }},~T^{\widehat{\beta }}=\frac 12T_{\quad <\mu ><\nu >}^{%
\widehat{\beta }}\delta u^{<\mu >}\bigwedge \delta u^{<\nu >}\},
$$
$$
\chi ^T=\{\chi _{\widehat{\alpha }}=\eta _{\widehat{\alpha }\widehat{\beta }%
}\chi ^{\widehat{\beta }},~\chi ^{\widehat{\beta }}=\chi _{\quad <\mu >}^{%
\widehat{\beta }}\delta u^{<\mu >}\},\qquad \widehat{{\cal D}}=d+\widehat{%
\Gamma }
$$
($\widehat{\Gamma }$ acts as $\Gamma _{\quad \widehat{\beta }<\mu >}^{%
\widehat{\alpha }}$ on indices $\widehat{\gamma },\widehat{\delta },...$ and
as $\Gamma _{\quad <\beta ><\mu >}^{<\alpha >}$ on indices $<\gamma
>,$ $<\delta >,...).$ In (49), $\tau $ defines the energy--momentum tensor of
the $S_{\left( \eta \right) }$--gauge gravitational field $\widehat{\Gamma }:$%
$$
\tau _{<\mu ><\nu >}\left( \widehat{\Gamma }\right) =\frac 12Tr\left( {\cal R%
}_{<\mu ><\alpha >}{\cal R}_{\quad <\nu >}^{<\alpha >}-\frac 14{\cal R}%
_{<\alpha ><\beta >}{\cal R}^{<\alpha ><\beta >}G_{<\mu ><\nu >}\right) .%
\eqno(50)
$$

Equations (47) (or equivalently (48),(49)) make up the complete system
of variational field equations for nonlinear de Sitter gauge gravity with
higher order anisotropy. They can be interpreted as a generalization of
gauge like equations for ha--gravity
\cite{VacaruGoncharenko} (equivalently, of gauge
gravitational equations (37)] to a system of gauge field equations with
dynamical torsion and corresponding spin-density source.

A. Tseytlin \cite{ts} presented a quantum analysis of the isotropic version
of equations (48) and (49). Of course, the problem of quantizing
gravitational interactions is unsolved for both variants of locally
anisotropic and isotropic gauge de Sitter gravitational theories, but we
think that the generalized Lagrange version of $S_{\left( \eta \right) }$%
-gravity is more adequate for studying quantum radiational and statistical
gravitational processes. This is a matter for further investigations.

Finally, we remark that we can obtain a nonvariational Poincare gauge
gravitational theory on ha--spaces if we consider the contraction of the
gauge potential (42) to a potential with values in the Poincare Lie
algebra
$$
\Gamma =\left(
\begin{array}{cc}
\Gamma _{\quad \widehat{\beta }}^{\widehat{\alpha }} & l_0^{-1}\chi ^{
\widehat{\alpha }} \\ l_0^{-1}\chi _{\widehat{\beta }} & 0
\end{array}
\right) \rightarrow \Gamma =\left(
\begin{array}{cc}
\Gamma _{\quad \widehat{\beta }}^{\widehat{\alpha }} & l_0^{-1}\chi ^{
\widehat{\alpha }} \\ 0 & 0
\end{array}
\right) .
$$
Isotropic Poincare gauge gravitational theories are studied in a number of
papers (see, for example, \cite{wal,ts,lue80,pon}). In a manner similar to
considerations presented in this work, we can generalize Poincare gauge
models for spaces with local anisotropy.

\section{Ha--Gravitational Gauge Instantons}

The existence of self-dual, or instanton, topologically nontrivial solutions
of Yang-Mills equations is a very important physical consequence of gauge
theories. All known instanton-type Yang-Mills and gauge gravitational
solutions (see, for example, \cite{ts,pon}) are locally isotropic. A
variational gauge-gravitational extension of la-gravity makes possible a
straightforward application of techniques of constructing solutions for
first order gauge equations for the definition of locally anisotropic
gravitational instantons. This section is devoted to the study of some
particular instanton solutions of the gauge gravitational theory on la-space.

Let us consider the Euclidean formulation of the $S_{\left( \eta \right) }$%
-gauge gravitational theory by changing gauge structural groups and flat
metric:%
$$
SO_{(\eta )}(n_E+1)\rightarrow SO(n_E+1),SO_{(\eta )}(n_E)\rightarrow
SO(n_E),\eta _{AB}\rightarrow -\delta _{AB}.
$$
Self-dual (anti-self-dual) conditions for the curvature (43)
$$
{\cal R}^{(\Gamma )}=*{\cal R}^{(\Gamma )}\quad (-*{\cal R}^{(\Gamma )})%
$$
can be written as a system of equations
$$
\left( {\cal R}_{\quad \widehat{\beta }}^{\widehat{\alpha }}-l_0^{-2}\pi
_{\quad \widehat{\beta }}^{\widehat{\alpha }}\right) =\pm *\left( {\cal R}%
_{\quad \widehat{\beta }}^{\widehat{\alpha }}-l_0^{-2}\pi _{\quad \widehat{%
\beta }}^{\widehat{\alpha }}\right) \eqno(51)
$$
$$
T^{\widehat{\alpha }}=\pm *T^{\widehat{\alpha }}\eqno(52)
$$
(the ''-'' refers to the anti-self-dual case), where the ''-'' before $%
l_0^{-2}$ appears because of the transition of the Euclidean negatively
defined metric $-\delta _{<\alpha ><\beta >},$ which leads to $\chi _{\quad
<\alpha >}^{\widehat{\alpha }}\rightarrow i\chi _{\quad <\alpha >|(E)}^{%
\widehat{\alpha }},\pi \rightarrow -\pi _{(E)}$ (we shall omit the index $%
(E) $ for Euclidean values).

For solutions of (51) and (52) the energy--momentum tensor (50) is
identically equal to zero. Vacuum equations (47) and (48), when source
is ${{\cal I}\equiv 0}$ (see (46)), are satisfied as a consequence of
generalized Bianchi identities for the curvature (43). The mentioned
solutions of (51) and (52) realize a local minimum of the Euclidean
action%
$$
S=\frac 1{8\lambda }\int \left| G^{1/2}\right| \delta ^{n_E}\{(R\left(
\Gamma \right) -l_0^{-2}\pi )^2+2T^2\},
$$
where $T^2=T_{\quad <\mu ><\nu >}^{\widehat{\alpha }}T_{\widehat{\alpha }%
}^{\quad <\mu ><\nu >}$ is extremal on the topological invariant (Pontryagin
index)%
$$
p_2=-\frac 1{8\pi ^2}\int Tr\left( {\cal R}^{(\Gamma )}\bigwedge {\cal R}%
^{(\Gamma )}\right) =--\frac 1{8\pi ^2}\int Tr\left( \widehat{{\cal R}}%
\bigwedge \widehat{{\cal R}}\right) .
$$

For the Euclidean de Sitter spaces, when%
$$
{\cal R}=0\quad \{T=0,R_{\quad <\mu ><\nu >}^{\widehat{\alpha }\widehat{%
\beta }}=-\frac 2{l_0^2}\chi _{\quad <\mu >}^{[\alpha }\chi _{\quad <\nu
>}^{\beta ]}\}\eqno(53)
$$
we obtain the absolute minimum, $S=0.$

We emphasize that for $R_{<\beta >\quad <\mu ><\beta >}^{\quad <\alpha
>}=\left( 2/l_0^2\right) \delta _{[<\mu >}^{<\alpha >}G_{<\nu >]<\beta >}$
torsion vanishes. Torsionless instantons also have another interpretation.\
For\\ $T_{\quad <\beta ><\gamma >}^{<\alpha >}=0$ contraction of equations
(51) leads to Einstein equations with cosmological $\lambda $-term (as a
consequence of generalized Ricci identities):%
$$
R_{<\alpha ><\beta ><\mu ><\nu >}-R_{<\mu ><\nu ><\alpha ><\beta >}=\frac
32\{R_{[<\alpha ><\beta ><\mu >]<\nu >}-
$$
$$
R_{[<\alpha ><\beta ><\nu >]<\mu >}+R_{[<\mu ><\nu ><\alpha >]<\beta
>}-R_{[<\mu ><\nu ><\beta >]<\alpha >}\}.
$$
So, in the Euclidean case the locally anisotropic vacuum Einstein equations
are a subset of instanton solutions.

Now, let us study the $SO\left( n_E\right) $ solution of equations (51)
and\ (52). We consider the spherically symmetric ansatz (in order to point
out the connection between high-dimensional gravity and ha--gravity the
N--connection structure is chosen to be trivial, i.e. ${N}_j^a\left( u\right)
\equiv 0):$%
$$
\Gamma _{\quad \widehat{\beta }<\mu >}^{\widehat{\alpha }}=a\left( u\right)
\left( u^{\widehat{\alpha }}\delta _{\widehat{\beta }<\mu >}-u_{\widehat{%
\beta }}\delta _{<\mu >}^{\widehat{\alpha }}\right) +q\left( u\right)
\epsilon _{\quad \widehat{\beta }<\mu ><\nu >}^{\widehat{\alpha }}u^{<\mu
>},
$$
$$
\chi _{<\alpha >}^{\widehat{\alpha }}=f\left( u\right) \delta _{\quad
<\alpha >}^{\widehat{\alpha }}+n\left( u\right) u^{\widehat{\alpha }%
}u_\alpha ,\eqno(54)
$$
where $u=u^{<\alpha >}u^{<\beta >}G_{<\alpha ><\beta >}=x^{\widehat{i}}x_{%
\widehat{i}}+y^{\widehat{a}}y_{\widehat{a}},$ and $a\left( u\right) ,q\left(
u\right) ,f\left( u\right) $ and $n\left( u\right) $ are some scalar
functions. Introducing (54) into (51) and (52), we obtain,
respectively,%
$$
u\left( \pm \frac{dq}{du}-a^2-q^2\right) +2\left( a\pm q\right) +l_0^{-1}f^2,%
\eqno(55)
$$
$$
2d\left( a\mp q\right) /du+\left( a\mp q\right) ^2-l_0^{-1}fn=0,\eqno(56)
$$
$$
2\frac{df}{du}+f\left( a\mp 2q\right) +n\left( au-1\right) =0.\eqno(57)
$$

The traceless part of the torsion vanishes because of the parametrization
(54), but in the general case the trace and pseudo-trace of the torsion
are not identical to zero:%
$$
T^\mu =q^{\left( 0\right) }u^\mu \left( -2df/du+n-a\left( f+un\right)
\right) ,
$$
$$
\overbrace{T}^\mu =q^{\left( 1\right) }u^\mu \left( 2qf\right) ,
$$
$q^{\left( 0\right) }$ and $q^{\left( 0\right) }$ are constant. Equation
(52) or (57) establishes the proportionality of $T^\mu $ and $\overbrace{%
T}^\mu .$ As a consequence we obtain that the $SO\left( n+m\right) $
solution of (52) is torsionless if $q\left( u\right) =0$ of $f\left(
u\right) =0.$

Let first analyze the torsionless instantons, $T_{\quad <\alpha ><\beta
>}^{<\mu >}=0.$ If $f=0,$ then from (56) one has two possibilities: (a) $%
n=0$ leads to nonsense because $\chi _\alpha ^{\widehat{\alpha }}=0$ or $%
G_{\alpha \beta }=0.$ b) $a=u^{-1}$ and $n\left( u\right) $ is an arbitrary
scalar function; we have from (56) $a\mp q=2/\left( a+C^2\right) $ or $%
q=\pm 2/u\left( u+C^2\right) ,$ where $C=const.$ If $q\left( u\right) =0,$
we obtain the de Sitter space (53) because equations (55) and (56)
impose vanishing of both self-dual and anti-self-dual parts of $\left( {\cal %
R}_{\quad \widehat{\beta }}^{\widehat{\alpha }}-l_0^2\pi _{\quad \widehat{%
\beta }}^{\widehat{\alpha }}\right) ,$ so, as a consequence, ${\cal R}%
_{\quad \widehat{\beta }}^{\widehat{\alpha }}-l_0^2\pi _{\quad \widehat{%
\beta }}^{\widehat{\alpha }}\equiv 0.$ There is an infinite number of $%
SO\left( n_E\right) $-symmetrical solutions of (53):%
$$
f=l_0\left[ a\left( 2-au\right) \right] ^{1/2},\quad n=l_0\{2\frac{da}{du}+%
\frac{a^2}{\left[ a\left( 2-au\right) \right] ^{1/2}}\},
$$
$a(u)$ is a scalar function.

To find instantons with torsion, $T_{\quad \beta \gamma }^\alpha \neq 0,$ is
also possible. We present the $SO\left( 4\right) $ one-instanton solution,
obtained in \cite{pon} (which in the case of $H^4$-space parametrized by
local coordinates $\left( x^{\widehat{1}},x^{\widehat{2}},y^{\widehat{1}},y^{%
\widehat{2}}\right) ,$ with $u=x^{\widehat{1}}x_{\widehat{1}}+x^{\widehat{2}%
}x_{\widehat{2}}+y^{\widehat{1}}y_{\widehat{1}}+y^{\widehat{2}}y_{\widehat{2}%
}):$%
$$
a=a_0\left( u+c^2\right) ^{-1},q=\mp q_0\left( u+c^2\right) ^{-1}
$$
$$
f=l_0\left( \alpha u+\beta \right) ^{1/2}/\left( u+c^2\right) ,n=c_0/\left(
u+c^2\right) \left( \gamma u+\delta \right) ^{1/2}
$$
where%
$$
a_0=-1/18,q_0=5/6, \alpha =266/81, \beta =8/9,%
$$
$$
\gamma =10773/11858,\delta = 1458/5929.
$$
We suggest that local regions with $T_{\ \beta \gamma }^\alpha \neq 0$ are
similarly to Abrikosov vortexes in superconductivity and the appearance of
torsion is a possible realization of the Meisner effect in gravity (for
details and discussions on the superconducting or Higgs-like interpretation
of gravity see \cite{ts,pon}).

\section{Nearly Autoparallel Maps of La--Spaces}

The aim of the section is to present a generalization of the
 nearly geodesic map (ng--map) theory \cite{sin} and nearly autoparallel map
 (na--map) theory \cite{vk,vb12,vrjp,vob12,vob13,vog}  by introducing
into consideration maps of vector bundles provided with compatible
N--connection, d--connection and metric structures. For simplicity, the
 basic definitions and theorems will be formulated only for locally
 anisotropic spaces. The transition to higher order anisotropies
 can be made in a strightforward manner by introducing higher order
 distinguishing of indices,e $\alpha \to <{\alpha}>,$ corresponding
 to a higher order distingushing of nonlinear connection and of basic
 geometric objects.

Our geometric arena consists from pairs of open regions $( U, {\underline U}%
) $ of la--spaces, $U{\subset}{\xi},\, {\underline U}{\subset}{\underline
{\xi}}$, and 1--1 local maps $f : U{\to}{\underline U}$ given by functions $%
f^{a}(u)$ of smoothly class $C^r( U) \, (r>2, $ or $r={\omega}$~ for
analytic functions) and their inverse functions $f^{\underline a}({%
\underline u})$ with corresponding non--zero Jacobians in every point $u{\in}
U$ and ${\underline u}{\in}{\underline U}.$

We consider that two open regions $U$~ and ${\underline U}$~ are attributed
to a common for f--map coordinate system if this map is realized on the
principle of coordinate equality $q(u^{\alpha}) {\to} {\underline q}
(u^{\alpha})$~ for every point $q {\in} U$~ and its f--image ${\underline q}
{\in} {\underline U}.$ We note that all calculations included in this work
will be local in nature and taken to refer to open subsets of mappings of
type
 ${\xi} {\supset} U {\stackrel {f}{\longrightarrow}} {\underline U}
 {\subset} {\underline {\xi}}.$
For simplicity, we suppose that in a fixed
common coordinate system for $U$ and ${\underline U}$ spaces $\xi$ and ${%
\underline {\xi}}$ are characterized by a common N--connection structure (in
consequence of (5) by a corresponding concordance of d--metric
structure), i.e.
$$
N^{a}_{j}(u)={\underline N}^{a}_{j}(u)={\underline N}^{a}_{j} ({\underline u}%
),%
$$
which leads to the possibility to establish common local bases, adapted to a
given N--connection, on both regions $U$ and ${\underline U.}$ We consider
that on $\xi$ it is defined the linear d--connection structure with
components ${\Gamma}^{\alpha}_{{.}{\beta}{\gamma}}.$ On the space $%
\underline {\xi}$ the linear d--connection is considered to be a general one
with torsion
$$
{\underline T}^{\alpha}_{{.}{\beta}{\gamma}}={\underline {\Gamma}}^{\alpha}_
{{.}{\beta}{\gamma}}-{\underline {\Gamma}}^{\alpha}_{{.}{\gamma}{\beta}}+
w^{\alpha}_{{.}{\beta}{\gamma}}
$$
and nonmetricity
$$
{\underline K}_{{\alpha}{\beta}{\gamma}}={{\underline D}_{\alpha}} {%
\underline G}_{{\beta}{\gamma}}. \eqno(58)
$$

Geometrical objects on ${\underline {\xi}}$ are specified by underlined
symbols (for example, ${\underline A}^{\alpha}, {\underline B}^{{\alpha}{%
\beta}})$~ or underlined indices (for example, $A^{\underline a}, B^{{%
\underline a}{\underline b}}).$

For our purposes it is convenient to introduce auxiliary sym\-met\-ric
d--con\-nec\-ti\-ons, ${\gamma}^{\alpha}_{{.}{\beta}{\gamma}}={\gamma}%
^{\alpha}_{{.}{\gamma}{\beta}} $~ on $\xi$ and ${\underline {\gamma}}%
^{\alpha}_{.{\beta}{\gamma}}= {\underline {\gamma}}^{\alpha}_{{.}{\gamma}{%
\beta}}$ on ${\underline {\xi}}$ defined, correspondingly, as
$$
{\Gamma}^{\alpha}_{{.}{\beta}{\gamma}}= {\gamma}^{\alpha}_{{.}{\beta}{\gamma}%
}+ T^{\alpha}_{{.}{\beta}{\gamma}}\quad {\rm \; and}\quad {\underline
{\Gamma}}^{\alpha}_{{.}{\beta}{\gamma}}= {\underline {\gamma}}^{\alpha}_{{.}{%
\beta}{\gamma}}+ {\underline T}^{\alpha}_{{.}{\beta}{\gamma}}.%
$$

We are interested in definition of local 1--1 maps from $U$ to ${\underline U%
}$ characterized by symmetric, $P^{\alpha}_{{.}{\beta}{\gamma}},$ and
antisymmetric, $Q^{\alpha}_{{.}{\beta}{\gamma}}$,~ deformations:
$$
{\underline {\gamma}}^{\alpha}_{{.}{\beta}{\gamma}} ={\gamma}^{\alpha}_{{.}{%
\beta}{\gamma}}+ P^{\alpha}_{{.}{\beta}{\gamma}} \eqno(59)%
$$
and
$$
{\underline T}^{\alpha}_{{.}{\beta}{\gamma}}= T^{\alpha}_{{.}{\beta}{\gamma}%
}+ Q^{\alpha}_{{.}{\beta}{\gamma}}. \eqno(60)
$$
The auxiliary linear covariant derivations induced by ${\gamma}^{\alpha}_{{.}%
{\beta}{\gamma}}$ and ${\underline {\gamma}}^{\alpha}_{{.}{\beta}{\gamma}}$~
are denoted respectively as $^{({\gamma})}D$~ and $^{({\gamma})}{\underline D%
}.$~

Let introduce this local coordinate parametrization of curves on $U$~:
$$
u^{\alpha}=u^{\alpha}({\eta})=(x^{i}({\eta}), y^{i}({\eta})),~{\eta}_{1}<{\eta%
}<{\eta}_{2},%
$$
where corresponding tangent vector field is defined as
$$
v^{\alpha}={\frac{{du^{\alpha}} }{d{\eta}}}= ({\frac{{dx^{i}({\eta})} }{{d{%
\eta}}}}, {\frac{{dy^{j}({\eta})} }{d{\eta}}}).%
$$

\begin{definition}
Curve $l$~ is called auto parallel, a--parallel, on $\xi $ if its tangent
vector field $v^\alpha $~ satisfies a--parallel equations:
$$
vDv^\alpha =v^\beta {^{({\gamma })}D}_\beta v^\alpha ={\rho }({\eta }%
)v^\alpha ,\eqno(61)
$$
where ${\rho }({\eta })$~ is a scalar function on $\xi $.
\end{definition}

Let curve ${\underline l} {\subset} {\underline {\xi}}$ is given in
parametric form as $u^{\alpha}=u^{\alpha}({\eta}),~{\eta}_1 < {\eta} <{\eta}%
_2$ with tangent vector field $v^{\alpha} = {\frac{{du^{\alpha}} }{{d{\eta}}}%
} {\ne} 0.$ We suppose that a 2--dimensional distribution $E_2({\underline l}%
)$ is defined along ${\underline l} ,$ i.e. in every point $u {\in} {%
\underline l}$ is fixed a 2-dimensional vector space $E_{2}({\underline l}) {%
\subset} {\underline {\xi}}.$ The introduced distribution $E_{2}({\underline
l})$~ is coplanar along ${\underline l}$~ if every vector ${\underline p}%
^{\alpha}(u^{b}_{(0)}) {\subset} E_{2}({\underline l}), u^{\beta}_{(0)} {%
\subset} {\underline l}$~ rests contained in the same distribution after
parallel transports along ${\underline l},$~ i.e. ${\underline p}%
^{\alpha}(u^{\beta}({\eta})) {\subset} E_{2} ({\underline l}).$

\begin{definition}
A curve ${\underline{l}}$~ is called nearly autoparallel, or in brief an
na--parallel, on space ${\underline{\xi }}$~ if a coplanar along ${%
\underline{l}}$~ distribution $E_2({\underline{l}})$ containing tangent to ${%
\underline{l}}$~ vector field $v^\alpha ({\eta })$,~ i.e. $v^\alpha ({\eta })%
{\subset }E_2({\underline{l}}),$~ is defined.
\end{definition}

We can define nearly autoparallel maps of la--spaces as an anisotropic
generalization (see also \cite{vodg,voa} ng--\cite{sin} and na--maps \cite
{vk,vob12,vog,vob13}):

\begin{definition}
Nearly autoparallel maps, na--maps, of la--spaces are defined as local 1--1
mappings of v--bundles, $\xi {\to }{\underline{\xi }},$ changing every
a--parallel on $\xi $ into a na--parallel on ${\underline{\xi }}.$
\end{definition}

Now we formulate the general conditions when deformations (59) and (60)
charac\-ter\-ize na-maps : Let a-parallel $l{\subset }U$~ is given by
func\-ti\-ons\\ $u^\alpha =u^{({\alpha })}({\eta }),v^\alpha ={\frac{{%
du^\alpha }}{d{\eta }}}$, ${\eta }_1<{\eta }<{\eta }_2$, satisfying
equations (61). We suppose that to this a--parallel corresponds a
na--parallel ${\underline{l}}\subset {\underline{U}}$ given by the same
parameterization in a common for a chosen na--map coordinate system on $U$~
and ${\underline{U}}.$ This condition holds for vectors ${\underline{v}}%
_{(1)}^\alpha =v{\underline{D}}v^\alpha $~ and $v_{(2)}^\alpha =v{\underline{%
D}}v_{(1)}^\alpha $ satisfying equality
$$
{\underline{v}}_{(2)}^\alpha ={\underline{a}}({\eta })v^\alpha +{\underline{b%
}}({\eta }){\underline{v}}_{(1)}^\alpha \eqno(62)
$$
for some scalar functions ${\underline{a}}({\eta })$~ and ${\underline{b}}({%
\eta })$~ (see Definitions 2 and 3). Putting splittings (59) and
(4.3) into expressions for
$$
{\underline{v}}_{(1)}^\alpha \mbox{ and }{\underline{v}}_{(2)}^\alpha
$$
in (62) we obtain:
$$
v^\beta v^\gamma v^\delta (D_\beta P_{{.}{\gamma }{\delta }}^\alpha +P_{{.}{%
\beta }{\tau }}^\alpha P_{{.}{\gamma }{\delta }}^\tau +Q_{{.}{\beta }{\tau }%
}^\alpha P_{{.}{\gamma }{\delta }}^\tau )=bv^\gamma v^\delta P_{{.}{\gamma }{%
\delta }}^\alpha +av^\alpha ,\eqno(63)
$$
where
$$
b({\eta },v)={\underline{b}}-3{\rho },\qquad \mbox{and}\qquad a({\eta },v)={%
\underline{a}}+{\underline{b}}{\rho }-v^b{\partial }_b{\rho }-{\rho }^2%
\eqno(64)
$$
are called the deformation parameters of na--maps.

The algebraic equations for the deformation of torsion $Q_{{.}{\beta }{\tau }%
}^\alpha $ should be written as the compatibility conditions for a given
nonmetricity tensor ${\underline{K}}_{{\alpha }{\beta }{\gamma }}$~ on ${%
\underline{\xi }}$ ( or as the metricity conditions if d--connection ${%
\underline{D}}_\alpha $~ on ${\underline{\xi }}$~ is required to be metric) :%
$$
D_\alpha G_{{\beta }{\gamma }}-P_{{.}{\alpha }({\beta }}^\delta G_{{{\gamma }%
)}{\delta }}-{\underline{K}}_{{\alpha }{\beta }{\gamma }}=Q_{{.}{\alpha }({%
\beta }}^\delta G_{{\gamma }){\delta }},\eqno(65)
$$
where $({\quad })$ denotes the symmetrical alternation.

So, we have proved this

\begin{theorem}
The na--maps from la--space $\xi $ to la--space ${\underline{\xi }}$~ with a
fixed common nonlinear connection $N_j^a(u)={\underline{N}}_j^a(u)$ and
given d--connections, ${\Gamma }_{{.}{\beta }{\gamma }}^\alpha $~ on $\xi $~
and ${\underline{\Gamma }}_{{.}{\beta }{\gamma }}^\alpha $~ on ${\underline{%
\xi }}$ are locally parametrized by the solutions of equations (63) and
(65) for every point $u^\alpha $~ and direction $v^\alpha $~ on $U{\subset }%
{\xi }.$
\end{theorem}

We call (63) and (65) the basic equations for na--maps of la--spaces. They
generalize the corresponding Sinyukov's equations \cite{sin} for isotropic
spaces provided with symmetric affine connection structure.

\section{ Classification of Na--Maps of La--Spaces}

Na--maps are classed on possible polynomial parametrizations on variables $%
v^{\alpha}$~ of deformations parameters $a$ and $b$ (see (63) and (64) ).
\begin{theorem}
There are four classes of na--maps characterized by cor\-res\-pond\-ing
deformation parameters and tensors and basic equations:

\begin{enumerate}
\item  for $na_{(0)}$--maps, ${\pi }_{(0)}$--maps,
$$
P_{{\beta }{\gamma }}^\alpha (u)={\psi }_{{(}{\beta }}{\delta }_{{\gamma }%
)}^\alpha
$$
(${\delta }_\beta ^\alpha $~ is Kronecker symbol and ${\psi }_\beta ={\psi }%
_\beta (u)$~ is a covariant vector d--field);

\item  for $na_{(1)}$--maps
$$
a(u,v)=a_{{\alpha }{\beta }}(u)v^\alpha v^\beta ,\quad b(u,v)=b_\alpha
(u)v^\alpha
$$
and $P_{{.}{\beta }{\gamma }}^\alpha (u)$~ is the solution of equations
$$
D_{({\alpha }}P_{{.}{\beta }{\gamma })}^\delta +P_{({\alpha }{\beta }}^\tau
P_{{.}{\gamma }){\tau }}^\delta -P_{({\alpha }{\beta }}^\tau Q_{{.}{\gamma })%
{\tau }}^\delta =b_{({\alpha }}P_{{.}{\beta }{\gamma })}^{{\delta }}+a_{({%
\alpha }{\beta }}{\delta }_{{\gamma })}^\delta ;\eqno(66)
$$

\item  for $na_{(2)}$--maps
$$
a(u,v)=a_\beta (u)v^\beta ,\quad b(u,v)={\frac{{b_{{\alpha }{\beta }%
}v^\alpha v^\beta }}{{{\sigma }_\alpha (u)v^\alpha }}},\quad {\sigma }%
_\alpha v^\alpha {\neq }0,
$$
$$
P_{{.}{\alpha }{\beta }}^\tau (u)={{\psi }_{({\alpha }}}{\delta }_{{\beta }%
)}^\tau +{\sigma }_{({\alpha }}F_{{\beta })}^\tau
$$
and $F_\beta ^\alpha (u)$~ is the solution of equations
$$
{D}_{({\gamma }}F_{{\beta })}^\alpha +F_\delta ^\alpha F_{({\gamma }}^\delta
{\sigma }_{{\beta })}-Q_{{.}{\tau }({\beta }}^\alpha F_{{\gamma })}^\tau ={%
\mu }_{({\beta }}F_{{\gamma })}^\alpha +{\nu }_{({\beta }}{\delta }_{{\gamma
})}^\alpha \eqno(67)
$$
$({\mu }_\beta (u),{\nu }_\beta (u),{\psi }_\alpha (u),{\sigma }_\alpha (u)$%
~ are covariant d--vectors) ;

\item  for $na_{(3)}$--maps
$$
b(u,v)={\frac{{{\alpha }_{{\beta }{\gamma }{\delta }}v^\beta v^\gamma
v^\delta }}{{{\sigma }_{{\alpha }{\beta }}v^\alpha v^\gamma }}},
$$
$$
P_{{.}{\beta }{\gamma }}^\alpha (u)={\psi }_{({\beta }}{\delta }_{{\gamma }%
)}^\alpha +{\sigma }_{{\beta }{\gamma }}{\varphi }^\alpha ,
$$
where ${\varphi }^\alpha $~ is the solution of equations

$$
D_\beta {\varphi }^\alpha ={\nu }{\delta }_\beta ^\alpha +{\mu }_\beta {%
\varphi }^\alpha +{\varphi }^\gamma Q_{{.}{\gamma }{\delta }}^\alpha ,%
\eqno(68)
$$
${\alpha }_{{\beta }{\gamma }{\delta }}(u),{\sigma }_{{\alpha }{\beta }}(u),{%
\psi }_\beta (u),{\nu }(u)$~ and ${\mu }_\beta (u)$~ are d--tensors.
\end{enumerate}
\end{theorem}

{\it Proof.} We sketch the proof respectively for every point in the theorem:

\begin{enumerate}
\item  It is easy to verify that a--parallel equations (61) on $\xi $
transform into similar ones on $\underline{\xi }$ if and only if
deformations (4.2) with deformation d--tensors of type ${P^\alpha }_{\beta
\gamma }(u)={\psi }_{(\beta }{\delta }_{\gamma )}^\alpha $ are considered.

\item  Using corresponding to $na_{(1)}$--maps parametrizations of $a(u,v)$
and $b(u,v)$ (see conditions of the theorem) for arbitrary $v^\alpha \neq 0$
on $U\in \xi $ and after a redefinition of deformation parameters we obtain
that equations (63) hold if and only if ${P^\alpha }_{\beta \gamma }$
satisfies (60).

\item  In a similar manner we obtain basic $na_{(2)}$--map equations (67)
from (63) by considering $na_{(2)}$--parametrizations of deformation
parameters and d--tensor.

\item  For $na_{(3)}$--maps we mast take into consideration deformations of
torsion (4.3) and introduce $na_{(3)}$--parametrizations for $b(u,v)$ and ${%
P^\alpha }_{\beta \gamma }$ into the basic na--equations (63). The last
ones for $na_{(3)}$--maps are equivalent to equations (68) (with a
corresponding redefinition of deformation parameters). \qquad $\Box $
\end{enumerate}

We point out that for ${\pi}_{(0)}$-maps we have not differential equations
on $P^{\alpha}_{{.}{\beta}{\gamma}}$ (in the isotropic case one considers a
first order system of differential equations on metric \cite{sin}; we omit
constructions with deformation of metric in this section).\

To formulate invariant conditions for reciprocal na--maps (when every
a-parallel on ${\underline {\xi}}$~ is also transformed into na--parallel on
$\xi$ ) it is convenient to introduce into consideration the curvature and
Ricci tensors defined for auxiliary connection ${\gamma}^{\alpha}_{{.}{\beta}%
{\gamma}}$~ :
$$
r^{{.}{\delta}} _{{\alpha}{.}{\beta}{\tau}}={\partial}_{[{\beta}}{\gamma}%
^{\delta}_ {{.}{\tau}]{\alpha}}+{\gamma}^{\delta}_{{.}{\rho}[{\beta}}{\gamma}%
^{\rho}_ {{.}{\tau}]{\alpha}} + {{\gamma}^{\delta}}_{\alpha \phi} {w^{\phi}}%
_{\beta \tau}%
$$
and, respectively, $r_{{\alpha}{\tau}}=r^{{.}{\gamma}} _{{\alpha}{.}{\gamma}{%
\tau}} $, where $[\quad ]$ denotes antisymmetric alternation of indices, and
to define values:
$$
^{(0)}T^{\mu}_{{.}{\alpha}{\beta}}= {\Gamma}^{\mu}_{{.}{\alpha}{\beta}} -
T^{\mu}_{{.}{\alpha}{\beta}}- {\frac{1 }{(n+m + 1)}}({\delta}^{\mu}_{({\alpha%
}}{\Gamma}^{\delta}_ {{.}{\beta}){\delta}}-{\delta}^{\mu}_{({\alpha}%
}T^{\delta}_ {{.}{\beta}){\gamma}}),
$$
$$
{}^{(0)}{W}^{\cdot \tau}_{\alpha \cdot \beta \gamma} = {r}^{\cdot
\tau}_{\alpha \cdot \beta \gamma} + {\frac{1}{n+m+1}} [ {\gamma}%
^{\tau}_{\cdot \varphi \tau} {\delta}^{\tau}_{( \alpha} {w^{\varphi}}_{\beta
) \gamma} - ( {\delta}^{\tau}_{\alpha}{r}_{[ \gamma \beta ]} + {\delta}%
^{\tau}_{\gamma} {r}_{[ \alpha \beta ]} - {\delta}^{\tau}_{\beta} {r}_{[
\alpha \gamma ]} )] -%
$$
$$
{\frac{1}{{(n+m+1)}^2}} [ {\delta}^{\tau}_{\alpha} ( 2 {\gamma}%
^{\tau}_{\cdot \varphi \tau} {w^{\varphi}}_{[ \gamma \beta ] } - {\gamma}%
^{\tau}_{\cdot \tau [ \gamma } {w^{\varphi}}_{\beta ] \varphi} ) + {\delta}%
^{\tau}_{\gamma} ( 2 {\gamma}^{\tau}_{\cdot \varphi \tau} {w^{\varphi}}%
_{\alpha \beta} -{\gamma}^{\tau}_{\cdot \alpha \tau} {w^{\varphi}}_{\beta
\varphi}) -
$$
$$
{\delta}^{\tau}_{\beta} ( 2 {\gamma}^{\tau}_{\cdot \varphi \tau} {w^{\varphi}%
}_{\alpha \gamma} - {\gamma}^{\tau}_{\cdot \alpha \tau} {w^{\varphi}}%
_{\gamma \varphi} ) ],%
$$
$$
{^{(3)}T}^{\delta}_{{.}{\alpha}{\beta}}= {\gamma}^{\delta}_{{.}{\alpha}{\beta%
}}+ {\epsilon}{\varphi}^{\tau}{^{({\gamma})}D}_{\beta}q_{\tau}+ {\frac{1 }{%
n+m}}({\delta}^{\gamma}_{\alpha}- {\epsilon}{\varphi}^{\delta}q_{\alpha})[{%
\gamma}^{\tau}_{{.}{\beta}{\tau}}+ {\epsilon}{\varphi}^{\tau}{^{({\gamma})}D}%
_{\beta}q_{\tau}+%
$$
$$
{\frac{1 }{{n+m -1}}}q_{\beta}({\epsilon}{\varphi}^{\tau}{\gamma}^{\lambda}_
{{.}{\tau}{\lambda}}+ {\varphi}^{\lambda}{\varphi}^{\tau}{^{({\gamma})}D}%
_{\tau}q_{\lambda})]- {\frac{1 }{n+m}}({\delta}^{\delta}_{\beta}-{\epsilon}{%
\varphi}^{\delta} q_{\beta})[{\gamma}^{\tau}_{{.}{\alpha}{\tau}}+
$$
$$ {\epsilon}{%
\varphi}^{\tau} {^{({\gamma})}D}_{\alpha}q_{\tau}%
+{\frac{1 }{{n+m -1}}}q_{\alpha}({\epsilon}{\varphi}^{\tau}{\gamma}%
^{\lambda}_ {{.}{\tau}{\lambda}}+ {\varphi}^{\lambda}{\varphi}^{\tau} {%
^{(\gamma)}D} _{\tau}q_{\lambda})],%
$$
$$
{^{(3)}W}^\alpha {{.}{\beta }{\gamma }{\delta }}={\rho }_{{\beta }{.}{\gamma
}{\delta }}^{{.}{\alpha }}+{\epsilon }{\varphi }^\alpha q_\tau {\rho }_{{%
\beta }{.}{\gamma }{\delta }}^{{.}{\tau }}+({\delta }_\delta ^\alpha -
$$
$$
{\epsilon }{\varphi }^\alpha q_\delta )p_{{\beta }{\gamma }}-({\delta }%
_\gamma ^\alpha -{\epsilon }{\varphi }^\alpha q_\gamma )p_{{\beta }{\delta }%
}-({\delta }_\beta ^\alpha -{\epsilon }{\varphi }^\alpha q_\beta )p_{[{%
\gamma }{\delta }]},
$$
$$
(n+m-2)p_{{\alpha }{\beta }}=-{\rho }_{{\alpha }{\beta }}-{\epsilon }q_\tau {%
\varphi }^\gamma {\rho }_{{\alpha }{.}{\beta }{\gamma }}^{{.}{\tau }}+{\frac
1{n+m}}[{\rho }_{{\tau }{.}{\beta }{\alpha }}^{{.}{\tau }}-{\epsilon }q_\tau
{\varphi }^\gamma {\rho }_{{\gamma }{.}{\beta }{\alpha }}^{{.}{\tau }}+{%
\epsilon }q_\beta {\varphi }^\tau {\rho }_{{\alpha }{\tau }}+
$$
$$
{\epsilon }q_\alpha (-{\varphi }^\gamma {\rho }_{{\tau }{.}{\beta }{\gamma }%
}^{{.}{\tau }}+{\epsilon }q_\tau {\varphi }^\gamma {\varphi }^\delta {\rho }%
_{{\gamma }{.}{\beta }{\delta }}^{{.}{\tau }}]),
$$
where $q_\alpha {\varphi }^\alpha ={\epsilon }=\pm 1,$
$$
{{\rho }^\alpha }_{\beta \gamma \delta }=r_{\beta \cdot \gamma \delta
}^{\cdot \alpha }+{\frac 12}({\psi }_{(\beta }{\delta }_{\varphi )}^\alpha +{%
\sigma }_{\beta \varphi }{\varphi }^\tau ){w^\varphi }_{\gamma \delta }
$$
( for a similar value on $\underline{\xi }$ we write ${\quad }{\underline{%
\rho }}_{\cdot \beta \gamma \delta }^\alpha ={\underline{r}}_{\beta \cdot
\gamma \delta }^{\cdot \alpha }-{\frac 12}({\psi }_{(\beta }{\delta }_{{%
\varphi })}^\alpha -{\sigma }_{\beta \varphi }{\varphi }^\tau ){w^\varphi }%
_{\gamma \delta }{\quad })$ and ${\rho }_{\alpha \beta }={\rho }_{\cdot
\alpha \beta \tau }^\tau .$

Similar values,
$$
^{(0)}{\underline T}^{\alpha}_{{.}{\beta}{\gamma}}, ^{(0)}{\underline W}%
^{\nu}_{{.}{\alpha}{\beta}{\gamma}}, {\hat T}^{\alpha} _{{.}{\beta}{\gamma}%
}, {\check T}^{\alpha}_ {{.}{\beta}{\tau}}, {\hat W}^{\delta}_{{.}{\alpha}{%
\beta}{\gamma}}, {\check W}^{\delta}_ {{.}{\alpha}{\beta}{\gamma}}, ^{(3)}{%
\underline T}^{\delta} _{{.}{\alpha}{\beta}},%
$$
and $^{(3)}{\underline W}^ {\alpha}_{{.}{\beta}{\gamma}{\delta}} $ are
given, correspondingly, by auxiliary connections ${\quad}{\underline {\Gamma}%
}^{\mu}_{{.}{\alpha}{\beta}},$~
$$
{\star {\gamma}}^{\alpha}_{{.}{\beta}{\lambda}}={\gamma}^{\alpha} _{{.}{\beta%
}{\lambda}} + {\epsilon}F^{\alpha}_{\tau}{^{({\gamma})}D}_{({\beta}}
F^{\tau}_{{\lambda})}, \quad {\check {\gamma}}^{\alpha}_{{.}{\beta}{\lambda}%
}= {\widetilde {\gamma}}^{\alpha}_{{.}{\beta}{\lambda}} + {\epsilon}%
F^{\lambda} _{\tau} {\widetilde D}_{({\beta}}F^{\tau}_{{\lambda})},%
$$
$$
{\widetilde {\gamma}}^{\alpha}_{{.}{\beta}{\tau}}={\gamma}^{\alpha} _{{.}{%
\beta}{\tau}}+ {\sigma}_{({\beta}}F^{\alpha}_{{\tau})}, \quad {\hat {\gamma}}%
^{\alpha}_{{.}{\beta}{\lambda}}={\star {\gamma}}^{\alpha}_ {{.}{\beta}{%
\lambda}} + {\widetilde {\sigma}}_{({\beta}}{\delta}^{\alpha}_ {{\lambda})},%
$$
where ${\widetilde {\sigma}}_{\beta}={\sigma}_{\alpha}F^{\alpha}_{\beta}.$

\begin{theorem}
Four classes of reciprocal na--maps of la--spaces are characterized by
corresponding invariant criterions:

\begin{enumerate}
\item  for a--maps $^{(0)}T_{{.}{\alpha }{\beta }}^\mu =^{(0)}{\underline{T}}%
_{{.}{\alpha }{\beta }}^\mu ,$
$$
{}^{(0)}W_{{.}{\alpha }{\beta }{\gamma }}^\delta =^{(0)}{\underline{W}}_{{.}{%
\alpha }{\beta }{\gamma }}^\delta ;\eqno(69)
$$

\item  for $na_{(1)}$--maps
$$
3({^{({\gamma })}D}_\lambda P_{{.}{\alpha }{\beta }}^\delta +P_{{.}{\tau }{%
\lambda }}^\delta P_{{.}{\alpha }{\beta }}^\tau )=r_{({\alpha }{.}{\beta }){%
\lambda }}^{{.}{\delta }}-{\underline{r}}_{({\alpha }{.}{\beta }){\lambda }%
}^{{.}{\delta }}+\eqno(70)
$$
$$
[T_{{.}{\tau }({\alpha }}^\delta P_{{.}{\beta }{\lambda })}^\tau +Q_{{.}{%
\tau }({\alpha }}^\delta P_{{.}{\beta }{\lambda })}^\tau +b_{({\alpha }}P_{{.%
}{\beta }{\lambda })}^\delta +{\delta }_{({\alpha }}^\delta a_{{\beta }{%
\lambda })}];
$$

\item  for $na_{(2)}$--maps ${\hat T}_{{.}{\beta }{\tau }}^\alpha ={\star T}%
_{{.}{\beta }{\tau }}^\alpha ,$
$$
{\hat W}_{{.}{\alpha }{\beta }{\gamma }}^\delta ={\star W}_{{.}{\alpha }{%
\beta }{\gamma }}^\delta ;\eqno(71)
$$

\item  for $na_{(3)}$--maps $^{(3)}T_{{.}{\beta }{\gamma }}^\alpha =^{(3)}{%
\underline{T}}_{{.}{\beta }{\gamma }}^\alpha ,$
$$
{}^{(3)}W_{{.}{\beta }{\gamma }{\delta }}^\alpha =^{(3)}{\underline{W}}_{{.}{%
\beta }{\gamma }{\delta }}^\alpha .\eqno(72)
$$
\end{enumerate}
\end{theorem}

{\it Proof. }

\begin{enumerate}
\item  Let us prove that a--invariant conditions (69) hold. Deformations
of d--connections of type
$$
{}^{(0)}{\underline{\gamma }}_{\cdot \alpha \beta }^\mu ={{\gamma }^\mu }%
_{\alpha \beta }+{\psi }_{(\alpha }{\delta }_{\beta )}^\mu \eqno(73)
$$
define a--applications. Contracting indices $\mu $ and $\beta $ we can write
$$
{\psi }_\alpha ={\frac 1{m+n+1}}({{\underline{\gamma }}^\beta }_{\alpha
\beta }-{{\gamma }^\beta }_{\alpha \beta }).\eqno(74)
$$
Introducing d--vector ${\psi }_\alpha $ into previous relation and
expressing
$$
{{\gamma }^\alpha }_{\beta \tau }=-{T^\alpha }_{\beta \tau }+{{\Gamma }%
^\alpha }_{\beta \tau }
$$
and similarly for underlined values we obtain the first invariant conditions
from (69).

Putting deformation (73) into the formula for
$$
{\underline{r}}_{\alpha \cdot \beta \gamma }^{\cdot \tau }\quad \mbox{and}%
\quad {\underline{r}}_{\alpha \beta }={\underline{r}}_{\alpha \tau \beta
\tau }^{\cdot \tau }
$$
we obtain respectively relations
$$
{\underline{r}}_{\alpha \cdot \beta \gamma }^{\cdot \tau }-r_{\alpha \cdot
\beta \gamma }^{\cdot \tau }={\delta }_\alpha ^\tau {\psi }_{[\gamma \beta
]}+{\psi }_{\alpha [\beta }{\delta }_{\gamma ]}^\tau +{\delta }_{(\alpha
}^\tau {\psi }_{\varphi )}{w^\varphi }_{\beta \gamma }\eqno(75)
$$
and
$$
{\underline{r}}_{\alpha \beta }-r_{\alpha \beta }={\psi }_{[\alpha \beta
]}+(n+m-1){\psi }_{\alpha \beta }+{\psi }_\varphi {w^\varphi }_{\beta \alpha
}+{\psi }_\alpha {w^\varphi }_{\beta \varphi },\eqno(76)
$$
where
$$
{\psi }_{\alpha \beta }={}^{({\gamma })}D_\beta {\psi }_\alpha -{\psi }%
_\alpha {\psi }_\beta .
$$
Putting (73) into (75) and (76) we can express ${\psi }_{[\alpha \beta ]}$ as
$$
{\psi }_{[\alpha \beta ]}={\frac 1{n+m+1}}[{\underline{r}}_{[\alpha \beta ]}+%
{\frac 2{n+m+1}}{\underline{\gamma }}_{\cdot \varphi \tau }^\tau {w^\varphi }%
_{[\alpha \beta ]}-{\frac 1{n+m+1}}{\underline{\gamma }}_{\cdot \tau [\alpha
}^\tau {w^\varphi }_{\beta ]\varphi }]
$$
$$
-{\frac 1{n+m+1}}[r_{[\alpha \beta ]}+{\frac 2{n+m+1}}{{\gamma }^\tau }%
_{\varphi \tau }{w^\varphi }_{[\alpha \beta ]}-{\frac 1{n+m+1}}{{\gamma }%
^\tau }_{\tau [\alpha }{w^\varphi }_{\beta ]\varphi }].\eqno(77)
$$
To simplify our consideration we can choose an a--transform,
pa\-ra\-met\-riz\-ed by corresponding $\psi $--vector from (73), (or fix a
local coordinate cart) the antisymmetrized relations (77) to be satisfied
by d--tensor
$$
{\psi }_{\alpha \beta }={\frac 1{n+m+1}}[{\underline{r}}_{\alpha \beta }+{%
\frac 2{n+m+1}}{\underline{\gamma }}_{\cdot \varphi \tau }^\tau {w^\varphi }%
_{\alpha \beta }-{\frac 1{n+m+1}}{\underline{\gamma }}_{\cdot \alpha \tau
}^\tau {w^\varphi }_{\beta \varphi }-%
$$
$$
r_{\alpha \beta }- {\frac 2{n+m+1}}{{\gamma }^\tau }_{\varphi \tau }{%
w^\varphi }_{\alpha \beta }+{\frac 1{n+m+1}}{{\gamma }^\tau }_{\alpha \tau }{%
w^\varphi }_{\beta \varphi }]\eqno(78)
$$
Introducing expressions (73),(77) and (78) into deformation of
curvature (74) we obtain the second conditions (69) of a-map invariance:
$$
^{(0)}W_{\alpha \cdot \beta \gamma }^{\cdot \delta }={}^{(0)}{\underline{W}}%
_{\alpha \cdot \beta \gamma }^{\cdot \delta },
$$
where the Weyl d--tensor on $\underline{\xi }$ (the extension of the usual
one for geodesic maps on (pseudo)--Riemannian spaces to the case of
v--bund\-les provided with N--connection structure) is defined as
$$
{}^{(0)}{\underline{W}}_{\alpha \cdot \beta \gamma }^{\cdot \tau }=$$
$$
{\underline{r}}_{\alpha \cdot \beta \gamma }^{\cdot \tau }+{\frac 1{n+m+1}}[{%
\underline{\gamma }}_{\cdot \varphi \tau }^\tau {\delta }_{(\alpha }^\tau {%
w^\varphi }_{\beta )\gamma }-({\delta }_\alpha ^\tau {\underline{r}}%
_{[\gamma \beta ]}+{\delta }_\gamma ^\tau {\underline{r}}_{[\alpha \beta ]}-{%
\delta }_\beta ^\tau {\underline{r}}_{[\alpha \gamma ]})]-
$$
$$
{\frac 1{{(n+m+1)}^2}}[{\delta }_\alpha ^\tau (2{\underline{\gamma }}_{\cdot
\varphi \tau }^\tau {w^\varphi }_{[\gamma \beta ]}-{\underline{\gamma }}%
_{\cdot \tau [\gamma }^\tau {w^\varphi }_{\beta ]\varphi })+{\delta }_\gamma
^\tau (2{\underline{\gamma }}_{\cdot \varphi \tau }^\tau {w^\varphi }%
_{\alpha \beta }-{\underline{\gamma }}_{\cdot \alpha \tau }^\tau {w^\varphi }%
_{\beta \varphi })-
$$
$$
{\delta }_\beta ^\tau (2{\underline{\gamma }}_{\cdot \varphi \tau }^\tau {%
w^\varphi }_{\alpha \gamma }-{\underline{\gamma }}_{\cdot \alpha \tau }^\tau
{w^\varphi }_{\gamma \varphi })].
$$

\item  To obtain $na_{(1)}$--invariant conditions we rewrite $na_{(1)}$%
--equations (66) as to consider in explicit form covariant derivation $^{({%
\gamma })}D$ and deformations (4.2) and (4.3):
$$
2({}^{({\gamma })}D_\alpha {P^\delta }_{\beta \gamma }+{}^{({\gamma }%
)}D_\beta {P^\delta }_{\alpha \gamma }+{}^{({\gamma })}D_\gamma {P^\delta }%
_{\alpha \beta }+{P^\delta }_{\tau \alpha }{P^\tau }_{\beta \gamma }+
$$
$$
{P^\delta }_{\tau \beta }{P^\tau }_{\alpha \gamma }+{P^\delta }_{\tau \gamma
}{P^\tau }_{\alpha \beta })={T^\delta }_{\tau (\alpha }{P^\tau }_{\beta
\gamma )}+
$$
$$
{H^\delta }_{\tau (\alpha }{P^\tau }_{\beta \gamma )}+b_{(\alpha }{P^\delta }%
_{\beta \gamma )}+a_{(\alpha \beta }{\delta }_{\gamma )}^\delta .\eqno(79)
$$
Alternating the first two indices in (79) we have
$$
2({\underline{r}}_{(\alpha \cdot \beta )\gamma }^{\cdot \delta }-r_{(\alpha
\cdot \beta )\gamma }^{\cdot \delta })=2({}^{(\gamma )}D_\alpha {P^\delta }%
_{\beta \gamma }+
$$
$$
{}^{(\gamma )}D_\beta {P^\delta }_{\alpha \gamma }-2{}^{(\gamma )}D_\gamma {%
P^\delta }_{\alpha \beta }+{P^\delta }_{\tau \alpha }{P^\tau }_{\beta \gamma
}+{P^\delta }_{\tau \beta }{P^\tau }_{\alpha \gamma }-2{P^\delta }_{\tau
\gamma }{P^\tau }_{\alpha \beta }).
$$
Substituting the last expression from (79) and rescalling the deformation
parameters and d--tensors we obtain the conditions (66).

\item  Now we prove the invariant conditions for $na_{(0)}$--maps satisfying
conditions
$$
\epsilon \neq 0\quad \mbox{and}\quad \epsilon -F_\beta ^\alpha F_\alpha
^\beta \neq 0
$$
Let define the auxiliary d--connection
$$
{\tilde \gamma }_{\cdot \beta \tau }^\alpha ={\underline{\gamma }}_{\cdot
\beta \tau }^\alpha -{\psi }_{(\beta }{\delta }_{\tau )}^\alpha ={{\gamma }%
^\alpha }_{\beta \tau }+{\sigma }_{(\beta }F_{\tau )}^\alpha \eqno(80)
$$
and write
$$
{\tilde D}_\gamma ={}^{({\gamma })}D_\gamma F_\beta ^\alpha +{\tilde \sigma }%
_\gamma F_\beta ^\alpha -{\epsilon }{\sigma }_\beta {\delta }_\gamma ^\alpha
,
$$
where ${\tilde \sigma }_\beta ={\sigma }_\alpha F_\beta ^\alpha ,$ or, as a
consequence from the last equality,
$$
{\sigma }_{(\alpha }F_{\beta )}^\tau ={\epsilon }F_\lambda ^\tau ({}^{({%
\gamma })}D_{(\alpha }F_{\beta )}^\alpha -{\tilde D}_{(\alpha }F_{\beta
)}^\lambda )+{\tilde \sigma }_{(}{\alpha }{\delta }_{\beta )}^\tau .
$$
Introducing auxiliary connections
$$
{\star {\gamma }}_{\cdot \beta \lambda }^\alpha ={\gamma }_{\cdot \beta
\lambda }^\alpha +{\epsilon }F_\tau ^\alpha {}^{({\gamma })}D_{(\beta
}F_{\lambda )}^\tau
$$
and
$$
{\check \gamma }_{\cdot \beta \lambda }^\alpha ={\tilde \gamma }_{\cdot
\beta \lambda }^\alpha +{\epsilon }F_\tau ^\alpha {\tilde D}_{(\beta
}F_{\lambda )}^\tau
$$
we can express deformation (80) in a form characteristic for a--maps:
$$
{\hat \gamma }_{\cdot \beta \gamma }^\alpha ={\star {\gamma }}_{\cdot \beta
\gamma }^\alpha +{\tilde \sigma }_{(\beta }{\delta }_{\lambda )}^\alpha .%
\eqno(81)
$$
Now it's obvious that $na_{(2)}$--invariant conditions (81) are equivalent
with a--invariant conditions (69) written for d--connection (81). As a
matter of principle we can write formulas for such $na_{(2)}$--invariants in
terms of ''underlined'' and ''non--underlined'' values by expressing
consequently all used auxiliary connections as deformations of ''prime''
connections on $\xi $ and ''final'' connections on $\underline{\xi }.$ We
omit such tedious calculations in this work.

\item  Finally, we prove the last statement, for $na_{(3)}$--maps, of this
theorem. Let
$$
q_\alpha {\varphi }^\alpha =e=\pm 1,\eqno(82)
$$
where ${\varphi }^\alpha $ is contained in
$$
{\underline{\gamma }}_{\cdot \beta \gamma }^\alpha ={{\gamma }^\alpha }%
_{\beta \gamma }+{\psi }_{(\beta }{\delta }_{\gamma )}^\alpha +{\sigma }%
_{\beta \gamma }{\varphi }^\alpha .\eqno(83)
$$
Acting with operator $^{({\gamma })}{\underline{D}}_\beta $ on (82) we
write
$$
{}^{({\gamma })}{\underline{D}}_\beta q_\alpha ={}^{({\gamma })}D_\beta
q_\alpha -{\psi }_{(\alpha }q_{\beta )}-e{\sigma }_{\alpha \beta }.%
\eqno(84)
$$
Contracting (84) with ${\varphi }^\alpha $ we can express
$$
e{\varphi }^\alpha {\sigma }_{\alpha \beta }={\varphi }^\alpha ({}^{({\gamma
})}D_\beta q_\alpha -{}^{({\gamma })}{\underline{D}}_\beta q_\alpha )-{%
\varphi }_\alpha q^\alpha q_\beta -e{\psi }_\beta .
$$
Putting the last formula in (83) contracted on indices $\alpha $ and $%
\gamma $ we obtain
$$
(n+m){\psi }_\beta ={\underline{\gamma }}_{\cdot \alpha \beta }^\alpha -{{%
\gamma }^\alpha }_{\alpha \beta }+e{\psi }_\alpha {\varphi }^\alpha q_\beta
+e{\varphi }^\alpha {\varphi }^\beta ({}^{({\gamma })}{\underline{D}}_\beta
-{}^{({\gamma })}D_\beta ).\eqno(85)
$$
From these relations, taking into consideration (82), we have%
$$
(n+m-1){\psi }_\alpha {\varphi }^\alpha =
$$
$$
{\varphi }^\alpha ({\underline{\gamma }}_{\cdot \alpha \beta }^\alpha -{{%
\gamma }^\alpha }_{\alpha \beta })+e{\varphi }^\alpha {\varphi }^\beta ({}^{(%
{\gamma })}{\underline{D}}_\beta q_\alpha -{}^{({\gamma })}D_\beta q_\alpha
)
$$

Using the equalities and identities (84) and (85) we can express
deformations (83) as the first $na_{(3)}$--invariant conditions from
(72).

To prove the second class of $na_{(3)}$--invariant conditions we introduce
two additional d--tensors:
$$
{{\rho }^\alpha }_{\beta \gamma \delta }=r_{\beta \cdot \gamma \delta
}^{\cdot \alpha }+{\frac 12}({\psi }_{(\beta }{\delta }_{\varphi )}^\alpha +{%
\sigma }_{\beta \varphi }{\varphi }^\tau ){w^\varphi }_{\gamma \delta }{%
\quad }
$$
and
$$
{\underline{\rho }}_{\cdot \beta \gamma \delta }^\alpha ={\underline{r}}%
_{\beta \cdot \gamma \delta }^{\cdot \alpha }-{\frac 12}({\psi }_{(\beta }{%
\delta }_{{\varphi })}^\alpha -{\sigma }_{\beta \varphi }{\varphi }^\tau ){%
w^\varphi }_{\gamma \delta }.\eqno(86)
$$
Using deformation (83) and (86) we write relation
$$
{\tilde \sigma }_{\cdot \beta \gamma \delta }^\alpha ={\underline{\rho }}%
_{\cdot \beta \gamma \delta }^\alpha -{\rho }_{\cdot \beta \gamma \delta
}^\alpha ={\psi }_{\beta [\delta }{\delta }_{\gamma ]}^\alpha -{\psi }_{[{%
\gamma }{\delta }]}{\delta }_\beta ^\alpha -{\sigma }_{\beta \gamma \delta }{%
\varphi }^\alpha ,\eqno(87)
$$
where
$$
{\psi }_{\alpha \beta }={}^{({\gamma })}D_\beta {\psi }_\alpha +{\psi }%
_\alpha {\psi }_\beta -({\nu }+{\varphi }^\tau {\psi }_\tau ){\sigma }%
_{\alpha \beta },
$$
and
$$
{\sigma }_{\alpha \beta \gamma }={}^{({\gamma })}D_{[\gamma }{\sigma }_{{%
\beta }]{\alpha }}+{\mu }_{[\gamma }{\sigma }_{{\beta }]{\alpha }}-{\sigma }%
_{{\alpha }[{\gamma }}{\sigma }_{{\beta }]{\tau }}{\varphi }^\tau .
$$
Let multiply (87) on $q_\alpha $ and write (taking into account relations
(82)) the relation
$$
e{\sigma }_{\alpha \beta \gamma }=-q_\tau {\tilde \sigma }_{\cdot \alpha
\beta \delta }^\tau +{\psi }_{\alpha [\beta }q_{\gamma ]}-{\psi }_{[\beta
\gamma ]}q_\alpha .\eqno(88)
$$
The next step is to express ${\psi }_{\alpha \beta }$ trough d--objects on ${%
\xi }.$ To do this we contract indices $\alpha $ and $\beta $ in (87) and
obtain
$$
(n+m){\psi }_{[\alpha \beta ]}=-{\sigma }_{\cdot \tau \alpha \beta }^\tau
+eq_\tau {\varphi }^\lambda {\sigma }_{\cdot \lambda \alpha \beta }^\tau -e{%
\tilde \psi }_{[\alpha }{\tilde \psi }_{\beta ]}.
$$
Then contracting indices $\alpha $ and $\delta $ in (87) and using (88)
we write
$$
(n+m-2){\psi }_{\alpha \beta }={\tilde \sigma }_{\cdot \alpha \beta \tau
}^\tau -eq_\tau {\varphi }^\lambda {\tilde \sigma }_{\cdot \alpha \beta
\lambda }^\tau +{\psi }_{[\beta \alpha ]}+e({\tilde \psi }_\beta q_\alpha -{%
\hat \psi }_{(\alpha }q_{\beta )}),\eqno(89)
$$
where ${\hat \psi }_\alpha ={\varphi }^\tau {\psi }_{\alpha \tau }.$ If the
both parts of (89) are contracted with ${\varphi }^\alpha ,$ it results
that
$$
(n+m-2){\tilde \psi }_\alpha ={\varphi }^\tau {\sigma }_{\cdot \tau \alpha
\lambda }^\lambda -eq_\tau {\varphi }^\lambda {\varphi }^\delta {\sigma }%
_{\lambda \alpha \delta }^\tau -eq_\alpha ,
$$
and, in consequence of ${\sigma }_{\beta (\gamma \delta )}^\alpha =0,$ we
have
$$
(n+m-1){\varphi }={\varphi }^\beta {\varphi }^\gamma {\sigma }_{\cdot \beta
\gamma \alpha }^\alpha .
$$
By using the last expressions we can write
$$
(n+m-2){\underline{\psi }}_\alpha ={\varphi }^\tau {\sigma }_{\cdot \tau
\alpha \lambda }^\lambda -eq_\tau {\varphi }^\lambda {\varphi }^\delta {%
\sigma }_{\cdot \lambda \alpha \delta }^\tau -e{(n+m-1)}^{-1}q_\alpha {%
\varphi }^\tau {\varphi }^\lambda {\sigma }_{\cdot \tau \lambda \delta
}^\delta .\eqno(90)
$$
Contracting (89) with ${\varphi }^\beta $ we have
$$
(n+m){\hat \psi }_\alpha ={\varphi }^\tau {\sigma }_{\cdot \alpha \tau
\lambda }^\lambda +{\tilde \psi }_\alpha
$$
and taking into consideration (90) we can express ${\hat \psi }_\alpha $
through ${\sigma }_{\cdot \beta \gamma \delta }^\alpha .$

As a consequence of (88)--(90) we obtain this formulas for d--tensor ${%
\psi }_{\alpha \beta }:$
$$
(n+m-2){\psi }_{\alpha \beta }={\sigma }_{\cdot \alpha \beta \tau }^\tau
-eq_\tau {\varphi }^\lambda {\sigma }_{\cdot \alpha \beta \lambda }^\tau +
$$
$$
{\frac 1{n+m}}\{-{\sigma }_{\cdot \tau \beta \alpha }^\tau +eq_\tau {\varphi
}^\lambda {\sigma }_{\cdot \lambda \beta \alpha }^\tau -q_\beta (e{\varphi }%
^\tau {\sigma }_{\cdot \alpha \tau \lambda }^\lambda -q_\tau {\varphi }%
^\lambda {\varphi }^\delta {\sigma }_{\cdot \alpha \lambda \delta }^\tau )+
eq_\alpha \times
$$
$$
[{\varphi }^\lambda {\sigma }_{\cdot \tau \beta \lambda }^\tau -eq_\tau {%
\varphi }^\lambda {\varphi }^\delta {\sigma }_{\cdot \lambda \beta \delta
}^\tau -{\frac e{n+m-1}}q_\beta ({\varphi }^\tau {\varphi }^\lambda {\sigma }%
_{\cdot \tau \gamma \delta }^\delta -eq_\tau {\varphi }^\lambda {\varphi }%
^\delta {\varphi }^\varepsilon {\sigma }_{\cdot \lambda \delta \varepsilon
}^\tau )]\}.
$$

Finally, putting the last formula and (88) into (87) and after a
rearrangement of terms we obtain the second group of $na_{(3)}$-invariant
conditions (72). If necessary we can rewrite these conditions in terms of
geometrical objects on $\xi $ and $\underline{\xi }.$ To do this we mast
introduce splittings (86) into (72). \qquad $\Box $
\end{enumerate}

For the particular case of $na_{(3)}$--maps when
$$
{\psi}_{\alpha}=0 , {\varphi}_{\alpha} = g_{\alpha \beta} {\varphi}^{\beta}
= {\frac{\delta }{\delta u^{\alpha}}} ( \ln {\Omega} ) , {\Omega}(u) > 0
$$
and
$$
{\sigma}_{\alpha \beta} = g_{\alpha \beta}%
$$
we define a subclass of conformal transforms ${\underline g}_{\alpha \beta}
(u) = {\Omega}^2 (u) g_{\alpha \beta}$ which, in consequence of the fact
that d--vector ${\varphi}_{\alpha}$ must satisfy equations (68),
generalizes the class of concircular transforms (see \cite{sin} for
references and details on concircular mappings of Riemannaian spaces) .

We emphasize that basic na--equations (66)--(68) are systems of first
order partial differential equations. The study of their geometrical
properties and definition of integral varieties, general and particular
solutions are possible by using the formalism of Pffaf systems \cite{vog}.
Here we point out that by using algebraic methods we can always verify if
systems of na--equations of type (66)--(68) are, or not, involute, even
to find their explicit solutions it is a difficult task (see more detailed
considerations for isotropic ng--maps in \cite{sin} and, on language of
Pffaf systems for na--maps, in \cite{vb12}). We can also formulate the
Cauchy problem for na--equations on $\xi $~ and choose deformation
parameters (64) as to make involute mentioned equations for the case of
maps to a given background space ${\underline{\xi }}$. If a solution, for
example, of $na_{(1)}$--map equations exists, we say that space $\xi $ is $%
na_{(1)}$--projective to space ${\underline{\xi }}.$ In general, we have to
introduce chains of na--maps in order to obtain involute systems of
equations for maps (superpositions of na-maps) from $\xi $ to ${\underline{%
\xi }}:$
 %%%%%%%%%%%%%%%%%%%%%%%%%%%%%%%%%%%%%%%%%%%%%%%%%%%
$$U \buildrel {ng<i_{1}>} \over \longrightarrow  {U_{\underline 1}}
\buildrel ng<i_2> \over \longrightarrow \cdots
\buildrel ng<i_{k-1}> \over \longrightarrow U_{\underline {k-1}}
\buildrel ng<i_k> \over \longrightarrow {\underline U} $$
where $U\subset {\xi },U_{\underline{1}}\subset {\xi }_{\underline{1}%
},\ldots ,U_{k-1}\subset {\xi }_{k-1},{\underline{U}}\subset {\xi }_k$ with
corresponding splittings of auxiliary symmetric connections
$$
{\underline{\gamma }}_{.{\beta }{\gamma }}^\alpha =_{<i_1>}P_{.{\beta }{%
\gamma }}^\alpha +_{<i_2>}P_{.{\beta }{\gamma }}^\alpha +\cdots +_{<i_k>}P_{.%
{\beta }{\gamma }}^\alpha
$$
and torsion
$$
{\underline{T}}_{.{\beta }{\gamma }}^\alpha =T_{.{\beta }{\gamma }}^\alpha
+_{<i_1>}Q_{.{\beta }{\gamma }}^\alpha +_{<i_2>}Q_{.{\beta }{\gamma }%
}^\alpha +\cdots +_{<i_k>}Q_{.{\beta }{\gamma }}^\alpha
$$
where cumulative indices $<i_1>=0,1,2,3,$ denote possible types of na--maps.

\begin{definition}
Space $\xi $~ is nearly conformally projective to space ${\underline{\xi }},{%
\quad }nc:{\xi }{\to }{\underline{\xi }},$~ if there is a finite chain of
na--maps from $\xi $~ to ${\underline{\xi }}.$
\end{definition}

For nearly conformal maps we formulate:

\begin{theorem}
For every fixed triples $(N_j^a,{\Gamma }_{{.}{\beta }{\gamma }}^\alpha
,U\subset {\xi }$ and $(N_j^a,{\underline{\Gamma }}_{{.}{\beta }{\gamma }%
}^\alpha $, ${\underline{U}}\subset {\underline{\xi }})$, components of
nonlinear connection, d--connection and d--metric being of class $C^r(U),C^r(%
{\underline{U}})$, $r>3,$ there is a finite chain of na--maps $nc:U\to {%
\underline{U}}.$
\end{theorem}

Proof is similar to that for isotropic maps \cite{vk,vob12,vrjp} (we have to
introduce a finite number of na-maps with corresponding components of
deformation parameters and deformation tensors in order to transform step by
step coefficients of d-connection ${\Gamma}^{\alpha}_{\gamma \delta}$ into ${%
\underline {\Gamma}}^{\alpha}_{\beta \gamma} ).$

Now we introduce the concept of the Category of la--spaces, ${\cal C}({\xi}%
). $ The elements of ${\cal C}({\xi})$ consist from $Ob{\cal C}({\xi})=\{{\xi%
}, {\xi}_{<i_{1}>}, {\xi}_{<i_{2}>},{\ldots} \}$ being la--spaces, for
simplicity in this work, having common N--connection structures, and\\ $Mor
{\cal C}({\xi})=\{ nc ({\xi}_{<i_{1}>}, {\xi}_{<i_{2}>})\}$ being chains of
na--maps interrelating la--spaces. We point out that we can consider
equivalent models of physical theories on every object of ${\cal C}({\xi})$
(see details for isotropic gravitational models in \cite
{vk,vrjp,vog,vb12,vob12,vob13} and anisotropic gravity in \cite
{vcl96,voa,vodg}). One of the main purposes of this section is to develop a
d--tensor and d--variational formalism on ${\cal C}({\xi}),$ i.e. on
la--multispaces, interrelated with nc--maps. Taking into account the
distinguished character of geometrical objects on la--spaces we call tensors
on ${\cal C}({\xi})$ as distinguished tensors on la--space Category, or
dc--tensors.

Finally, we emphasize that presented in this section definitions and
theorems can be generalized for v--bundles with arbitrary given structures
of nonlinear connection, linear d--connection and metric structures. Proofs
are similar to those from \cite{vth,sin}.

\section{Na-Tensor-Integral on La-Spaces}

The aim of this section is to define tensor integration not only for
bitensors, objects defined on the same curved space, but for dc--tensors,
defined on two spaces, $\xi$ and ${\underline {\xi}}$, even it is necessary
on la--multispaces. A. Mo\'or tensor--integral formalism having a lot of
applications in classical and quantum gravity \cite{syn,dewd,goz} was
extended for locally isotropic multispaces in \cite{vog,vob12}. The
unispacial locally anisotropic version is given in \cite{v295a,gv}.

Let $T_{u}{\xi}$~ and $T_{\underline u}{\underline {\xi}}$ be tangent spaces
in corresponding points $u {\in} U {\subset} {\xi}$ and ${\underline u} {\in}
{\underline U} {\subset} {\underline {\xi}}$ and, respectively, $T^{\ast}_{u}%
{\xi}$ and $T^{\ast}_{\underline u}{\underline {\xi}} $ be their duals (in
general, in this section we shall not consider that a common
coordinatization is introduced for open regions $U$ and ${\underline U}$ ).
We call as the dc--tensors on the pair of spaces (${\xi}, {\underline {\xi}}$
) the elements of distinguished tensor algebra
$$
( {\otimes}_{\alpha} T_{u}{\xi}) {\otimes} ({\otimes}_{\beta} T^{\ast}_{u} {%
\xi}) {\otimes}({\otimes}_{\gamma} T_{\underline u}{\underline {\xi}}) {%
\otimes} ({\otimes}_{\delta} T^{\ast}_{\underline u} {\underline {\xi}})%
$$
defined over the space ${\xi}{\otimes} {\underline {\xi}}, $ for a given $nc
: {\xi} {\to} {\underline {\xi}} $.

We admit the convention that underlined and non--underlined indices refer,
respectively, to the points ${\underline u}$ and $u$. Thus $Q_{{.}{%
\underline {\alpha}}}^{\beta}, $ for instance, are the components of
dc--tensor $Q{\in} T_{u}{\xi} {\otimes} T_{\underline u}{\underline {\xi}}.$

Now, we define the transport dc--tensors. Let open regions $U$ and ${%
\underline U}$ be homeomorphic to sphere ${\cal R}^{2n}$ and introduce
isomorphism ${\mu}_{{u},{\underline u}}$ between $T_{u}{\xi}$ and $%
T_{\underline u}{\underline {\xi}}$ (given by map $nc : U {\to} {\underline U%
}).$ We consider that for every d--vector $v^{\alpha} {\in} T_{u}{\xi}$
corresponds the vector ${\mu}_{{u},{\underline u}}
(v^{\alpha})=v^{\underline {\alpha}} {\in} T_{\underline u}{\underline {\xi}}%
,$ with components $v^{\underline {\alpha}}$ being linear functions of $%
v^{\alpha}$:
$$
v^{\underline {\alpha}}=h^{\underline {\alpha}}_{\alpha}(u, {\underline u})
v^{\alpha}, \quad v_{\underline {\alpha}}= h^{\alpha}_{\underline {\alpha}}({%
\underline u}, u)v_{\alpha},%
$$
where $h^{\alpha}_{\underline {\alpha}}({\underline u}, u)$ are the
components of dc--tensor associated with ${\mu}^{-1}_{u,{\underline u}}$. In
a similar manner we have
$$
v^{\alpha}=h^{\alpha}_{\underline {\alpha}}({\underline u}, u) v^{\underline
{\alpha}}, \quad v_{\alpha}=h^{\underline {\alpha}}_{\alpha} (u, {\underline
u})v_{\underline {\alpha}}.%
$$

In order to reconcile just presented definitions and to assure the identity
for trivial maps ${\xi }{\to }{\xi },u={\underline{u}},$ the transport
dc-tensors must satisfy conditions:
$$
h_\alpha ^{\underline{\alpha }}(u,{\underline{u}})h_{\underline{\alpha }%
}^\beta ({\underline{u}},u)={\delta }_\alpha ^\beta ,h_\alpha ^{\underline{%
\alpha }}(u,{\underline{u}})h_{\underline{\beta }}^\alpha ({\underline{u}}%
,u)={\delta }_{\underline{\beta }}^{\underline{\alpha }}
$$
and ${\lim }_{{({\underline{u}}{\to }u})}h_\alpha ^{\underline{\alpha }}(u,{%
\underline{u}})={\delta }_\alpha ^{\underline{\alpha }},\quad {\lim }_{{({%
\underline{u}}{\to }u})}h_{\underline{\alpha }}^\alpha ({\underline{u}},u)={%
\delta }_{\underline{\alpha }}^\alpha .$

Let ${\overline S}_{p} {\subset} {\overline U} {\subset} {\overline {\xi}}$
is a homeomorphic to $p$-dimensional sphere and suggest that chains of
na--maps are used to connect regions:
%%%%%%%%%%%%%%%%%%%%%%%%%%%%%%%%%%%%%%%%%%%%%%%%%%%%%%%%%%%%%%%%%
$$ U \buildrel nc_{(1)} \over \longrightarrow {\overline S}_p
     \buildrel nc_{(2)} \over \longrightarrow {\underline U}.$$

\begin{definition}
The tensor integral in ${\overline{u}}{\in }{\overline{S}}_p$ of a
dc--tensor $N_{{\varphi }{.}{\underline{\tau }}{.}{\overline{\alpha }}_1{%
\cdots }{\overline{\alpha }}_p}^{{.}{\gamma }{.}{\underline{\kappa }}}$ $({%
\overline{u}},u),$ completely antisymmetric on the indices ${{\overline{%
\alpha }}_1},{\ldots },{\overline{\alpha }}_p,$ over domain ${\overline{S}}%
_p,$ is defined as
$$
N_{{\varphi }{.}{\underline{\tau }}}^{{.}{\gamma }{.}{\underline{\kappa }}}({%
\underline{u}},u)=I_{({\overline{S}}_p)}^{\underline{U}}N_{{\varphi }{.}{%
\overline{\tau }}{.}{\overline{\alpha }}_1{\ldots }{\overline{\alpha }}_p}^{{%
.}{\gamma }{.}{\overline{\kappa }}}({\overline{u}},{\underline{u}})dS^{{%
\overline{\alpha }}_1{\ldots }{\overline{\alpha }}_p}=
$$
$$
{\int }_{({\overline{S}}_p)}h_{\underline{\tau }}^{\overline{\tau }}({%
\underline{u}},{\overline{u}})h_{\overline{\kappa }}^{\underline{\kappa }}({%
\overline{u}},{\underline{u}})N_{{\varphi }{.}{\overline{\tau }}{.}{%
\overline{\alpha }}_1{\cdots }{\overline{\alpha }}_p}^{{.}{\gamma }{.}{%
\overline{\kappa }}}({\overline{u}},u)d{\overline{S}}^{{\overline{\alpha }}_1%
{\cdots }{\overline{\alpha }}_p},\eqno(91)
$$
where $dS^{{\overline{\alpha }}_1{\cdots }{\overline{\alpha }}_p}={\delta }%
u^{{\overline{\alpha }}_1}{\land }{\cdots }{\land }{\delta }u_p^{\overline{%
\alpha }}$.
\end{definition}

Let suppose that transport dc--tensors $h_\alpha ^{\underline{\alpha }}$~
and $h_{\underline{\alpha }}^\alpha $~ admit covariant derivations of
or\-der two and pos\-tu\-la\-te ex\-is\-ten\-ce of de\-for\-ma\-ti\-on
dc--ten\-sor\\ $B_{{\alpha }{\beta }}^{{..}{\gamma }}(u,{\underline{u}})$~
satisfying relations
$$
D_\alpha h_\beta ^{\underline{\beta }}(u,{\underline{u}})=B_{{\alpha }{\beta
}}^{{..}{\gamma }}(u,{\underline{u}})h_\gamma ^{\underline{\beta }}(u,{%
\underline{u}})\eqno(92)
$$
and, taking into account that $D_\alpha {\delta }_\gamma ^\beta =0,$

$$
D_\alpha h_{\underline{\beta }}^\beta ({\underline{u}},u)=-B_{{\alpha }{%
\gamma }}^{{..}{\beta }}(u,{\underline{u}})h_{\underline{\beta }}^\gamma ({%
\underline{u}},u).
$$
By using formulas for torsion and, respectively, curvature of connection ${%
\Gamma }_{{\beta }{\gamma }}^\alpha $~ we can calculate next commutators:
$$
D_{[{\alpha }}D_{{\beta }]}h_\gamma ^{\underline{\gamma }}=-(R_{{\gamma }{.}{%
\alpha }{\beta }}^{{.}{\lambda }}+T_{{.}{\alpha }{\beta }}^\tau B_{{\tau }{%
\gamma }}^{{..}{\lambda }})h_\lambda ^{\underline{\gamma }}.\eqno(93)
$$
On the other hand from (92) one follows that
$$
D_{[{\alpha }}D_{{\beta }]}h_\gamma ^{\underline{\gamma }}=(D_{[{\alpha }}B_{%
{\beta }]{\gamma }}^{{..}{\lambda }}+B_{[{\alpha }{|}{\tau }{|}{.}}^{{..}{%
\lambda }}B_{{\beta }]{\gamma }{.}}^{{..}{\tau }})h_\lambda ^{\underline{%
\gamma }},\eqno(94)
$$
where ${|}{\tau }{|}$~ denotes that index ${\tau }$~ is excluded from the
action of antisymmetrization $[{\quad }]$. From (93) and (94) we obtain
$$
D_{[{\alpha }}B_{{\beta }]{\gamma }{.}}^{{..}{\lambda }}+B_{[{\beta }{|}{%
\gamma }{|}}B_{{\alpha }]{\tau }}^{{..}{\lambda }}=(R_{{\gamma }{.}{\alpha }{%
\beta }}^{{.}{\lambda }}+T_{{.}{\alpha }{\beta }}^\tau B_{{\tau }{\gamma }}^{%
{..}{\lambda }}).\eqno(95)
$$

Let ${\overline{S}}_p$~ be the boundary of ${\overline{S}}_{p-1}$. The
Stoke's type formula for tensor--integral (91) is defined as
$$
I_{{\overline{S}}_p}N_{{\varphi }{.}{\overline{\tau }}{.}{\overline{\alpha }}%
_1{\ldots }{\overline{\alpha }}_p}^{{.}{\gamma }{.}{\overline{\kappa }}}dS^{{%
\overline{\alpha }}_1{\ldots }{\overline{\alpha }}_p}=
I_{{\overline{S}}_{p+1}}{^{{\star }{(p)}}{\overline{D}}}_{[{\overline{\gamma
}}{|}}N_{{\varphi }{.}{\overline{\tau }}{.}{|}{\overline{\alpha }}_1{\ldots }%
{{\overline{\alpha }}_p]}}^{{.}{\gamma }{.}{\overline{\kappa }}}dS^{{%
\overline{\gamma }}{\overline{\alpha }}_1{\ldots }{\overline{\alpha }}_p},%
$$
where
$$
{^{{\star }{(p)}}D}_{[{\overline{\gamma }}{|}}N_{{\varphi }{.}{\overline{%
\tau }}{.}{|}{\overline{\alpha }}_1{\ldots }{\overline{\alpha }}_p]}^{{.}{%
\gamma }{.}{\overline{\kappa }}}=
$$
$$
D_{[{\overline{\gamma }}{|}}N_{{\varphi }{.}{\overline{\tau }}{.}{|}{%
\overline{\alpha }}_1{\ldots }{\overline{\alpha }}_p]}^{{.}{\gamma }{.}{%
\overline{\kappa }}}+pT_{{.}[{\overline{\gamma }}{\overline{\alpha }}_1{|}}^{%
\underline{\epsilon }}N_{{\varphi }{.}{\overline{\tau }}{.}{\overline{%
\epsilon }}{|}{\overline{\alpha }}_2{\ldots }{\overline{\alpha }}_p]}^{{.}{%
\gamma }{.}{\overline{\kappa }}}-B_{[{\overline{\gamma }}{|}{\overline{\tau }%
}}^{{..}{\overline{\epsilon }}}N_{{\varphi }{.}{\overline{\epsilon }}{.}{|}{%
\overline{\alpha }}_1{\ldots }{\overline{\alpha }}_p]}^{{.}{\gamma }{.}{%
\overline{\kappa }}}+B_{[{\overline{\gamma }}{|}{\overline{\epsilon }}}^{..{%
\overline{\kappa }}}N_{{\varphi }{.}{\overline{\tau }}{.}{|}{\overline{%
\alpha }}_1{\ldots }{\overline{\alpha }}_p]}^{{.}{\gamma }{.}{\overline{%
\epsilon }}}.
$$
We define the dual element of the hypersurfaces element $dS^{{j}_1{\ldots }{j%
}_p}$ as
$$
d{\cal S}_{{\beta }_1{\ldots }{\beta }_{q-p}}={\frac 1{{p!}}}{\epsilon }_{{%
\beta }_1{\ldots }{\beta }_{k-p}{\alpha }_1{\ldots }{\alpha }_p}dS^{{\alpha }%
_1{\ldots }{\alpha }_p},\eqno(96)
$$
where ${\epsilon }_{{\gamma }_1{\ldots }{\gamma }_q}$ is completely
antisymmetric on its indices and
$$
{\epsilon }_{12{\ldots }(n+m)}=\sqrt{{|}G{|}},G=det{|}G_{{\alpha }{\beta }{|}%
},
$$
$G_{{\alpha }{\beta }}$ is taken as the d--metric (5).
 The dual of dc--tensor $N_{{%
\varphi }{.}{\overline{\tau }}{.}{\overline{\alpha }}_1{\ldots }{\overline{%
\alpha }}_p}^{{.}{\gamma }{\overline{\kappa }}}$ is defined as the
dc--tensor  ${\cal N}_{{\varphi }{.}{\overline{\tau }}}^{{.}{\gamma }{.}{%
\overline{\kappa }}{\overline{\beta }}_1{\ldots }{\overline{\beta }}%
_{n+m-p}} $ satisfying
$$
N_{{\varphi }{.}{\overline{\tau }}{.}{\overline{\alpha }}_1{\ldots }{%
\overline{\alpha }}_p}^{{.}{\gamma }{.}{\overline{\kappa }}}={\frac 1{{p!}}}%
{\cal N}_{{\varphi }{.}{\overline{\tau }}}^{{.}{\gamma }{.}{\overline{\kappa
}}{\overline{\beta }}_1{\ldots }{\overline{\beta }}_{n+m-p}}{\epsilon }_{{%
\overline{\beta }}_1{\ldots }{\overline{\beta }}_{n+m-p}{\overline{\alpha }}%
_1{\ldots }{\overline{\alpha }}_p}.\eqno(97)
$$
Using (73), (96) and (97) we can write
$$
I_{{\overline{S}}_p}N_{{\varphi }{.}{\overline{\tau }}{.}{\overline{\alpha }}%
_1{\ldots }{\overline{\alpha }}_p}^{{.}{\gamma }{.}{\overline{\kappa }}}dS^{{%
\overline{\alpha }}_1{\ldots }{\overline{\alpha }}_p}={\int }_{{\overline{S}}%
_{p+1}}{^{\overline{p}}D}_{\overline{\gamma }}{\cal N}_{{\varphi }{.}{%
\overline{\tau }}}^{{.}{\gamma }{.}{\overline{\kappa }}{\overline{\beta }}_1{%
\ldots }{\overline{\beta }}_{n+m-p-1}{\overline{\gamma }}}d{\cal S}_{{%
\overline{\beta }}_1{\ldots }{\overline{\beta }}_{n+m-p-1}},\eqno(98)
$$
where
$$
{^{\overline{p}}D}_{\overline{\gamma }}{\cal N}_{{\varphi }{.}{\overline{%
\tau }}}^{{.}{\gamma }{.}{\overline{\kappa }}{\overline{\beta }}_1{\ldots }{%
\overline{\beta }}_{n+m-p-1}{\overline{\gamma }}}=
{\overline{D}}_{\overline{\gamma }}{\cal N}_{{\varphi }{.}{\overline{\tau }}%
}^{{.}{\gamma }{.}{\overline{\kappa }}{\overline{\beta }}_1{\ldots }{%
\overline{\beta }}_{n+m-p-1}{\overline{\gamma }}}+ $$ $$
(-1)^{(n+m-p)}(n+m-p+1)T_{{%
.}{\overline{\gamma }}{\overline{\epsilon }}}^{[{\overline{\epsilon }}}{\cal %
N}_{{\varphi }{.}{{\overline{\tau }}}}^{{.}{|}{\gamma }{.}{\overline{\kappa }%
}{|}{\overline{\beta }}_1{\ldots }{\overline{\beta }}_{n+m-p-1}]{\overline{%
\gamma }}}-
$$
$$
B_{{\overline{\gamma }}{\overline{\tau }}}^{{..}{\overline{\epsilon }}}{\cal %
N}_{{\varphi }{.}{\overline{\epsilon }}}^{{.}{\gamma }{.}{\overline{\kappa }}%
{\overline{\beta }}_1{\ldots }{\overline{\beta }}_{n+m-p-1}{\overline{\gamma
}}}+B_{{\overline{\gamma }}{\overline{\epsilon }}}^{{..}{\overline{\kappa }}}%
{\cal N}_{{\varphi }{.}{\overline{\tau }}}^{{.}{\gamma }{.}{\overline{%
\epsilon }}{\overline{\beta }}_1{\ldots }{\overline{\beta }}_{n+m-p-1}{%
\overline{\gamma }}}.
$$
To verify the equivalence of (97) and (98) we must take in consideration
that
$$
D_\gamma {\epsilon }_{{\alpha }_1{\ldots }{\alpha }_k}=0\ \mbox{and}\ {%
\epsilon }_{{\beta }_1{\ldots }{\beta }_{n+m-p}{\alpha }_1{\ldots }{\alpha }%
_p}{\epsilon }^{{\beta }_1{\ldots }{\beta }_{n+m-p}{\gamma }_1{\ldots }{%
\gamma }_p}=p!(n+m-p)!{\delta }_{{\alpha }_1}^{[{\gamma }_1}{\cdots }{\delta
}_{{\alpha }_p}^{{\gamma }_p]}.
$$
The developed tensor integration formalism will be used in the next
section for definition of conservation laws on spaces with local
anisotropy.

\section{On Conservation Laws on La--Spaces}

To define conservation laws on locally anisotropic spaces is a challenging
task because of absence of global and local groups of automorphisms of such
spaces. Our main idea is to use chains of na--maps from a given, called
hereafter as the fundamental, la--space to an auxiliary one with trivial
curvatures and torsions admitting a global group of automorphisms. The aim
of this section is to formulate conservation laws for la-gravitational
fields by using dc--objects and tensor--integral values, na--maps and
variational calculus on the Category of la--spaces.

\subsection{Nonzero divergence of the energy--mo\-men\-tum d--tensor}

R. Miron and M. Anastasiei
\cite{MironAnastasiei1987,MironAnastasiei1994}
 pointed to this specific form of
conservation laws of matter on la--spaces: They calculated the divergence of
the energy--momentum d--tensor on la--space $\xi ,$%
$$
D_\alpha {E}_\beta ^\alpha ={\frac 1{\ {\kappa }_1}}U_\alpha ,\eqno(99)
$$
and concluded that d--vector
$$
U_\alpha ={\frac 12}(G^{\beta \delta }{{R_\delta }^\gamma }_{\phi \beta }{T}%
_{\cdot \alpha \gamma }^\phi -G^{\beta \delta }{{R_\delta }^\gamma }_{\phi
\alpha }{T}_{\cdot \beta \gamma }^\phi +{R_\phi ^\beta }{T}_{\cdot \beta
\alpha }^\phi )
$$
vanishes if and only if d--connection $D$ is without torsion.

No wonder that conservation laws, in usual physical theories being a
consequence of global (for usual gravity of local) automorphisms of the
fundamental space--time, are more sophisticate on the spaces with local
anisotropy. Here it is important to emphasize the multiconnection character
of la--spaces. For example, for a d--metric (5) on $\xi $ we can
equivalently introduce another  metric linear connection $\tilde
D.$ The Einstein equations
$$
{\tilde R}_{\alpha \beta }-{\frac 12}G_{\alpha \beta }{\tilde R}={\kappa }_1{%
\tilde E}_{\alpha \beta }\eqno(100)
$$
constructed by using connection (80) have vanishing divergences
$$
{\tilde D}^\alpha ({{\tilde R}_{\alpha \beta }}-{\frac 12}G_{\alpha \beta }{%
\tilde R})=0\mbox{ and }{\tilde D}^\alpha {\tilde E}_{\alpha \beta }=0,
$$
similarly as those on (pseudo)Riemannian spaces. We conclude that by using
the connection ${\gamma}^{\alpha}_{\quad \beta \gamma}$
 we construct a model of la--gravity which looks like
locally isotropic on the total space $E.$ More general gravitational models
with local anisotropy can be obtained by using deformations of connection ${%
\tilde \Gamma }_{\cdot \beta \gamma }^\alpha ,$
$$
{{\Gamma }^\alpha }_{\beta \gamma }={\tilde \Gamma }_{\cdot \beta \gamma
}^\alpha +{P^\alpha }_{\beta \gamma }+{Q^\alpha }_{\beta \gamma },
$$
were, for simplicity, ${{\Gamma }^\alpha }_{\beta \gamma }$ is chosen to be
also metric and satisfy Einstein equations (100). We can consider
deformation d--tensors ${P^\alpha }_{\beta \gamma }$ generated (or not) by
deformations of type (66)--(68) for na--maps. In this case d--vector $%
U_\alpha $ can be interpreted as a generic source of local anisotropy on $%
\xi $ satisfying generalized conservation laws (99).

\subsection{\qquad
 Deformation d--tensors and tensor--integ\-ral con\-servation laws}

From (91) we obtain a tensor integral on ${\cal C}({\xi})$ of a d--tensor:
$$
N_{{\underline {\tau}}}^{{.}{\underline {\kappa}}}(\underline u)= I_{{%
\overline S}_{p}}N^{{..}{\overline {\kappa}}}_ {{\overline {\tau}}{..}{%
\overline {\alpha}}_{1}{\ldots}{\overline {\alpha}}_{p}} ({\overline u})h^{{%
\overline {\tau}}}_{{\underline {\tau}}}({\underline u}, {\overline u})h^{{%
\underline {\kappa}}}_{{\overline {\kappa}}} ({\overline u}, {\underline u}%
)dS^{{\overline {\alpha}}_{1}{\ldots} {\overline {\alpha}}_{p}}.%
$$

We point out that tensor--integral can be defined not only for dc--tensors
but and for d--tensors on $\xi $. Really, suppressing indices ${\varphi }$~
and ${\gamma }$~ in (97) and (98), considering instead of a deformation
dc--tensor a deformation tensor
$$
B_{{\alpha }{\beta }}^{{..}{\gamma }}(u,{\underline{u}})=B_{{\alpha }{\beta }%
}^{{..}{\gamma }}(u)=P_{{.}{\alpha }{\beta }}^\gamma (u)\eqno(101)
$$
(we consider deformations induced by a nc--transform) and integration\\ $%
I_{S_p}{\ldots }dS^{{\alpha }_1{\ldots }{\alpha }_p}$ in la--space $\xi $ we
obtain from (91) a tensor--integral on ${\cal C}({\xi })$~ of a d--tensor:
$$
N_{{\underline{\tau }}}^{{.}{\underline{\kappa }}}({\underline{u}}%
)=I_{S_p}N_{{\tau }{.}{\alpha }_1{\ldots }{\alpha }_p}^{.{\kappa }}(u)h_{{%
\underline{\tau }}}^\tau ({\underline{u}},u)h_\kappa ^{\underline{\kappa }%
}(u,{\underline{u}})dS^{{\alpha }_1{\ldots }{\alpha }_p}.
$$
Taking into account (95) we can calculate that curvature
$$
{\underline{R}}_{{\gamma }{.}{\alpha }{\beta }}^{.{\lambda }}=D_{[{\beta }%
}B_{{\alpha }]{\gamma }}^{{..}{\lambda }}+B_{[{\alpha }{|}{\gamma }{|}}^{{..}%
{\tau }}B_{{\beta }]{\tau }}^{{..}{\lambda }}+T_{{.}{\alpha }{\beta }}^{{%
\tau }{..}}B_{{\tau }{\gamma }}^{{..}{\lambda }}
$$
of connection ${\underline{\Gamma }}_{{.}{\alpha }{\beta }}^\gamma (u)={%
\Gamma }_{{.}{\alpha }{\beta }}^\gamma (u)+B_{{\alpha }{\beta }{.}}^{{..}{%
\gamma }}(u),$ with $B_{{\alpha }{\beta }}^{{..}{\gamma }}(u)$~ taken from
(101), vanishes, ${\underline{R}}_{{\gamma }{.}{\alpha }{\beta }}^{{.}{%
\lambda }}=0.$ So, we can conclude that la--space $\xi $ admits a tensor
integral structure on ${\cal {C}}({\xi })$ for d--tensors associated to
deformation tensor $B_{{\alpha }{\beta }}^{{..}{\gamma }}(u)$ if the
nc--image ${\underline{\xi }}$~ is locally parallelizable. That way we
generalize the one space tensor integral constructions in \cite{goz,gv,v295a}%
, were the possibility to introduce tensor integral structure on a curved
space was restricted by the condition that this space is locally
parallelizable. For $q=n+m$~ relations (98), written for d--tensor ${\cal N%
}_{\underline{\alpha }}^{{.}{\underline{\beta }}{\underline{\gamma }}}$ (we
change indices ${\overline{\alpha }},{\overline{\beta }},{\ldots }$ into ${%
\underline{\alpha }},{\underline{\beta }},{\ldots })$ extend the Gauss
formula on ${\cal {C}}({\xi })$:
$$
I_{S_{q-1}}{\cal N}_{\underline{\alpha }}^{{.}{\underline{\beta }}{%
\underline{\gamma }}}d{\cal S}_{\underline{\gamma }}=I_{{\underline{S}}_q}{^{%
\underline{q-1}}D}_{{\underline{\tau }}}{\cal N}_{{\underline{\alpha }}}^{{.}%
{\underline{\beta }}{\underline{\tau }}}d{\underline{V}},\eqno(102)
$$
where $d{\underline{V}}={\sqrt{{|}{\underline{G}}_{{\alpha }{\beta }}{|}}}d{%
\underline{u}}^1{\ldots }d{\underline{u}}^q$ and
$$
{^{\underline{q-1}}D}_{{\underline{\tau }}}{\cal N}_{\underline{\alpha }}^{{.%
}{\underline{\beta }}{\underline{\tau }}}=D_{{\underline{\tau }}}{\cal N}_{%
\underline{\alpha }}^{{.}{\underline{\beta }}{\underline{\tau }}}-T_{{.}{%
\underline{\tau }}{\underline{\epsilon }}}^{{\underline{\epsilon }}}{\cal N}%
_{{\underline{\alpha }}}^{{\underline{\beta }}{\underline{\tau }}}-B_{{%
\underline{\tau }}{\underline{\alpha }}}^{{..}{\underline{\epsilon }}}{\cal N%
}_{{\underline{\epsilon }}}^{{.}{\underline{\beta }}{\underline{\tau }}}+B_{{%
\underline{\tau }}{\underline{\epsilon }}}^{{..}{\underline{\beta }}}{\cal N}%
_{{\underline{\alpha }}}^{{.}{\underline{\epsilon }}{\underline{\tau }}}.%
\eqno(103)
$$

Let consider physical values $N_{{\underline{\alpha }}}^{{.}{\underline{%
\beta }}}$ on ${\underline{\xi }}$~ defined on its density ${\cal N}_{{%
\underline{\alpha }}}^{{.}{\underline{\beta }}{\underline{\gamma }}},$ i. e.
$$
N_{{\underline{\alpha }}}^{{.}{\underline{\beta }}}=I_{{\underline{S}}_{q-1}}%
{\cal N}_{{\underline{\alpha }}}^{{.}{\underline{\beta }}{\underline{\gamma }%
}}d{\cal S}_{{\underline{\gamma }}}\eqno(104)
$$
with this conservation law (due to (102)):%
$$
{^{\underline{q-1}}D}_{{\underline{\gamma }}}{\cal N}_{{\underline{\alpha }}%
}^{{.}{\underline{\beta }}{\underline{\gamma }}}=0.\eqno(105)
$$
We note that these conservation laws differ from covariant conservation laws
for well known physical values such as density of electric current or of
energy-- momentum tensor. For example, taking density ${E}_\beta ^{{.}{%
\gamma }},$ with corresponding to (103) and (105) conservation law,
$$
{^{\underline{q-1}}D}_{{\underline{\gamma }}}{E}_{{\underline{\beta }}}^{{%
\underline{\gamma }}}=D_{{\underline{\gamma }}}{E}_{{\underline{\beta }}}^{{%
\underline{\gamma }}}-T_{{.}{\underline{\epsilon }}{\underline{\tau }}}^{{%
\underline{\tau }}}{E}_{{\underline{\beta }}}^{{.}{\underline{\epsilon }}%
}-B_{{\underline{\tau }}{\underline{\beta }}}^{{..}{\underline{\epsilon }}}{E%
}_{\underline{\epsilon }}^{{\underline{\tau }}}=0,\eqno(106)
$$
we can define values (see (102) and (104))
$$
{\cal P}_\alpha =I_{{\underline{S}}_{q-1}}{E}_{{\underline{\alpha }}}^{{.}{%
\underline{\gamma }}}d{\cal S}_{{\underline{\gamma }}}.
$$
Defined conservation laws (106) for ${E}_{{\underline{\beta }}}^{{.}{%
\underline{\epsilon }}}$ have nothing to do with those for energy--momentum
tensor $E_\alpha ^{{.}{\gamma }}$ from Einstein equations for the almost
Hermitian gravity \cite{MironAnastasiei1987,MironAnastasiei1994}
 or with ${\tilde E}_{\alpha \beta }$ from
(100) with vanishing divergence $D_\gamma {\tilde E}_\alpha ^{{.}{\gamma }%
}=0.$ So ${\tilde E}_\alpha ^{{.}{\gamma }}{\neq }{E}_\alpha ^{{.}{\gamma }%
}. $ A similar conclusion was made in \cite{goz} for unispacial locally
isotropic tensor integral. In the case of multispatial tensor integration we
have another possibility (firstly pointed in \cite{vog,v295a} for
Einstein-Cartan spaces), namely, to identify ${E}_{{\underline{\beta }}}^{{.}%
{\underline{\gamma }}}$ from (106) with the na-image of ${E}_\beta ^{{.}{%
\gamma }}$ on la--space $\xi .$ We shall consider this construction in the
next section.

\section{Na--Conservation Laws in La--Gravity}

Let us consider a fixed background la--space $\underline {\xi}$ with given
metric ${\underline G}_{\alpha \beta} = ({\underline g}_{ij} , {\underline h}%
_{ab} )$ and d--connection ${\underline {\tilde {\Gamma}}}^{\alpha}_{\cdot
\beta\gamma}.$ For simplicity, we suppose that metric is
 compatible  and that connections are torsionless and with
 vanishing curvatures. Introducing  an
nc--transform from the fundamental la--space $\xi$ to an auxiliary one $%
\underline {\xi} $ we are interested in the equivalents of the Einstein
equations (100) on $\underline {\xi} .$

We suppose that a part of gravitational degrees of freedom is "pumped out"
into the dynamics of deformation d--tensors for d--connection, ${P^{\alpha}}%
_{\beta \gamma},$ and metric, $B^{\alpha \beta} = ( b^{ij} , b^{ab} ) .$ The
remained part of degrees of freedom is coded into the metric ${\underline G}%
_{\alpha \beta}$ and d--connection ${\underline {\tilde {\Gamma}}}%
^{\alpha}_{\cdot \beta \gamma} .$

Following \cite{gri,vrjp} we apply the first order formalism and consider $%
B^{\alpha \beta }$ and ${P^\alpha }_{\beta \gamma }$ as independent
variables on $\underline{\xi }.$ Using notations
$$
P_\alpha ={P^\beta }_{\beta \alpha },\quad {\Gamma }_\alpha ={{\Gamma }%
^\beta }_{\beta \alpha },
$$
$$
{\hat B}^{\alpha \beta }=\sqrt{|G|}B^{\alpha \beta },{\hat G}^{\alpha \beta
}=\sqrt{|G|}G^{\alpha \beta },{\underline{\hat G}}^{\alpha \beta }=\sqrt{|%
\underline{G}|}{\underline{G}}^{\alpha \beta }
$$
and making identifications
$$
{\hat B}^{\alpha \beta }+{\underline{\hat G}}^{\alpha \beta }={\hat G}%
^{\alpha \beta },{\quad }{\underline{\Gamma }}_{\cdot \beta \gamma }^\alpha -%
{P^\alpha }_{\beta \gamma }={{\Gamma }^\alpha }_{\beta \gamma },
$$
we take the action of la--gravitational field on $\underline{\xi }$ in this
form:
$$
{\underline{{\cal S}}}^{(g)}=-{(2c{\kappa }_1)}^{-1}\int {\delta }^qu{}{%
\underline{{\cal L}}}^{(g)},\eqno(107)
$$
where
$$
{\underline{{\cal L}}}^{(g)}={\hat B}^{\alpha \beta }(D_\beta P_\alpha
-D_\tau {P^\tau }_{\alpha \beta })+({\underline{\hat G}}^{\alpha \beta }+{%
\hat B}^{\alpha \beta })(P_\tau {P^\tau }_{\alpha \beta }-{P^\alpha }%
_{\alpha \kappa }{P^\kappa }_{\beta \tau })
$$
and the interaction constant is taken ${\kappa }_1={\frac{4{\pi }}{{c^4}}}k,{%
\quad }(c$ is the light constant and $k$ is Newton constant) in order to
obtain concordance with the Einstein theory in the locally isotropic limit.

We construct on $\underline{\xi }$ a la--gravitational theory with matter
fields (denoted as ${\varphi }_A$ with $A$ being a general index)
interactions by postulating this Lagrangian density for matter fields
$$
{\underline{{\cal L}}}^{(m)}={\underline{{\cal L}}}^{(m)}[{\underline{\hat G}%
}^{\alpha \beta }+{\hat B}^{\alpha \beta };{\frac \delta {\delta u^\gamma }}(%
{\underline{\hat G}}^{\alpha \beta }+{\hat B}^{\alpha \beta });{\varphi }_A;{%
\frac{\delta {\varphi }_A}{\delta u^\tau }}].\eqno(108)
$$

Starting from (107) and (108) the total action of la--gravity on $%
\underline{\xi }$ is written as
$$
{\underline{{\cal S}}}={(2c{\kappa }_1)}^{-1}\int {\delta }^qu{\underline{%
{\cal L}}}^{(g)}+c^{-1}\int {\delta }^{(m)}{\underline{{\cal L}}}^{(m)}.%
\eqno(109)
$$
Applying variational procedure on $\underline{\xi },$ similar to that
presented in \cite{gri} but in our case adapted to N--connection by using
derivations (3) instead of partial derivations (1), we derive from (109) the
la--gravitational field equations
$$
{\bf {\Theta }}_{\alpha \beta }={{\kappa }_1}({\underline{{\bf t}}}_{\alpha
\beta }+{\underline{{\bf T}}}_{\alpha \beta })\eqno(110)
$$
and matter field equations
$$
{\frac{{\triangle }{\underline{{\cal L}}}^{(m)}}{\triangle {\varphi }_A}}=0,%
\eqno(111)
$$
where $\frac{\triangle }{\triangle {\varphi }_A}$ denotes the variational
derivation.

In (110) we have introduced these values: the energy--momentum d--tensor
for la--gravi\-ta\-ti\-on\-al field
$$
{\kappa }_1{\underline{{\bf t}}}_{\alpha \beta }=({\sqrt{|G|}})^{-1}{\frac{%
\triangle {\underline{{\cal L}}}^{(g)}}{\triangle G^{\alpha \beta }}}%
=K_{\alpha \beta }+{P^\gamma }_{\alpha \beta }P_\gamma -{P^\gamma }_{\alpha
\tau }{P^\tau }_{\beta \gamma }+
$$
$$
{\frac 12}{\underline{G}}_{\alpha \beta }{\underline{G}}^{\gamma \tau }({%
P^\phi }_{\gamma \tau }P_\phi -{P^\phi }_{\gamma \epsilon }{P^\epsilon }%
_{\phi \tau }),\eqno(112)
$$
(where
$$
K_{\alpha \beta }={\underline{D}}_\gamma K_{\alpha \beta }^\gamma ,
$$
$$
2K_{\alpha \beta }^\gamma =-B^{\tau \gamma }{P^\epsilon }_{\tau (\alpha }{%
\underline{G}}_{\beta )\epsilon }-B^{\tau \epsilon }{P^\gamma }_{\epsilon
(\alpha }{\underline{G}}_{\beta )\tau }+
$$
$$
{\underline{G}}^{\gamma \epsilon }h_{\epsilon (\alpha }P_{\beta )}+{%
\underline{G}}^{\gamma \tau }{\underline{G}}^{\epsilon \phi }{P^\varphi }%
_{\phi \tau }{\underline{G}}_{\varphi (\alpha }B_{\beta )\epsilon }+{%
\underline{G}}_{\alpha \beta }B^{\tau \epsilon }{P^\gamma }_{\tau \epsilon
}-B_{\alpha \beta }P^\gamma {\quad }),
$$
$$
2{\bf \Theta }={\underline{D}}^\tau {\underline{D}}_{tau}B_{\alpha \beta }+{%
\underline{G}}_{\alpha \beta }{\underline{D}}^\tau {\underline{D}}^\epsilon
B_{\tau \epsilon }-{\underline{G}}^{\tau \epsilon }{\underline{D}}_\epsilon {%
\underline{D}}_{(\alpha }B_{\beta )\tau }
$$
and the energy--momentum d--tensor of matter
$$
{\underline{{\bf T}}}_{\alpha \beta }=2{\frac{\triangle {\cal L}^{(m)}}{%
\triangle {\underline{\hat G}}^{\alpha \beta }}}-{\underline{G}}_{\alpha
\beta }{\underline{G}}^{\gamma \delta }{\frac{\triangle {\cal L}^{(m)}}{%
\triangle {\underline{\hat G}}^{\gamma \delta }}}.\eqno(113)
$$
As a consequence of (111)--(113) we obtain the d--covariant on $\underline{%
\xi }$ conservation laws
$$
{\underline{D}}_\alpha ({\underline{{\bf t}}}^{\alpha \beta }+{\underline{%
{\bf T}}}^{\alpha \beta })=0.\eqno(114)
$$
We have postulated the Lagrangian density of matter fields (108) in a form
as to treat ${\underline{{\bf t}}}^{\alpha \beta }+{\underline{{\bf T}}}%
^{\alpha \beta }$ as the source in (110).

Now we formulate the main results of this section:

\begin{proposition}
The dynamics of the Einstein la--gravitational fields, modeled as solutions
of equations (100) and matter fields on la--space $\xi ,$ can be
equivalently locally modeled on a background la--space $\underline{\xi }$
provided with a trivial d-connection and metric structures having zero
d--tensors of torsion and curvature by field equations (110) and (111) on
condition that deformation tensor ${P^\alpha }_{\beta \gamma }$ is a
solution of the Cauchy problem posed for basic equations for a chain of
na--maps from $\xi $ to $\underline{\xi }.$
\end{proposition}

\begin{proposition}
Local, d--tensor, conservation laws for Einstein la--gravita\-ti\-on\-al
fields can be written in form (114) for la--gravita\-ti\-on\-al (112) and
matter (113) energy--momentum d--tensors. These laws are d--covariant on
the background space $\underline{\xi }$ and must be completed with invariant
conditions of type (69)--(72) for every deformation parameters of a
chain of na--maps from $\xi $ to $\underline{\xi }.$
\end{proposition}

The above presented considerations consist proofs of both propositions.

We emphasize that nonlocalization of both locally an\-i\-sot\-rop\-ic and
isot\-rop\-ic gravitational energy--momentum values on the fundamental
(locally an\-i\-sot\-rop\-ic or isotropic) space $\xi $ is a consequence of
the absence of global group automorphisms for generic curved spaces.
Considering gravitational theories from view of multispaces and their mutual
maps (directed by the basic geometric structures on $\xi $ such as
N--connection, d--connection, d--torsion and d--curvature components, see
coefficients for basic na--equations (66)--(68)), we can formulate local
d--tensor conservation laws on auxiliary globally automorphic spaces being
related with space $\xi $ by means of chains of na--maps. Finally, we remark
that as a matter of principle we can use d--connection deformations in order
to modelate the la--gravitational interactions with nonvanishing torsion and
nonmetricity. In this case we must introduce a corresponding source in
(114) and define generalized conservation laws as in (99) \quad (see
similar details for locally isotropic generalizations of the Einstein
gravity in Refs \cite{vob13,vog,vb12}).

\section{Concluding Remarks}

In this paper we have reformulated the fiber bundle formalism for both
Yang-Mills and gravitational fields in order to include into consideration
space-times with higher order anisotropy. We have argued that our approach
has the advantage of making manifest the relevant structures of the theories
with local anisotropy and putting greater emphasis on the analogy with
anisotropic models than the standard coordinate formulation in Finsler
geometry and on higher dimension (Kaluza--Klein) spaces.

Our  models of higher order anisotropic gauge and gravitational
interactions are refined in such a way as to shed light on some of the more
specific properties and common and distinguishing features of the Yang-Mills
and Einstein higher order anisotropic fields.  As we have shown,
 it is possible  a gauge
like treatment for both models with local anisotropy (by using
correspondingly defined linear connections in bundle spaces with semisimple
structural groups, with variants of nonlinear realization and extension to
semisimple structural groups, for gravitational fields).

Another main results of this paper are the formulation of the theory of
 nearly autoparallel maps of locally  anisotropic spaces and a corresponding
 classification of such type spaces by using chains of nearly autoparallel
 maps (generalizing the class of conformal transforms). We have also
 analyzed in detail  two variants of solution of the problem of formulation
 of conservation laws for field interactions with local anisotropy.

 Finally, we emphasize that
 there are various possible developments of the ideas presented here.
 For instance, we  point to a possible  extension of the Ashtekar
 approch to gravity
 \cite{Ashtekar}  for  higher order an\-isot\-rop\-ic Kaluza--Klein models,
 possible applications of exact solutions for la--gravity in modern
  cosmology and astrophysics \cite{lodz,vmir},
 as well to a study
  of an\-isot\-rop\-ic low energy limits of string theories
 and  a generalization functional integration
 formalism for locally an\-isot\-rop\-ic theories.
 Such problems will require our attention in the future.

\vskip10pt
{\bf Acknoweledgments}
\vskip5pt
The author is grateful to Igor Kanatchikov for valuabe discussions
and support during his visit to Warsaw.

%\newpage

\end{document}